\documentclass[a4paper,fleqn]{cas-dc}
\usepackage{listings}
\usepackage{xcolor}
\usepackage{tikz}
\usepackage{flafter}
\usetikzlibrary{positioning, shapes.multipart, arrows.meta, decorations.pathreplacing, calc, automata}
\usepackage{amsmath, amssymb, url}

\usetikzlibrary{shapes.geometric, arrows.meta}

\usepackage{colortbl}

\usepackage{array}
\usepackage{graphicx}

\usepackage{pifont} 

\usepackage{colortbl}

\newcommand{\cmark}{\textcolor{green!50!black}{\ding{51}}} 
\newcommand{\xmark}{\textcolor{red!70!black}{\ding{55}}}   

\usepackage{algorithm}         
\usepackage{algpseudocode}     

\usepackage{pgfplots}
\pgfplotsset{compat=1.18} 
\usepackage{mathptmx} 
\pgfplotsset{
    every axis/.append style={
        font=\normalsize\rmfamily,
        tick label style={font=\normalsize\rmfamily},
        label style={font=\normalsize\rmfamily},
        title style={font=\normalsize\rmfamily},
        legend style={font=\normalsize\rmfamily},
        colorbar style={font=\normalsize\rmfamily}
    }
}
\usepackage{pgfplotstable}
\usepackage{array}
\usetikzlibrary{matrix, calc}
\usepgfplotslibrary{colorbrewer}
\pgfplotsset{compat = 1.15, cycle list/Set1-8} 
\usetikzlibrary{pgfplots.statistics, pgfplots.colorbrewer, plotmarks, arrows.meta, automata, positioning}

\lstdefinestyle{customsolidity}{
    language=Solidity,
    basicstyle=\ttfamily\small, 
    keywordstyle=\color{black},    
    commentstyle=\color{gray},    
    stringstyle=\color{purple},    
    numbers=left,                 
    numberstyle=\tiny\color{gray}, 
    stepnumber=1,                 
    numbersep=5pt,                
    frame=single,                 
    showspaces=false,             
    showstringspaces=false,       
    tabsize=2                     
}

\usepackage[authoryear,longnamesfirst]{natbib}
\usepackage{caption}
\usepackage{float}
\restylefloat{table}
\usepackage{listings}
\usepackage{color}
\usepackage{hyperref}
\usepackage{enumitem}

\hypersetup{hidelinks}

\lstdefinestyle{solidity}{
  language=Java,
  morekeywords={pragma, solidity, contract, mapping, public, event, function, returns, require, emit, address, uint256, string, view, internal, if, else, for, constructor},
  keywordstyle=\color{black}\bfseries,
  commentstyle=\color{gray},
  stringstyle=\color{red},
  basicstyle=\ttfamily\small,
  breaklines=true,
  frame=tb,
  columns=fullflexible,
}

\begin{document}
\let\WriteBookmarks\relax
\def\floatpagepagefraction{0.8}  
\def\textpagefraction{0.2}       
\newcommand{\red}[1]{{\color{red}#1}}
\newcommand{\blue}[1]{{\color{black}#1}}
\newcommand{\magenta}[1]{{\color{magenta}#1}}
\shorttitle{Securing the Future of Peer-to-Peer Network with Blockchain}
\shortauthors{S. Ghosh et~al.}

\title [mode = title]{Incentive-Driven Access Control Framework with Smart Contracts for Secure P2P Networks}

\author[1]{Saurav Ghosh}
\author[1]{Batuhan Tanrikulu}
\author[1]{Reshmi Mitra}
\author[1]{Indranil Roy}
\author[2]{Bidyut Gupta}

\cormark[1]
\fnmark[1]

\affiliation[1]{organization={Southeast Missouri State University},
                city={Cape Girardeau},
                country={USA}}
\affiliation[2]{organization={Southern Illinois University},
                city={Carbondale},
                country={USA}}

\begin{abstract}
Access control in P2P network is extremely challenging due to several reasons such as decentrailized network, anonymous connectivity, high churn, resource constraints and large attack surface. The number of attributes to be tracked grows exponentially as P2P networks scale, exacerbating  issues in traditional approaches such as RBAC, ABAC, RuBAC. 
This paper presents a novel blockchain-based access control framework for P2P networks using Ethereum smart contracts to offer an adaptable, decentralized, and scalable solution. Our token economy framework incentivizes nodes that demonstrate consistent compliance and penalize malicious actions using fungible ERC-20 tokens. Hierarchical static role-based structure consisting of regular members, primary and secondary group is the foundational layer of role management. Dynamic state management through a finite state machine classifies peers as benign, suspicious, or malicious mapping their behavior to the STRIDE threat model. We have used standard blockchain tool stack consisting of OpenZeppelin (contract library), Mocha (testing), and Web3.js (deployment). 
Our framework is tested with eight test cases, such as privilege escalation and denial-of-service attacks, and built using three types of Ethereum smart contracts. Our focus is on collecting four metrics gas cost, latency, execution time, simultaneous requests that provides a comprehensive view of realistic network activity. This study goes beyond conceptual nature of state-of-art
architecture to showcase the realistic mapping between
resources, peer roles and standard threats in a P2P network. The experimental results shows low gas costs (upto 1.4 million gas) and fast execution times (80–550 ms), confirming the design efficiency. The framework handles up to five concurrent requests with minimal performance impact, demonstrating strong scalability across roles and threat conditions. This framework enables scalable, behavior-driven governance in P2P networks, enhancing IoT and Web3 security. 
\end{abstract}


\maketitle

\section{Introduction} \label{Sec:introduction}
Access control policies regulate who can access certain resources, data, or services within a system. It involves both authentication (verifying identity) and authorization (granting permissions based on identity or other factors). In any network, especially in peer-to-peer (P2P) systems, access control is crucial for securing resources and preventing unauthorized interactions. For the diverse and rampant security threats in P2P networks such as Sybil attacks, man-in-the-middle attacks, and data breaches, the need for effective access control is critical~\cite{maesa2019blockchain, Xu2021, ding2019novelabac}. Moreover, with multiple peers acting as both clients and servers, there is a significant expansion in the attack surface~\cite{Wang2017}. 
In addition, variable levels of control ranging from lenient access for low-risk interactions to stringent controls for high-risk or sensitive operations are crucial. However, the limited resources of lightweight peer devices often make it difficult to orchestrate and deploy complex access policies, requiring efficient mechanisms that balance security with resource constraints \cite{yaqoob2017iotarchitecture,palattella2016iot5g,Lin2017,Chopra2019}. 

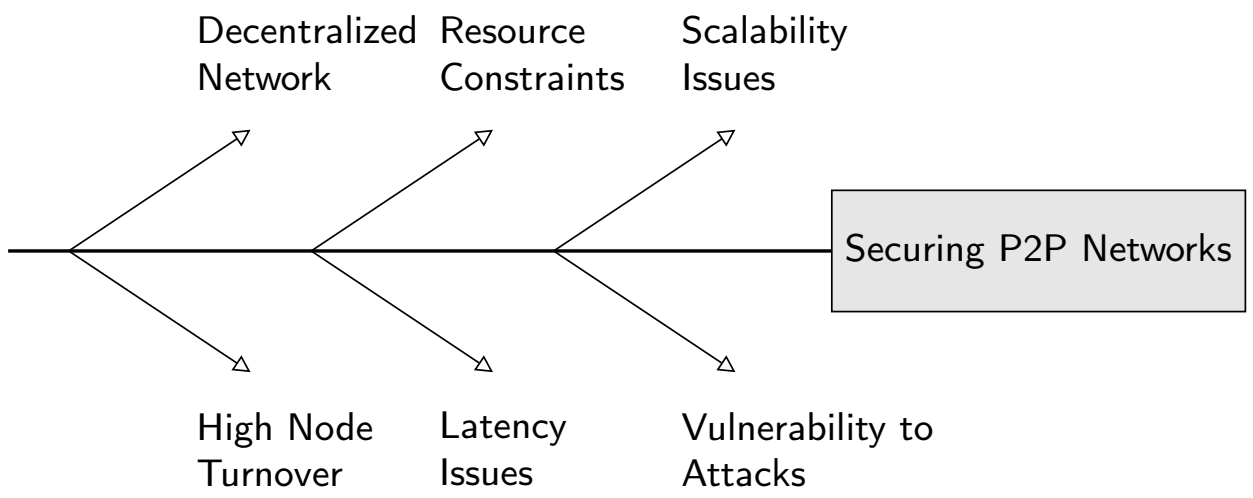
\begin{figure}[h]
    \centering
    \resizebox{\columnwidth}{!}{ 
        \begin{tikzpicture}[
            >=latex,
            node distance=2cm and 2cm,
            every node/.style={font=\sffamily},
            fishbone/.style={thick},
            cause/.style={thin, -{Triangle[open,length=4pt]}},
            branch/.style={anchor=north, text width=3cm, align=left, font=\small\sffamily} 
        ]

        \draw[fishbone] (-5.5,0) -- (3,0);

        \draw[cause] (-5,0) -- ++(1.5,1) node[branch, pos=1, anchor=south, xshift=30pt,  yshift=6pt] {Decentralized \\ Network};
        \draw[cause] (-3,0) -- ++(1.5,1) node[branch, pos=1, anchor=south, xshift=30pt,  yshift=6pt] {Resource \\ Constraints};
        \draw[cause] (-1,0) -- ++(1.5,1) node[branch, pos=1, anchor=south, xshift=30pt, yshift=6pt] {Scalability \\ Issues};

        \draw[cause] (-5,0) -- ++(1.5,-1) node[branch, pos=1, anchor=north, xshift=30pt, yshift=-6pt] {High Node \\ Turnover};
        \draw[cause] (-3,0) -- ++(1.5,-1) node[branch, pos=1, anchor=north, xshift=30pt, yshift=-6pt] {Latency \\ Issues};
        \draw[cause] (-1,0) -- ++(1.5,-1) node[branch, pos=1, anchor=north, xshift=30pt, yshift=-6pt] {Vulnerability to \\ Attacks};

        \node[draw, rectangle, minimum width=3cm, minimum height=1cm, fill=gray!20] at (3,0) {Securing P2P Networks};

        \end{tikzpicture}
    }
    \caption{Challenges in Securing P2P Networks}
    \label{fig:fishbone}
\end{figure}

Managing access control in P2P networks is particularly challenging due to the limitations of traditional models such as role-based access control (RBAC), attribute-based access control (ABAC), and rule-based access control (RuBAC)~\cite{Novo2018,Zhang2018}. The key motivating factors are summarized in Figure \ref{fig:fishbone}. RBAC relies on predefined, static roles that are impractical in high-churn P2P environment~\cite{sandhu1998role,yavari2017scalable}. It leads to \textit{role explosion} with the proliferation of different scenarios with respective trust levels, application and device types. Moreover, this rigid structure compromises anonymity since role definitions require identifiable information~\cite{Cruz2018}.

ABAC assigns flexible access based on dynamic attributes, but managing attributes in real-time across a evolving distributed network is inefficient \cite{Ma2019}. Moreover, attributes linking back to individual peers creates a potential privacy vulnerability~\cite{Reyna2018,Hu2018BlockchainSC}. RuBAC requires predefined rules that must be consistently updated and enforced~\cite{Novo2018}. As the network interaction grows, RuBAC suffers from \textit{rule explosion} leading to deployment overhead and bandwidth throttling~\cite{ding2019novelabac}. Mandatory Access Control (MAC) enforces strict, centrally defined policies that are difficult to adapt in the distributed systems context~\cite{Dorri2017}. Although Discretionary Access Control (DAC) allows individual peers to set access policies, it can lead to inconsistent enforcement~\cite{Xu2021}. Both MAC and DAC models struggle with scalability, as lightweight devices lack the resources to enforce such rigid, policy-driven frameworks in real-time \cite{Chopra2019}.

Access control enforcement in lightweight devices is a hard problem primarily due to the lack the computational power to handle complex policy evaluations \cite{Xu2021}. Secondly, the high churn rate in the network leads to performance degradation, increased latency, and security gaps \cite{ding2019novelabac}. Scalability further exacerbates these challenges, as lightweight devices struggle to enforce complex dynamic policies under real-time security threats. The continuous evaluation of policies on resource-constrained peers introduces significant overhead, making traditional access control models impractical for large-scale, dynamic P2P networks \cite{Novo2018}.

\textit{Overall, lightweight and adaptive solutions are required to ensure secure, scalable, and anonymous connectivity without overwhelming limited device resources.} 
Past works~\cite{ning2014aggregated, bera2021, han2022}, have focused on isolated aspects of blockchain-based solutions, emphasizing features on decentralized trust model, immutability and ability to enforce transparent, verifiable policies. However, there is a lack of detailed implementation, reproducible code, or rigorous security evaluations~\cite{ding2019novelabac, Novo2018}.

Our primary objective is to develop a  realistic blockchain-based access control for P2P network that \textit{incentivizes consistent compliance and penalize malicious behavior}. This level of network activity is represented by benign, suspicious, and malicious nodes in the state transition machine that is mapped to the STRIDE threat model.  Our Web 3.0 token-based system utilizes a hierarchical Ethereum smart contract architecture to manage regular operations and mitigate threats, thereby maintaining network security and privacy. 

The network architecture handles scalability by creating subgroups. Each group has an assigned roles of Primary Group Head (PGH), Secondary Group Head (SGH), or Regular Member (RM) based on their status~\cite{roy2022efficient}. Each role comes with a corresponding allocation of tokens, reflecting the responsibilities associated with that role. Dynamic policy enforcement is carried out through series of smart contracts, which verify the identity and permissions of peers attempting to access the system. If a request is authorized, it is approved and recorded on the blockchain. Unauthorized actions are detected and result in penalties, maintaining the integrity of the system. The system also includes continuous behavior monitoring and state transitions to present realistic threat scenario. We use Byzantine consensus algorithm to accept/reject access rights across peers.

Compliant and beneficial behavior is rewarded with additional tokens, while state upgrades are granted to recognize positive contributions. Based on the STRIDE threat model, violations lead to token deductions and stricter access restrictions. Although past work (~\cite{Zhang2018, bera2021, han2022, wang2021, ning2014aggregated}) have used hierarchical smart contract architecture, our work significantly contributes to the detailed implementation of an ERC-20\footnote{\url{https://ethereum.org/en/developers/docs/standards/tokens/erc-20/}} token-based incentive system. We have used Solidity to write smart contracts, where OpenZeppelin\footnote{\url{https://openzeppelin.com/}} provided pre-tested libraries for secure ERC-20 token management. Truffle\footnote{\url{https://trufflesuite.com/}} has been used to streamline contract development, deployment, and testing, while Ganache\footnote{\url{https://trufflesuite.com/ganache/}} was utilized for multiple accounts scenarios. Remix IDE\footnote{\url{https://remix.ethereum.org/}} was employed for prototyping and debugging, and testing is structured with Mocha\footnote{\url{https://mochajs.org/}} and Chai\footnote{\url{https://www.chaijs.com/}}. Automated deployment and interaction scripts are built in Node.js\footnote{\url{https://nodejs.org/en}} and Web3.js\footnote{\url{https://docs.web3js.org/}}.

Our framework is implemented with three types of Ethereum smart contracts, for a total of seven contract deployments. We are rigorously testing with eight realistic scenario-based test cases, including privilege escalation and denial-of-service (DoS) attacks. State-of-art research focuses on analyzing few metric groups in isolation. We are presenting a unified view with four critical metrics that essential for identifying performance bottlenecks in blockchain-based systems. They are listed as: \textit{gas cost, execution time, latency, and simultaneous request handling}. This study goes beyond conceptual nature of state-of-art architecture to showcase the realistic mapping between resources, peer roles and standard threats in a P2P network.

Gas cost is a key indicator of computational efficiency on Ethereum and is analyzed across routine and defense scenarios, even under simultaneous requests. The cost remains low (1.3 to 1.4 million gas units) even for intensive operations such as defense with penalizing nodes. The execution time, ranging from 80 to 550 milliseconds depending on the role and policy complexity, reflects the responsiveness of our smart contract logic. Scalability is assessed through the system ability to handle up to five concurrent access requests without performance degradation, highlighting its robustness under load. We have validated on 10, 25, and 100 peer nodes without compromising performance. Latency captures the real-time viability of access control enforcement, a crucial factor in network responsiveness and overall P2P performance. Finally, our comprehensive sensitivity analysis on latency across roles, defense states, and concurrent operations validates the effectiveness of our token-based incentivization mechanism.

Our framework uses a simple hierarchical structure and modular design to ensure low resource usage, making it suitable for devices with limited capacity. By efficiently managing access through role-based assignments (PGH–SGH–RM) and dynamically adapting to node behavior using token-based incentives, thereby ensuring both responsiveness and resilience. It supports real-time policy enforcement without compromising performance, even under simultaneous requests or defense-triggering conditions. The depth of its evaluation ranging from gas cost, execution time, latency, and scalability distinguishes it from prior work, which often lacks such implementation-backed analysis. Overall, this is a fully operational smart contract framework for decentralized access control that combines a structured hierarchy with incentive-driven mechanisms, offering a practical foundation for emerging blockchain-centric security solutions. 
The main contributions of the paper are:
\begin{itemize}
\item We present a realistic blockchain based access control framework for P2P network that incentivizes compliant behavior and penalizes malicious activity for dynamic policy enforcement. 
\item We design a Web 3.0 system that uses a hierarchical Ethereum smart contract architecture leveraging ERC-20 fungible token allocations and on-chain verification. 
\item We integrate STRIDE threat model to represent variable levels of malicious behavior with 8 expandable test-cases. 
\item We present complete tool stack and detailed code snippets to ensure complete trace-ability. 
\item We demonstrate continuous behavior monitoring with four key performance metrics: gas cost, latency, execution time and simultaneous requests to highlight normal, attack and defense scenario. 
\end{itemize}

The remainder of this paper is organized as follows: Section~\ref{Sec:related_work} reviews prior work in access control solutions. Section~\ref{Sec:system_arch} showases preliminary informaiton about the system architecture. Implementation details are presented in Section~\ref{Sec:sys_engr}. We are presenting the evaluation methodology and result visualizations in Section~\ref{Sec:perf_eval}. Finally, Section~\ref{Sec:conclusion} concludes the paper with the key findings.

\section{Related Work} \label{Sec:related_work}

\subsection{Access Control Mechanisms in IoT}

We begin with a brief overview of IoT security challenges, highlighting decentralization, resource constraints, and scalability-induced throttling. In contrast, P2P networks emphasize privacy-preserving data sharing and file distribution. Section~\ref{Sec:introduction} outlines the limitations of traditional access control in decentralized, high-churn, and resource-constrained environments, drawing parallels between IoT and P2P systems.

The rise of communication technologies such as Wi-Fi, ZigBee, and Bluetooth has allowed widespread connectivity across devices such as sensors, actuators, and smart gadgets leading to hyper-connected IoT paradigm in~\cite{yaqoob2017iotarchitecture}. The ecosystem has powered innovative applications in smart healthcare, intelligent transportation, and home automation \cite{xu2014iotindustries}. Attackers often exploit this huge attack surface to gain unauthorized data access and misuse the computational services \cite{darazio2017dataexfiltration, bertino2017botnets}.

A comprehensive survey article by~\cite{ahsan2025comprehensive} on the access control mechanisms in IoT recommends that conventional models set access rights based on roles, attributes, capabilities, usages, trusts, risks, protocols, and relations. In earliest works on RBAC~\cite{sandhu1998role}, permissions are assigned based on user roles, creating clear and easy-to-manage relationships between users and IoT resources. This simplicity makes RBAC popular, especially with the increasing number of IoT devices. Recent research has improved RBAC by introducing frameworks specifically designed for secure data handling in IoT~\cite{rbac_data_processing_2024}. Additionally, priority-based RBAC has been developed, proving beneficial in sensitive areas such as medical IoT systems \cite{thakare2020parbac}. Hybrid models combining RBAC with ABAC have also been studied to enhance flexibility and security in smart home IoT scenarios~\cite{hybrid_abac_rbac_2024}. 

ABAC makes access decisions based on detailed attributes of users, resources, and the surrounding environment, allowing context-sensitive control \cite{hu2015abac, servos2017current}. Although ABAC offers more precise control, it often struggles with complicated policy management and trustworthiness of attribute verification \cite{Drame2021ACMSurv}, which lends to its unsuitability for the distributed systems. Thus, ABAC has been extended to support delegation, particularly in healthcare IoT, providing more flexible yet secure control \cite{choksy2023attribute}. Further improvements include blockchain-based ABAC implementations, enhancing security and robustness \cite{yu2022blockchain}.

Susequent works have focused on overcoming these limitations in ABAC.~\cite{Wang2017} introduced a trust-based ABAC model that evaluates dynamic trust attributes alongside static attributes. \cite{putra2021trust} proposed a blockchain-powered ABAC model integrating a Trust and Reputation System (TRS), allowing more flexible and adaptive authorization decisions for IoT.~\cite{clark2022enhancing} have focused on strengthening security at the network edge, combining trust-based behavioral analysis with ABAC. The work uses Information-Centric Networks (ICNs) combined with trust and attribute-based access control on edge devices to better understand device behaviors and communication patterns. The auditing scheme in recent works~\cite{zhang2025efficient} supports data integrity and privacy for cloud-based IoT systems which can complement access control functions. Blockchain based access control system ensures traceability of access events along with encryption capabilities compared to the local control in ICN.~\cite{shammar2022attribute} designed an ABAC system built on Hyperledger Fabric blockchain by utilizing smart contracts for detailed, flexible access control without traditional fixed roles. 

Capability-Based Access Control (CapBAC) relies on capability tokens, which securely represent transferable access rights~\cite{gusmeroli2013capbac}. Initially, CapBAC faced issues such as poor token revocation methods and interoperability problems \cite{fotiou2022capabilities}. Recent advancements solved these issues by introducing verifiable credentials, proof-of-possession mechanisms, and dynamic policy tokens called CAPLets \cite{bakir2021caplets}. CapBAC has also been effectively used to prevent unauthorized IoT data transfers \cite{capbac_flow_control_2020}, and Ethereum smart contracts have been employed to deliver scalable and distributed security for IoT \cite{nakamura2019capability}. Although the IoT domain faces similar scalability and trust issues as P2P systems, our work targets the additional complexity of fully decentralized P2P architectures. In these network architecture, anonymity, high node turnover, and the absence of centralized control exacerbate the shortcomings of existing models.


Compared to these approaches, our framework does not merely adapt ABAC or CapBAC for IoT. Instead, we combine a hierarchical role structure with a dynamic state machine and an ERC-20 token economy to continuously reward compliant behavior and penalize malicious actions. This integration enables real-time, behavior-driven policy adaptation without central control, a feature not addressed in the state-of-art. Moreover, unlike most blockchain-based IoT models that operate within relatively static attribute sets, our design explicitly handles high churn and role reassignment in volatile P2P networks while maintaining low latency and scalability across heterogeneous devices.

\subsection{Blockchain-Based Access Control Mechanisms}

Blockchain-based access control mechanisms offer a decentralized, tamper-resistant alternative to traditional access control systems, eliminating the need for centralized authorities. By leveraging smart contracts, policies can be transparently defined, automatically enforced, and verifiably audited. This approach is particularly valuable in dynamic, trustless environments such as IoT and peer-to-peer networks, where devices frequently join or leave the system and centralized control is impractical. The inherent transparency, immutability, and distributed trust of blockchain make it a promising foundation for scalable, fine-grained, and accountable access control in modern decentralized applications. Recent surveys~\cite{abdulrahman2021survey, ragothaman2023access} highlight the potential and challenges of blockchain-enabled access control.

Early works~\cite{Dorri2017} used blockchain primarily to store IoT access control rules in private ledgers. Other studies leveraged blockchain platforms such as Hyperledger Fabric to enhance security and manage IoT access rights \cite{liang2019secure}. However, these initial methods mainly focused on blockchain as a storage medium and the secured computational potential in the dynamic distributed system remained limited. More advanced frameworks, such as FairAccess~\cite{ouaddah2016fairaccess} use blockchain for managing access rights through tokens and enforcing policies with smart contracts. These developments significantly improved computational efficiency in IoT environments. 

Additionally, researchers proposed ABAC using blockchain to achieve better security and scalability due to blockchain's distributed design \cite{ding2019novelabac, zaidi2021attribute, lu2021fine}. Smart contracts were also employed to implement RBAC specifically tailored for IoT \cite{Cruz2018}. All these solutions aim to overcome limitations of traditional access control issues of role explosion and dynamic adaptation in IoT devices with limited resources as explained in the seminal work~\cite{Novo2018}.


\cite{okian2020borderchain} introduced BorderChain, a token-based public blockchain framework for IoT endpoints, which improved decentralization but suffered from central trusted approvers.~\cite{xu2020bdssfa} developed a privacy-preserving data sharing platform with fine-grained access control using Hyperledger Fabric.~\cite{liu2020fabricoT} implemented Fabric-IoT, an ABAC system leveraging smart contracts for dynamic IoT permissions. 
~\cite{qin2021lbac} proposed a lightweight ciphertext-policy attribute-based encryption scheme for its one-to-many characteristics using cloud services.~\cite{pal2019delegation} designed an access right delegation model combining public and private Ethereum chains to support dynamic IoT coalitions.

Kumar and Tripathi \cite{kumar2021eblp} improved healthcare IoT security by integrating an enhanced Bell-LaPadula model with Hyperledger Fabric.~\cite{liu2020capbac} suggested a decentralized capability-based IoT model using Decentralized Identifier (DID) and Ethereum smart contracts, though reliance on edge servers limited decentralization.~\cite{ali2020xdbauth} proposed xDBAuth, a hybrid blockchain framework for cross-domain IoT authentication and delegation, overcoming storage limits via off-chain MongoDB. In contrast to these IoT-focused models, our framework is designed for P2P networks, where decentralization, anonymity, and churn are even more challenging. By leveraging smart contracts for dynamic tokenized access rights and incentivization, our approach aims to bridge the scalability and trust gaps that remain open in existing work.



Our work advances beyond these blockchain-based models in three key ways: (1) Operational Implementation: we provide a complete, deployable Ethereum smart contract suite (including role assignment, behavior monitoring, and policy enforcement) rather than a conceptual architecture. (2) Incentive-Driven Governance: we integrate a fungible token economy that directly links security compliance to tangible rewards or penalties, strengthening long-term adherence to policies. (3) Dynamic Threat-Responsive Control: by embedding a finite state machine mapped to the STRIDE threat model, our system dynamically modifies access privileges based on real-time peer behavior—enabling proactive mitigation of evolving threats, a capability not explicitly addressed in earlier frameworks. 

\section{Framework Design} \label{Sec:system_arch}
This section provides preliminary information about the network architecture and Ethereum smart contract architecture that enforces access policies. It provides a conceptual overview of the static and dynamic management system, along with the incentivization mechanism.


\subsection{Network Architecture Preliminaries} \label{Sec:network_arch}

At the highest level of our framework, the network \cite{roy2022efficient} is structured as a Transit Ring Network, where PGHs are interconnected in a ring topology. The ring topology enables direct communication between PGHs, reducing the lookup time to \(O(n)\), where \(n\) represents the total number of PGHs. By minimizing communication delays, this arrangement enhances network responsiveness, making it well-suited for rapidly evolving P2P environments with frequent peer additions or departures. 

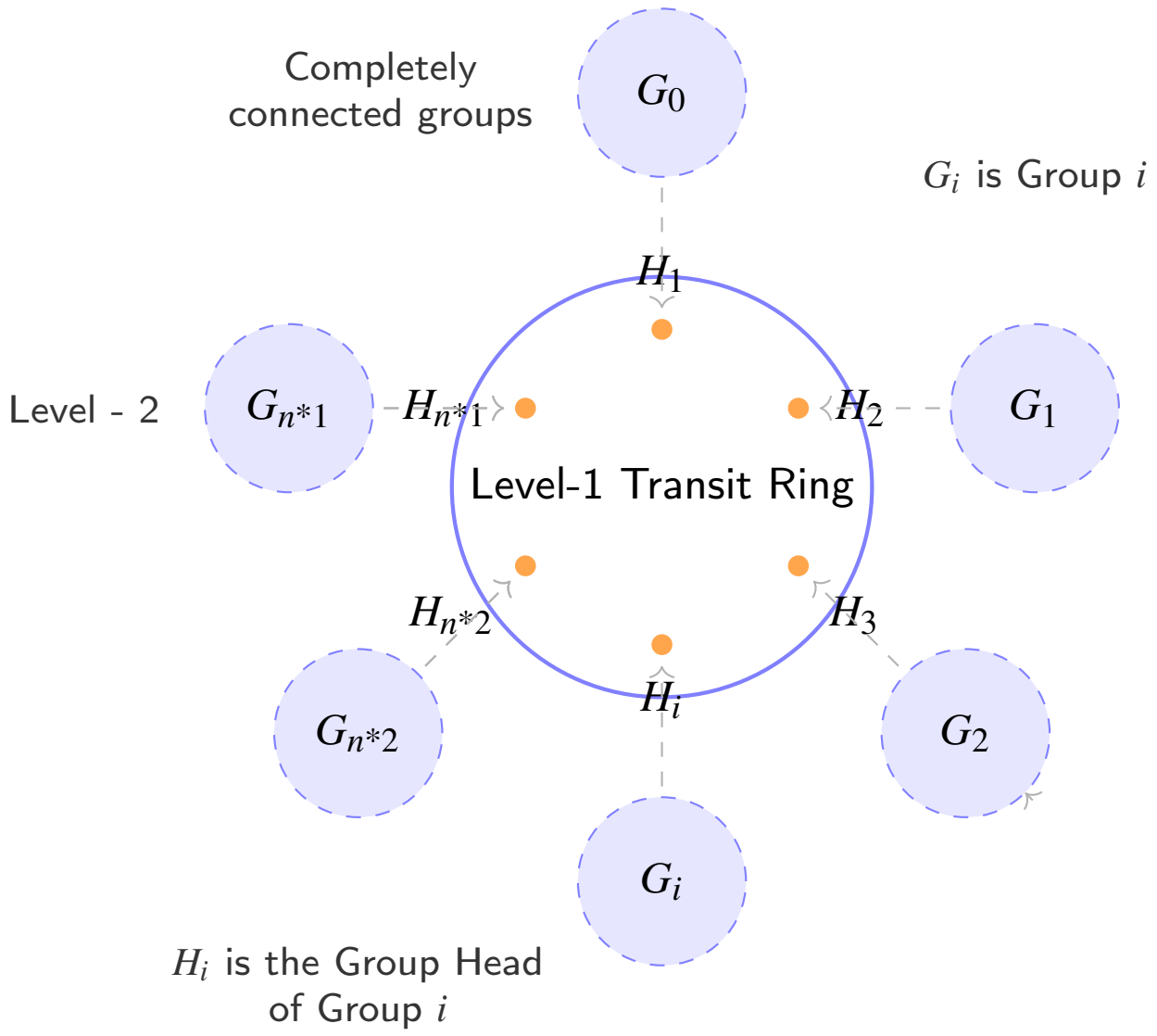
\begin{figure}[!ht]
\centering
\begin{tikzpicture}[scale=0.75, every node/.style={font=\sffamily},  
    node distance=1.8cm,
    group/.style={draw, circle, dashed, minimum size=1.2cm, align=center, font=\small, fill=blue!10, draw=blue!50}, 
    head/.style={circle, fill=orange!70, inner sep=1.5pt}, 
    arrow/.style={->, dashed, shorten <=2pt, shorten >=2pt, color=gray!60}, 
    label/.style={font=\footnotesize, align=center, color=black!80} 
]

\node[circle, draw=blue!50, minimum size=3cm, thick] (ring) {};
\node at (ring) {Level-1 Transit Ring};

\node[head] (H1) at (90:1.5cm) {};
\node[head] (H2) at (30:1.5cm) {};
\node[head] (H3) at (-30:1.5cm) {};
\node[head] (Hi) at (-90:1.5cm) {};
\node[head] (Hn2) at (-150:1.5cm) {};
\node[head] (Hn1) at (150:1.5cm) {};

\node[above=2pt of H1] {$H_1$};
\node[right=2pt of H2] {$H_2$};
\node[below right=2pt of H3] {$H_3$};
\node[below=2pt of Hi] {$H_i$};
\node[below left=2pt of Hn2] {$H_{n-2}$};
\node[left=2pt of Hn1] {$H_{n-1}$};

\node[group, above=1cm of H1] (G0) {$G_0$};
\node[group, right=1cm of H2] (G1) {$G_1$};
\node[group, below right=1cm of H3] (G2) {$G_2$};
\node[group, below=1cm of Hi] (Gi) {$G_i$};
\node[group, below left=1cm of Hn2] (Gn2) {$G_{n-2}$};
\node[group, left=1cm of Hn1] (Gn1) {$G_{n-1}$};

\node[label, left=0.2cm of G0] {Completely \\ connected groups};
\draw[arrow] (G2) -- ++(0.5,-0.5);

\node[label, left=0.2cm of Gn1] {Level - 2};
\draw[arrow] (Gn1) -- (Hn1);

\node[label, below=0.8cm of Gn2] {$H_i$ is the Group Head \\ of Group $i$};
\draw[arrow] (Gn2) -- (Hn2);

\node[label, above=0.8cm of G1] {$G_i$ is Group $i$};

\draw[arrow] (Gn1) -- (Hn1);
\draw[arrow] (Gn2) -- (Hn2);
\draw[arrow] (G1) -- (H2);
\draw[arrow] (G2) -- (H3);
\draw[arrow] (G0) -- (H1);
\draw[arrow] (Gi) -- (Hi);

\end{tikzpicture}
\caption{Double layer P2P architecture showing grouphead and regular members.} \label{Fig:network_arch}
\end{figure}

A hierarchical structure ensures scalability and robust security. The overall architecture is illustrated in Figure \ref{Fig:network_arch}. At the top, each PGH manages a subset of the larger network that includes a single SGH and multiple RMs. PGHs manage permissions, oversee inter- and intra-group communication, and monitor overall network activity. They act as the gatekeepers of the network, ensuring that only authorized nodes can interact with resources.

SGH provides essential support in managing group-level activities and maintaining operational continuity. If a PGH becomes malicious, SGHs seamlessly take over the responsibilities of the PGH, ensuring uninterrupted enforcement of access control policies. This built-in redundancy strengthens fault tolerance, a critical feature for maintaining the reliability and stability of the network \cite{ghosh2024securing}. RMs represent general peers or nodes within each group. Their access rights and permissions are determined by their compliance and behavior. Unlike PGHs and SGHs, RMs lack administrative authority and do not enforce access policies, focusing instead on using network resources within their defined permissions.

A pivotal component of this architecture is: Global Resource Table (GRT) and the Local Resource Table (LRT). At PGH level, GRT maintains a directory that maps resource types to their respective group heads. By including details such as logical addresses, IP addresses, and public keys, the GRT facilitates secure and efficient \textit{inter-group} communication. This table ensures that any resource request can be quickly directed to the appropriate group, enabling smooth and reliable data exchanges across the network.

Each PGH maintains a LRT to facilitate \textit{intra-group} communication. It contains information about all group members, including their logical addresses and public keys. SGH and RM have partial access about LRT as explained in static policy management schema in Section~\ref{Sec:static}. The integration of these components creates a cohesive network structure that addresses the inherent limitations of decentralized systems. The hierarchical organization ensures scalability, allowing the network to accommodate increasing numbers of peers without sacrificing performance. The architecture has increased resilience due to role-based redundancy of SGHs, and the robust management of resource tables despite high churn scenario.

\begin{figure*}[!ht]
\centering
\includegraphics[width=\textwidth,height=0.78\textheight]{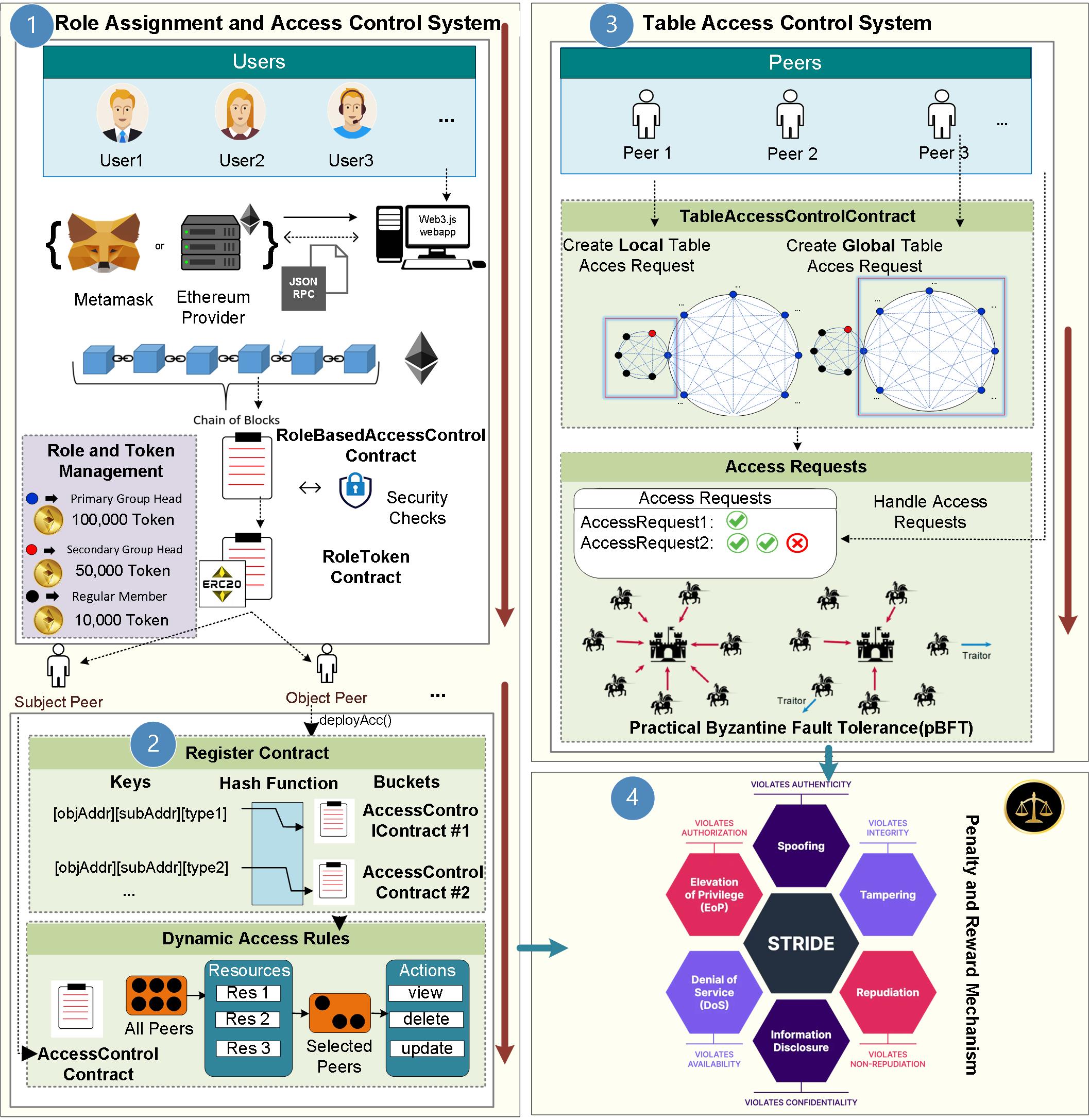} 
\caption{(1) Role Assignment: This is the part where the role assignment process is done, peers log in to the system as peers after the role assignment. (2) Access Control System: In the access control mechanism created by the object peer, the object peer adds some rules for its resources. The subject peer tries to access these resources. (3) Table Access Control System: Peers can create requests for operations to be performed on LRTs and GRTs, and if they are approved, they can perform these operations on these tables. (4) Penalty and Reward Mechanism: The operations made by peers in other contracts are monitored, and evaluated as misbehavior or reward according to their behavior, and tokens are minted/burned if necessary.}
\label{fig:workflow0}
\end{figure*}

\subsection{System Architecture}
\label{subsec:sysarch}

Our proposed framework integrates a blockchain-enabled token-based incentive model to govern access control in decentralized P2P networks. At the core is a collection of Ethereum-based smart contracts that handle peer roles, permissions, and resource management without relying on any central authority. Peers interact through a Web3-enabled web application and tools such as MetaMask, ensuring that all transactions, including role assignments and policy updates, are recorded immutably on the blockchain.

Participants in the network assume roles within a structured hierarchy: PGH, SGH, or RM. As explained in Section~\ref{Sec:network_arch}, Peer receives token allocation based on their roles and responsibilities. PGH receives the highest due to their coordination responsibilities and RM being the lowest as they form the bulk of network membership (Figure~\ref{Fig:network_arch}). 

All policy enforcement is carried out by Access Control Contracts (ACCs), which are dynamically deployed through a Register Contract. This register uses a key-hash mapping to link subject peers, object peers, and resource types to the appropriate ACC “bucket.” By consulting the Register Contract, the system can locate or create the relevant ACC that enforces the correct policies over specific resources. These policies define which actions such as view, delete, or update. These are permissible for particular peers, \textit{ensuring fine-grained and context-specific access control}. The workflow diagram is illustrated in Figure \ref{fig:workflow0}.

\begin{figure*}[!ht]
\centering
\includegraphics[width=\textwidth]{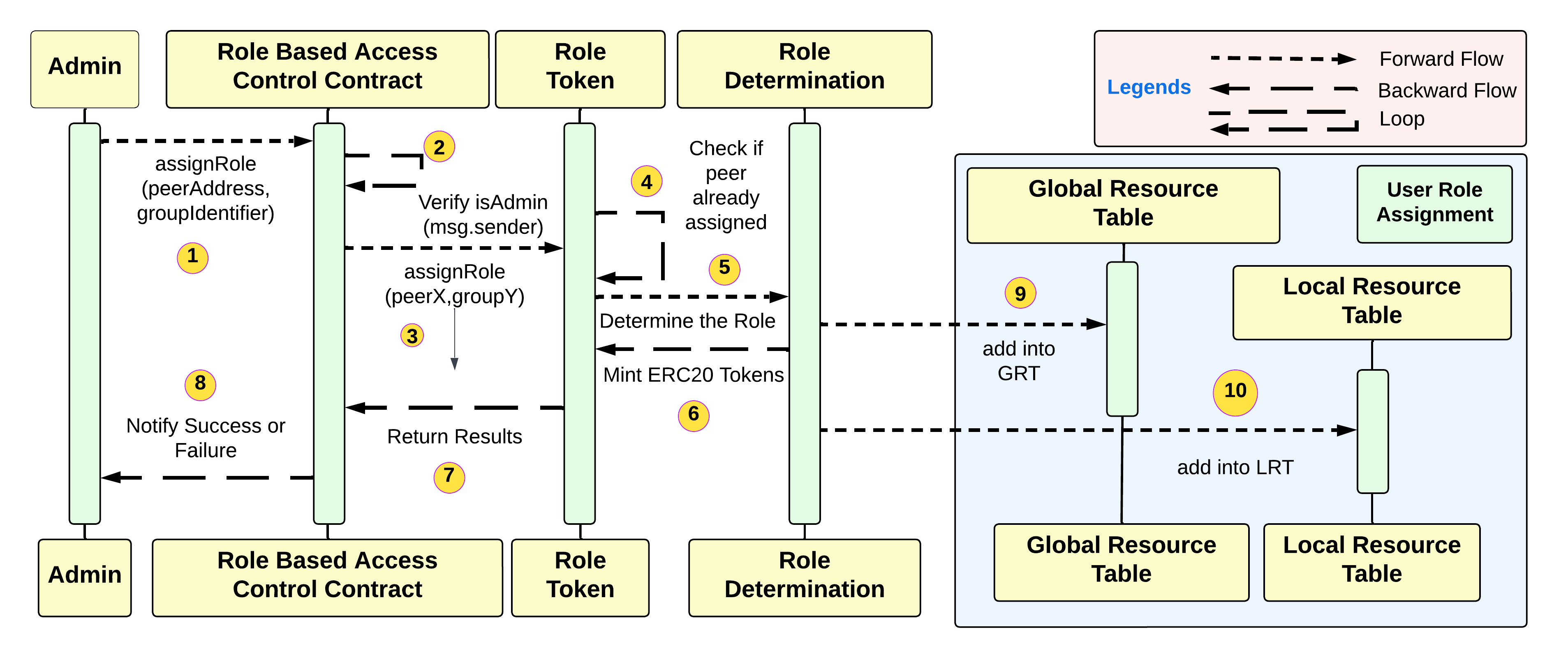} 
\caption{Sequence diagram of the role assignment process. The Admin initiates the \texttt{assignRole} function, which verifies permissions, assigns roles, mints ERC20 tokens, and updates the Global Resource Table (GRT) and Local Resource Table (LRT) to maintain network structure.}
\label{fig:communication_sequence1}
\end{figure*}

The network manages both GRT and LRT requests through a dedicated \texttt{TableAccessControlContract}. Peers submit requests to read or modify resources; the system evaluates these requests and determines whether to grant or deny them. Dynamic state management, aided by a finite state machine, allows the framework to classify peers into \textit{benign, suspicious, or malicious} states based on their actions. This adaptability ensures that access can be continuously tuned. Peers demonstrating repeated compliance remain benign, retaining or even earning increased privileges. In contrast, those engaged in questionable or harmful behavior slide into restricted states and face reduced access. 

Underpinning this dynamic environment is a penalty and reward mechanism informed by comprehensive threat modeling. The framework references the STRIDE security model, identifying categories of violations and applying measured responses. Compliant behavior is rewarded with additional tokens, strengthening a peer’s standing. Malicious acts trigger token deductions and potential blocks from resources, thus discouraging misconduct. Over time, as peers improve their behavior, their state can revert, and privileges are restored, thereby balancing security with a path to redemption.

This architecture embodies an evolving and fair system of governance. By combining on-chain role definitions, flexible resource mapping, dynamic policy enforcement, and incentivized behavior, the framework maintains both the security and the fluidity needed in modern P2P and Web3 environments.

\subsection{Peer Role Assignment}

In a P2P network, peer role assignment process  ensures proper network governance and compliance. Figure \ref{fig:communication_sequence1} shows the step-by-step process of peer role assignment in the network. Step 1 begins when the Admin calls the \texttt{assignRole} function in the Role Based Access Control Contract, providing the peer’s address and group identifier. Step 2 verifies if the caller is an Admin by checking \texttt{msg.sender}. If the verification is successful, Step 3 executes the \texttt{assignRole} function, sending the request to the Role Token contract. Step 4 checks whether the peer is already assigned a role. If the peer does not have a role, Step 5 determines the appropriate role based on network conditions. Step 6 mints ERC20 tokens as a reward for the assigned role. In Step 7, the Role Token contract returns the results of the role assignment process. Step 8 notifies the Admin whether the role assignment was successful or failed. Step 9 adds the new role information to the GRT for network-wide reference. Step 10 updates the LRT to store role details within the peer’s local network group. This structured workflow ensures that roles are assigned securely, tokens are distributed correctly, and the resource tables are updated for efficient peer management in the network.

\begin{table*}[ht]
\centering
\caption{Static Access Permissions of Different Subjects}
\normalsize\sffamily
\setlength{\tabcolsep}{3pt} 
\begin{tabular}{|p{3cm}|p{2.2cm}|p{2.2cm}|p{2.2cm}|p{2.2cm}|p{3.5cm}|}
\hline
\rowcolor{blue!15} 
\textbf{Roles / Access} & \textbf{Global Resource Table} & \textbf{Local Resource Table} & \textbf{Malicious Group Head} & \textbf{Malicious Member} & \textbf{Communication Access} \\ \hline
Primary Group Head & full (view, edit, create, delete) & full (view, edit, create, delete) & full (view, edit, create, delete) & full (view, edit, create, delete) & other group head, own members \\ \hline
Secondary Group Head & view & view & view, edit & deny & own group head, own members \\ \hline
Regular Members & deny & view & deny & deny & own group head \\ \hline
\end{tabular}
\label{tab:access_control}
\end{table*}

\begin{table*}[ht]
\centering
\caption{Comprehensive Security Event Responses, Impacts, and Token Adjustments}
\normalsize\sffamily
\setlength{\tabcolsep}{3pt} 
\begin{tabular}{|p{3.5cm}|p{1.5cm}|p{2cm}|p{3.5cm}|p{3cm}|p{2cm}|}
\hline
\rowcolor{blue!15} 
\textbf{Security Event} & \textbf{Severity} & \textbf{CIA Impact} & \textbf{STRIDE Impact} & \textbf{Response} & \textbf{Token Adjustment} \\ \hline
Too many Access Attempts & High & Integrity & Elevation of Privilege & 24-hour ban & -5 tokens \\ \hline
Data Tampering & High & Integrity & Tampering & Cancel access permanently & -20 tokens \\ \hline
Unauthorized Access & Medium & Confidentiality & Information Disclosure & Temporary suspension & -10 tokens \\ \hline
Disruption of Service & High & Availability & Denial of Service & 30 days suspension & -15 tokens \\ \hline
Misrepresentation of Identity & Low & Accountability & Spoofing & Warning, penalties & 0 tokens \\ \hline
Successful Access & Low & Integrity & None & None & +5 tokens \\ \hline
\end{tabular}
\label{tab:comprehensive_security_responses}
\end{table*}

Next up, to understand how state management works in our architecture, let’s explore the main components and roles:

\subsubsection{Static Access Management} \label{Sec:static}


Static access management establishes predefined permission levels for PGH, SGH, RM within the network, which is guided by the least privilege security principle. Table~\ref{tab:access_control} shows the details of permissions granted to each role. The subjects listed in the first column are all considered benign. The headers in the first row represent different objects. These objects include GRT, LRT, and other malicious peers. A benign PGH has full access rights. This includes complete control over their own benign and malicious group members. A benign SGH has partial access to malicious PGH. This access is provided because SGH is registered to take the role of PGH during threat situations. RM have very limited access. They can only communicate through a dedicated broadcast protocol with their own PGH. In most other cases, their access is denied.

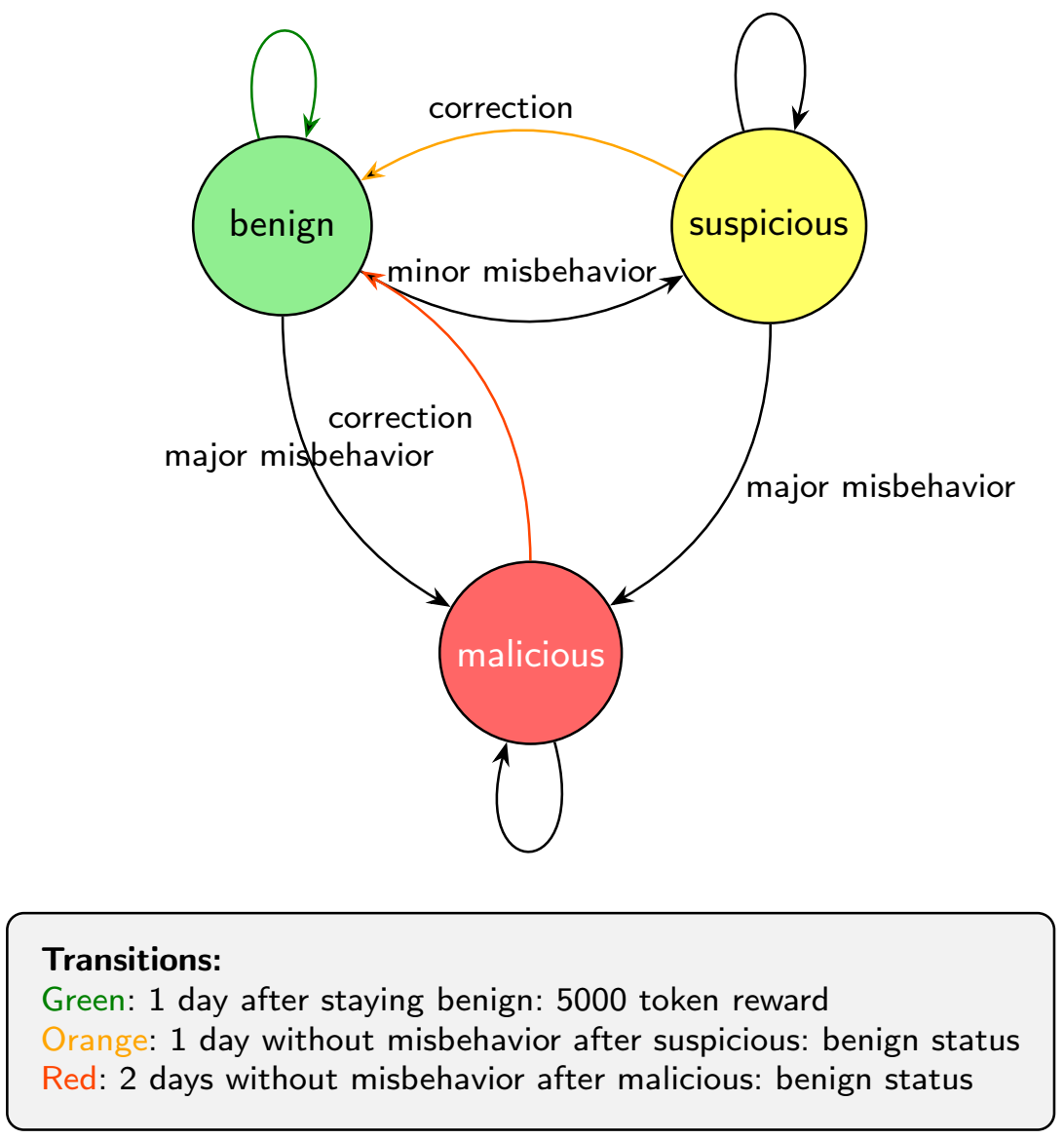
\begin{figure}[!ht] 
\centering
\resizebox{0.42\textwidth}{!}{ 
\begin{tikzpicture}[->, >=Stealth, node distance=2.5cm, auto, semithick, every node/.style={font=\sffamily}]
    
    \definecolor{benignColor}{RGB}{144, 238, 144} 
    \definecolor{suspiciousColor}{RGB}{255, 255, 102} 
    \definecolor{maliciousColor}{RGB}{255, 102, 102} 
    
    \definecolor{rewardColor}{RGB}{0, 128, 0}          
    \definecolor{firstCorrectionColor}{RGB}{255, 165, 0}     
    \definecolor{secondCorrectionColor}{RGB}{255, 69, 0}     

    \node[state, fill=benignColor, minimum size=1.5cm, font=\small] (benign) {benign};
    \node[state, fill=suspiciousColor, right=of benign, minimum size=1.5cm, font=\small] (suspicious) {suspicious};
    \node[state, fill=maliciousColor, text=white, below right=2.5cm and 1cm of benign, minimum size=1.5cm, font=\small] (malicious) {malicious};

    \draw[draw=rewardColor, loop above, looseness=8] (benign) edge[->] node {} (benign);
    \draw[loop above, looseness=8] (suspicious) edge[->] node {} (suspicious);
    \draw[loop below, looseness=8] (malicious) edge[->] node {} (malicious);

    \draw (benign) edge[->, bend right] node[above, yshift=6pt, font=\footnotesize] {minor misbehavior} (suspicious);
    \draw (benign) edge[->, bend right] node[above, xshift=-6pt, font=\footnotesize] {major misbehavior} (malicious);
    \draw[draw=firstCorrectionColor] (suspicious) edge[->, bend right] node[left, xshift=16pt, yshift=6pt, font=\footnotesize] {correction} (benign);
    \draw (suspicious) edge[->, bend left] node[right, font=\footnotesize] {major misbehavior} (malicious);
    \draw[draw=secondCorrectionColor] (malicious) edge[->, bend right] node[below left, font=\footnotesize] {correction} (benign);

    \node[align=left, below=1.4cm of malicious, font=\footnotesize, draw, fill=gray!10, rounded corners, inner sep=8pt] (legend) {
        \textbf{Transitions:}\\
        \textcolor{rewardColor}{Green}: 1 day after staying benign: 5000 token reward\\
        \textcolor{firstCorrectionColor}{Orange}: 1 day without misbehavior after suspicious: benign status\\
        \textcolor{secondCorrectionColor}{Red}: 2 days without misbehavior after malicious: benign status
    };

\end{tikzpicture}
}
\caption{State transition diagram showing peer states and transitions based on behavior and penalties.}
\label{fig:state_transitions}
\end{figure}

\subsubsection{Dynamic Access Management} \label{Sec:dynamic_access}

Static policies are incapable of dynamic adjustments in the access rights based on real-time peer actions or compliance levels. Moreover, it renders them ineffective for repeat offenders or evolving behaviors of malicious peers. 
Table \ref{tab:comprehensive_security_responses} provides an overview of STRIDE security events, their associated severity levels, the nature of their impact, and the corresponding responses. We begin by describing our state transition machine that represents a realistic peer status. 
Each peer’s behavior is evaluated according to a function that considers their current state, recent actions, and any penalties or rewards they have received. The state transitions are shown in Figure \ref{fig:state_transitions}. 

Peers in the network are categorized into one of three states based on their behavior and compliance. In the \textit{benign} state, peers are considered trusted due to their history of adhering to network policies. They enjoy normal access to resources with minimal restrictions and remain in this state unless they violate access policies. The \textit{suspicious} state is assigned to peers after minor policy violations. In this state, peers face partial restrictions, such as limited access to specific resources, to deter further violations and encourage compliance without imposing severe penalties. The \textit{malicious} state is reserved for peers who commit repeated or severe violations. Peers in this state face strict limitations, including being blocked from accessing most resources, effectively isolating them to protect network security. 

\begin{figure*}[!ht]
\centering
\begin{tikzpicture}[
    scale=0.9,
    every node/.style={font=\sffamily, transform shape}, 
    contract/.style = {draw, rectangle, minimum width=1.5cm, minimum height=1.2cm, align=center, font=\small},
    table/.style = {draw, rectangle, minimum width=2cm, minimum height=3.2cm, align=center, fill=gray!20}, 
    arrow style/.style = {->, >=Stealth},
    double arrow/.style = {<->, >=Stealth},
    label/.style = {font=\footnotesize, align=center}
]

\node[contract, fill=blue!20] (RC) {Register \\ Contract (RC)};

\node[table, right=1.5cm of RC] (Lookup) {};
\node[below=0.4cm of Lookup.north] {ACC 1};
\node[below=0.7cm of Lookup.north] {ACC 2};
\node[below=1cm of Lookup.north] {ACC 3};
\node[below=1.6cm of Lookup.north] {\vdots};
\node[below=0.8cm of Lookup] {Lookup Table};

\node[contract, fill=green!20, right=2cm of Lookup.north east] (ACC1) {ACC 1};
\node[contract, fill=green!20, below=1.3cm of ACC1] (ACC2) {ACC 2};
\node[contract, fill=green!20, below=1.3cm of ACC2] (ACC3) {ACC 3};

\node[contract, fill=orange!20, right=3cm of ACC2] (JC) {Judge \\ Contract (JC)};

\node[contract, fill=green!30, right=2cm of JC] (TACC) {Table Access \\Control Contract}; 

\draw[arrow style, dotted] (RC.east) -- (Lookup.west);
\draw[arrow style] ([yshift=0.2cm]Lookup.east) -- (ACC1.west);
\draw[arrow style] (Lookup.east) -- (ACC2.west);
\draw[arrow style] ([yshift=-0.2cm]Lookup.east) -- (ACC3.west);

\draw[double arrow] (ACC1.east) to[bend left] node[above, sloped, label] {Misbehavior report} (JC.west);
\draw[double arrow] (ACC2.east) -- node[above, sloped, label] {Penalty decision} (JC.west);
\draw[double arrow] (ACC3.east) to[bend right=20] node[below, sloped, label] {Correction process} (JC.west);
\draw[double arrow] (TACC.west) -- node[above, sloped, label] {Behavior \\Evaluation} (JC.east); 

\node[above=0.35cm of ACC2, xshift=0.3cm] {Access Control Contract (ACC)};

\end{tikzpicture}
\caption{Access control architecture showing interactions between RC, ACCs, JC, and Lookup Table for policy enforcement. The \texttt{TableAccessControlContract}  interacts with the JudgeContract to report and penalize malicious activities or reward benign behavior during access requests.}
\label{fig:double_column_architecture}
\end{figure*}
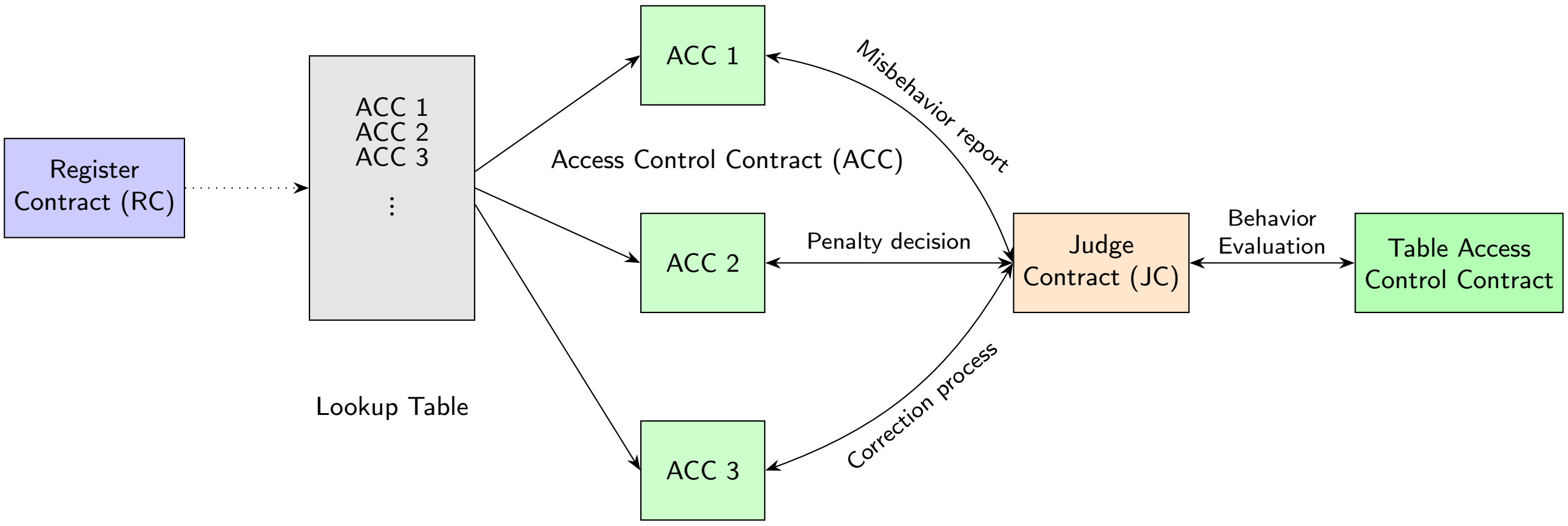



The hierarchical structure, with PGHs and SGHs overseeing groups, is essential for efficient state management in a large, dynamic P2P network. Firstly, the structure helps the network grow efficiently by distributing responsibilities. Each PGH oversees its group, and SGHs act as backups. The state also allows peers to be promoted or demoted, representing realistic user behaviour. This dynamic role assignment is crucial for regular operations and strict mitigation of threats.

\subsection{Token-Based Incentive Mechanism}  
Our framework uses the ERC-20 fungible token standard because of their identical nature and easy aggregation. Moreover, it is compatible with Ethereum's existing infrastructure including wallets. It also has well-defined functions such as \texttt{transfer()}, \texttt{balanceOf()}, and \texttt{approve()}. This simplifies  token transfers, check balances, and approve transactions.


To improve access control, we have added custom extensions to the ERC-20 standard. A role-aware token minting function dynamically assigns tokens based on peer roles. PGHs get a higher initial allocation than SGHs and RMs. We have also introduced automated state-linked deductions to apply penalties for policy violations, ensuring compliance with the system’s dynamic state management. These additions enhance flexibility while keeping the system stable within the ERC-20 structure.  

The system follows a token-based economy where tokens act as both currency and a measure of reputation. When a peer joins, they receive an initial token allocation based on their role. PGHs get the highest allocation, followed by SGHs and RMs. Peers earn tokens for following policies, such as making valid access requests or contributing new resources. If a peer violates policies, they lose tokens. The severity of the violation decides the penalty amount. Serious violations, such as data tampering or unauthorized deletion of critical resources, result in higher penalties compared to minor infractions.  

A peer’s token balance affects their state and access privileges. A higher balance signifies benign state with more access, while a lower balance can move them to a suspicious or malicious state, restricting their access. This incentive mechanism encourages peers to follow the rules to maintain or improve their access, ensuring a secure and self-regulating system.

\begin{table}[ht]
\centering
\caption{\small State Transitions on Various Network Interactions and Subsequent Token Balance Changes}
\label{tab:alice_state_transitions}
\renewcommand{\arraystretch}{1.2} 
\large\sffamily
\resizebox{\columnwidth}{!}{ 
\begin{tabular}{|m{3cm}|m{2cm}|m{3cm}|m{3cm}|}
\hline
\rowcolor{blue!15} 
\textbf{Action} & \textbf{Token Change} & \textbf{State Transition} & \textbf{Resulting State and Token Balance} \\ \hline
Peer Registration & +30 & Start & benign (30 tokens) \\ \hline
Access Granted to Resource & +5 & None & benign (35 tokens) \\ \hline
Unauthorized Resource Modification & -10 & benign to suspicious & suspicious (25 tokens) \\ \hline
Severe Violation Attempt & -20 & suspicious to malicious & malicious (5 tokens) \\ \hline
Compliance and Recovery & Varies & malicious to benign & benign (Recovery dependent on actions) \\ \hline
\end{tabular}
}
\end{table}

\begin{figure*}[!ht]
\centering
\includegraphics[width=\textwidth]{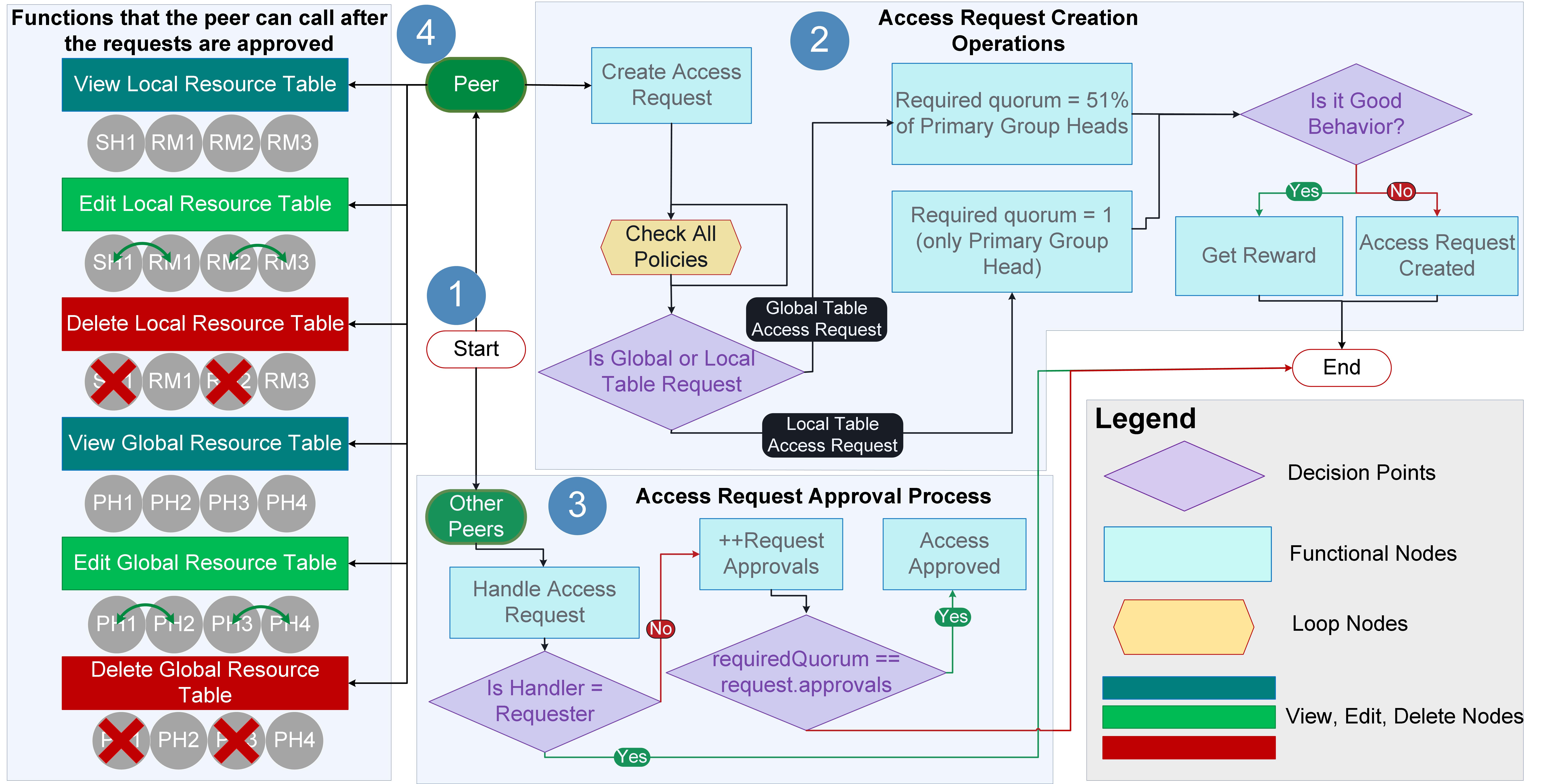} 
\caption{(1): Here the Start can branch to Peer or Other Peers. (2): This shows how the Table request operation is performed by the peer. The required quorum varies depending on the operation and which table it will access. (3): This shows how the request that created by a peer is approved by other peers. (4): Here, after the Peer's request is approved, it shows what Peer can do in which table, depending on which operation it was approved for.}
\label{fig:workflow2}
\end{figure*}

\textbf{Life Cycle}
\noindent This following writing portrays a typical scenario of how things work within the system, as detailed in Table \ref{tab:alice_state_transitions}. Each action by sample RM triggers specific transitions and token adjustments. \textit{peer Registration:} An RM joins the network with an initial token balance of \(T_{\text{RM}} = 30\), with a ``benign'' state. \textit{Access Request:} The RM successfully accesses ``File1,'' earning 5 tokens. This compliant action increases their token balance to \(T_{\text{RM}} = 35\). \textit{Unauthorized Attempt:} An unauthorized attempt to edit a resource results in a deduction of 10 tokens, transitioning the RM to a ``suspicious'' state with a new token balance of \(T_{\text{RM}} = 25\). \textit{Severe Violation:} Following a significant policy violation, the RM loses 20 tokens, moving into a ``malicious'' state with a balance of \(T_{\text{RM}} = 5\). \textit{Recovery:} 2 days without misbehavior in the malicious status, the RM's will revert to the ``benign'' state.

In actual architecture however, we are managing 3 hierarchical roles (PGH, SGH, RM) with token allocations of 1,000,000, 500,000, and 100,000, respectively. We define 3 behavioral states (benign, suspicious, malicious), and apply policies through 2 tables (GRT and LRT).

\begin{figure*}[!ht]
\centering
\includegraphics[width=\textwidth]{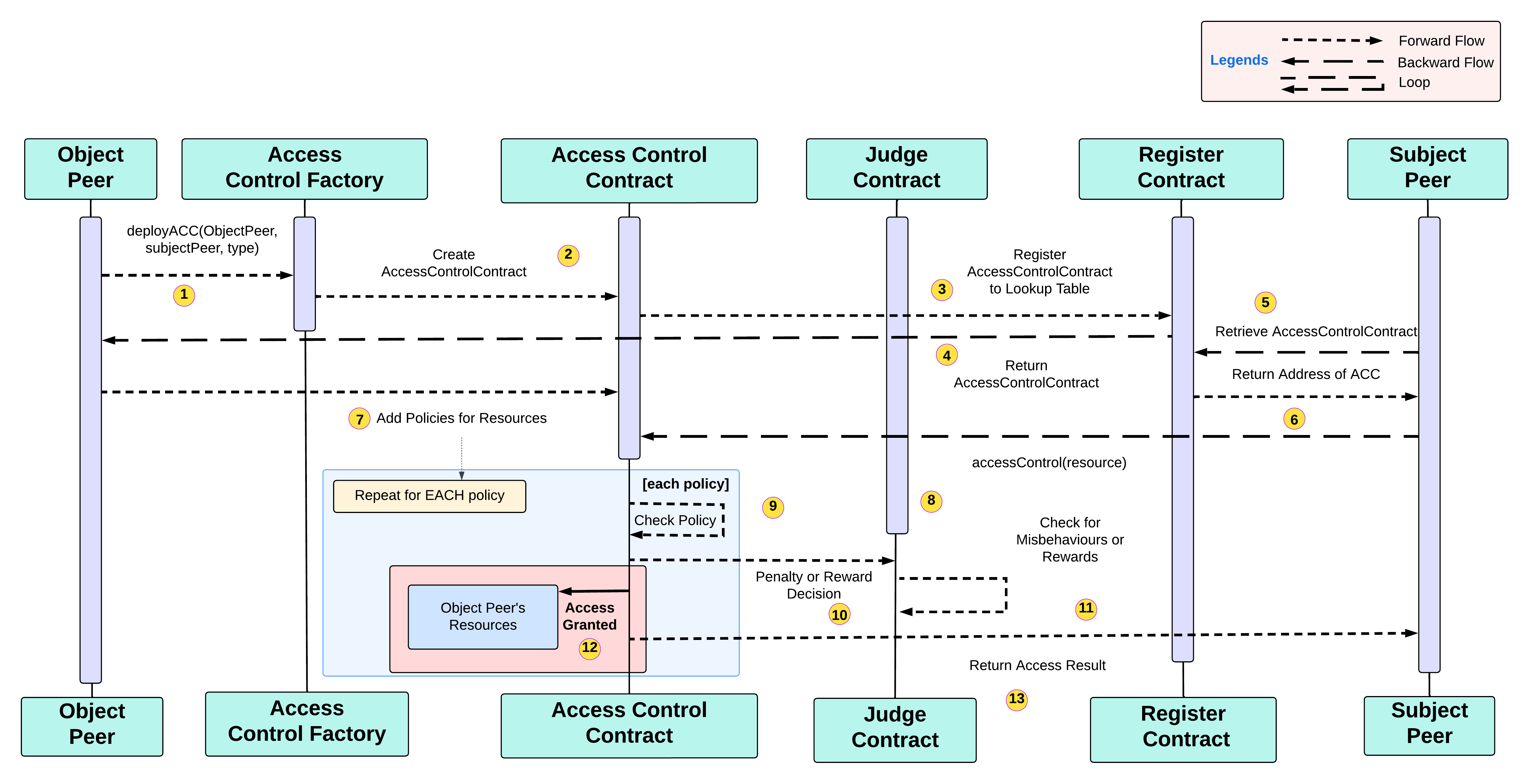} 
\caption{Sequence diagram showing the process of peer role assignment and access control, specifically how different contracts interact to assign roles, enforce policies, and verify access requests.}

\label{fig:communication_sequence2}
\end{figure*}

\subsection{Contract Hirearchy}
The system enforces access control through smart contracts deployed on the Ethereum blockchain. The main entities in this process are the ACCs, JC, and RC. 

The architecture of this system, including the interaction between RC, ACCs, JC, and the lookup table, is depicted in Figure \ref{fig:double_column_architecture}.

ACCs are smart contracts that enforce access policies on the blockchain, which also allows or restrict access based on role and behavior. The JC monitors peer actions, records misbehaviors, and applies penalties in response to each violation. The RC manages access policy registrations, and links subjects with the correct ACCs within the system. The sequence of operations for enforcing access control through the framework's smart contracts is portrayed in Figure \ref{fig:communication_sequence2}, which shows how peer roles are assigned and access is controlled in the network. Step 1 begins when the Object Peer requests the Access Control Factory to create an Access Control Contract (ACC). Step 2 follows as the factory generates the ACC and sends it to the Access Control Contract. Step 3 registers the ACC in the Register Contract, which keeps a record for future lookups. In Step 4, the Register Contract confirms the registration and returns the ACC details. Step 5 allows the Subject Peer to retrieve the ACC from the Register Contract, and Step 6 ensures the contract’s address is sent back to the Subject Peer. Step 7 involves the Access Control Contract, which adds policies for different resources based on the assigned roles. When a Subject Peer tries to access a resource in Step 8, the Access Control Contract processes the request. Step 9 checks the requested policy to see if the action is allowed. In Step 10, the Judge Contract evaluates the request and decides if the Subject Peer should receive a penalty or a reward. Step 11 ensures further checks for any misbehavior or compliance with the rules. Step 12 grants access if the request is valid, allowing the Object Peer to share its resource. Finally, in Step 13, the access result is sent back to the Subject Peer, completing the process.

When a peer attempts to access a resource, the ACC authenticates the peer's identity, role, and current state, and determines whether the requested action complies with predefined policies. If access is granted, the action is executed; otherwise, penalties are applied as necessary. Every action, permitted or denied, is recorded immutably on the blockchain, thereby providing audit trail for accountability. JC monitors peer activity on the network, finds misbehavior, and applies penalties proportional to the violations. The system discourages malicious actions by the automation of penalties enforcement while encouraging peers to respect the set policies. Register Contract keeps track of all access control policies, ensuring consistent enforcement of policies across the network.

The \texttt{TableAccessControlContract} is a specialized contract that governs access and management of both GRT and LRT, serving as a central authority for resource-related actions across all peers in the system. Unsuch as the AccessControlContract, which focuses exclusively on defining access control between specific subjects and objects, the \texttt{TableAccessControlContract}  is designed to handle broader operations involving resource tables. Each peer in the network can interact with this contract to create requests for actions such as viewing, modifying, swapping roles, or revoking roles in their local resource table. Importantly, the contract deploys with static rules that define the operational policies for resource table management. These rules are embedded during deployment and can only be modified by the owner of the contract, ensuring a controlled and consistent framework. Local table requests are processed based on these static rules, which streamline approval and enforcement while maintaining security and efficiency.

For operations involving the GRT, which spans the network and impacts all primary group heads, the contract enforces a robust Byzantine-such as consensus mechanism to safeguard against unauthorized changes. When a peer initiates a request for global table modifications, it must pass through a quorum-based approval process, requiring agreement from at least 51\% of the primary group heads. This ensures that no single entity can exert undue influence or introduce malicious changes to critical global resources. The contract’s static rules also guide the evaluation of requests in both local and global contexts, ensuring consistent decision-making. In this dual framework, local table requests are streamlined for individual peer resources, while global table requests foster decentralized collaboration and secure governance. The \texttt{TableAccessControlContract}  thus bridges localized resource management with system-wide coordination, enabling flexible, secure, and accountable operations.

The detailed workflow of access control is presented in Figure \ref{fig:workflow2}. 
\begin{figure}[htbp]
\centering
\begin{tikzpicture}[scale=0.40, every node/.style={font=\normalsize\rmfamily}]
    \definecolor{darkred}{rgb}{0.55,0,0}
    \definecolor{red}{rgb}{1,0,0}
    \definecolor{white}{rgb}{1,1,1}

    \begin{axis}[
        colormap={custom}{color(0cm)=(white); color(0.5cm)=(red); color(1cm)=(darkred)},
        colorbar,
        point meta min=0,
        point meta max=25000,
        xlabel={Roles},
        ylabel={Threat Types},
        title={Threat Token Deduction Map},
        xlabel style={font=\normalsize\rmfamily},
        ylabel style={font=\normalsize\rmfamily},
        title style={font=\normalsize\rmfamily},
        xticklabel style={font=\normalsize\rmfamily},
        yticklabel style={font=\normalsize\rmfamily},
        tick label style={font=\normalsize\rmfamily},
        colorbar style={
            ytick={0, 5000, 10000, 15000, 20000, 25000},
            yticklabels={0, 5000, 10000, 15000, 20000, 25000},
            ticklabel style={font=\normalsize\rmfamily},
        },
        grid=both,
        enlargelimits=true,
        width=15cm,
        height=17cm,
        xtick={1.7,2.78,3.75},
        xticklabels={PGH, SGH, RM},
        ytick={0,1,2,3,4,5,6,7},
        yticklabels={
            {Denial of Service},
            {Tampering with Data},
            {Repudiation Attempt},
            {Spoofing Attempt},
            {Privilege Escalation},
            {Data Leakage Attempt},
            {Too Frequent Access},
            {\shortstack{Unauthorized Access\\Attempt}}
        },
        every axis x label/.style={at={(0.5,-0.08)}, anchor=north},
        every axis y label/.style={at={(-0.33,0.6)}, anchor=east, rotate=90},
        anchor=west,
        x tick label style={xshift=-20mm},
    ]

    \addplot[matrix plot*, point meta=explicit, mesh/rows=5] 
    coordinates {
        (0,0) [0] (0,1) [0] (0,2) [0] (0,3) [0] (0,4) [0] (0,5) [0] (0,6) [0] (0,7) [0]
        (1,0) [24000] (1,1) [12000] (1,2) [7000] (1,3) [6000] (1,4) [14000] (1,5) [11000] (1,6) [5000] (1,7) [9000] 
        (2,0) [22000] (2,1) [10000] (2,2) [5000] (2,3) [4000] (2,4) [12000] (2,5) [9000] (2,6) [3000] (2,7) [7000] 
        (3,0) [20000] (3,1) [8000] (3,2) [3000] (3,3) [2000] (3,4) [10000] (3,5) [7000] (3,6) [1000] (3,7) [5000]  
        (4,0) [0] (4,1) [0] (4,2) [0] (4,3) [0] (4,4) [0] (4,5) [0] (4,6) [0] (4,7) [0]
    };

    \node[font=\normalsize\rmfamily, text=white] at (axis cs:1,0) {24000};
    \node[font=\normalsize\rmfamily, text=white] at (axis cs:2,0) {22000};
    \node[font=\normalsize\rmfamily, text=white] at (axis cs:3,0) {20000};

    \node[font=\normalsize\rmfamily, text=black] at (axis cs:1,1) {12000};
    \node[font=\normalsize\rmfamily, text=black] at (axis cs:2,1) {10000};
    \node[font=\normalsize\rmfamily, text=black] at (axis cs:3,1) {8000};

    \node[font=\normalsize\rmfamily, text=black] at (axis cs:1,2) {7000};
    \node[font=\normalsize\rmfamily, text=black] at (axis cs:2,2) {5000};
    \node[font=\normalsize\rmfamily, text=black] at (axis cs:3,2) {3000};

    \node[font=\normalsize\rmfamily, text=black] at (axis cs:1,3) {6000};
    \node[font=\normalsize\rmfamily, text=black] at (axis cs:2,3) {4000};
    \node[font=\normalsize\rmfamily, text=black] at (axis cs:3,3) {2000};

    \node[font=\normalsize\rmfamily, text=black] at (axis cs:1,4) {14000};
    \node[font=\normalsize\rmfamily, text=black] at (axis cs:2,4) {12000};
    \node[font=\normalsize\rmfamily, text=black] at (axis cs:3,4) {10000};

    \node[font=\normalsize\rmfamily, text=black] at (axis cs:1,5) {11000};
    \node[font=\normalsize\rmfamily, text=black] at (axis cs:2,5) {9000};
    \node[font=\normalsize\rmfamily, text=black] at (axis cs:3,5) {7000};

    \node[font=\normalsize\rmfamily, text=black] at (axis cs:1,6) {5000};
    \node[font=\normalsize\rmfamily, text=black] at (axis cs:2,6) {3000};
    \node[font=\normalsize\rmfamily, text=black] at (axis cs:3,6) {1000};

    \node[font=\normalsize\rmfamily, text=black] at (axis cs:1,7) {9000};
    \node[font=\normalsize\rmfamily, text=black] at (axis cs:2,7) {7000};
    \node[font=\normalsize\rmfamily, text=black] at (axis cs:3,7) {5000};

    \end{axis}
\end{tikzpicture}
\caption{Threat Token Deduction Heatmap by Roles}
\label{fig:threat_token_deduction_heatmap}
\end{figure}
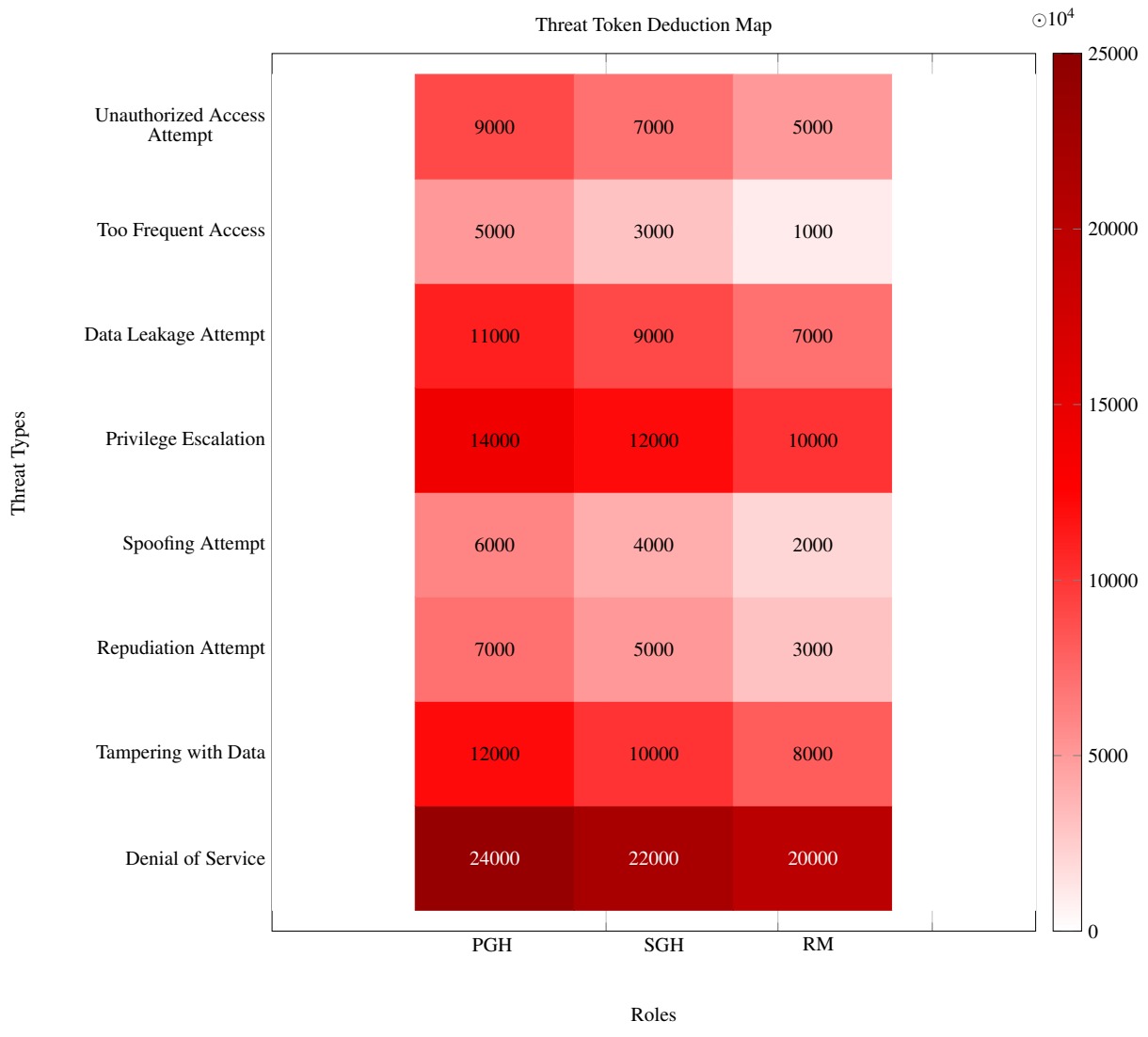

\textit{Token-Based Threat Management System}
A key innovation in our framework is the Threat Token Deduction Heatmap, which visually represents the token deduction mechanism across different threat types and peer roles. This system maintains network integrity by penalizing malicious behavior proportionally to the severity of the violation and the hierarchy of roles. Severe violations, such as Denial of Service or Privilege Escalation, result in significant deductions, particularly for higher-level roles such as PGHs, ensuring accountability at all levels.

In Figure \ref{fig:threat_token_deduction_heatmap}, The color gradient—from white to dark red—indicates severity, with darker shades reflecting higher penalties. PGHs incur the highest deductions for critical threats such as denial of Service, while RMs face relatively lower penalties for similar actions.

The token deduction system leverages dynamic state management and real-time monitoring. Violations are identified and processed through Judge Contracts (JC), which apply penalties using the Token Contract. Each penalty is carefully calibrated to the severity of the threat and the violator’s role, with atomic enforcement ensuring consistency.

The heatmap serves as a concise summary of these deductions. It showcases the system’s ability to deter malicious behavior and ensures equitable enforcement, offering a benchmark for decentralized networks aiming for both security and accountability.

\section{System Engineering and Deployment} \label{Sec:sys_engr}
This section provides a comprehensive overview of the framework’s implementation, including the hardware and software stack, testing methodologies, and performance evaluations.

\subsection{Algorithms and Processes}
This subsection presents the core algorithms enabling the functionality of the blockchain-based access control framework. It details essential processes such as role assignment, policy enforcement, and behavior tracking. 

\subsubsection{Role and State Management}
Algorithms in this section are designed to dynamically manage roles and peer states, ensuring the network hierarchy adapts efficiently to changes.

\textbf{Algorithm 1 assignRole} 

\begin{figure}[!ht]
\centering
\includegraphics[width=0.50\textwidth]{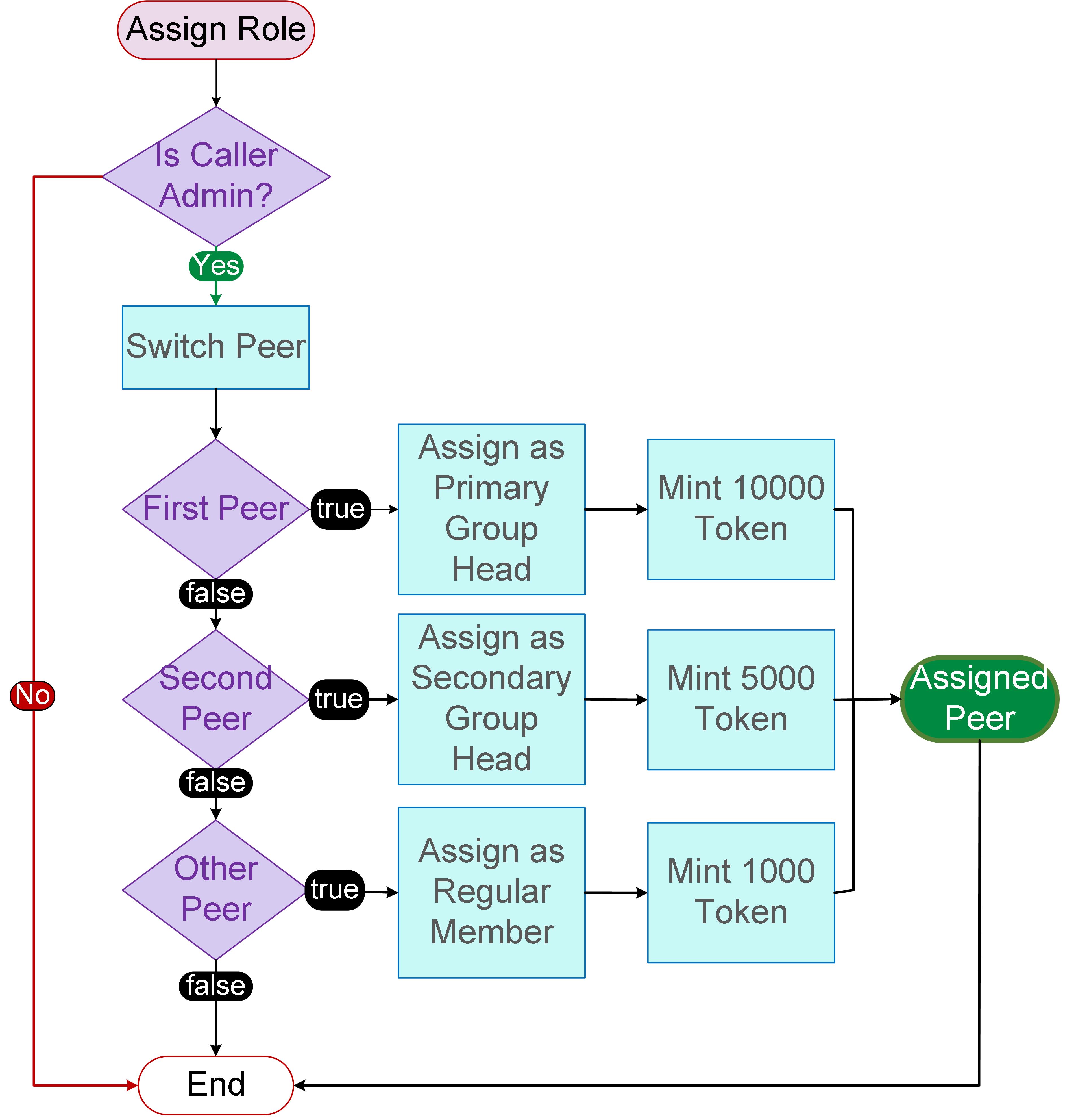} 
\caption{ Flowchart illustrating the role assignment and token minting process in the blockchain-based access control framework. The system first checks the administrative privileges of the caller. If authorized, peers are classified and assigned roles—Primary Group Head (10,000 tokens), Secondary Group Head (5,000 tokens), or Regular Member (1,000 tokens). Unauthorized requests terminate immediately. }\label{fig:flow-1}
\end{figure}

Figure \ref{fig:flow-1} illustrates the initialization algorithm for assigning roles to new peers. Step 1 checks if the caller is an admin. If the caller is not an admin, the process ends immediately. Step 2 switches the peer to be evaluated for role assignment. Step 3 checks if the peer is the first one in the group. If true, Step 4 assigns the peer as the Primary Group Head (PGH) and Step 5 mints 10,000 tokens as a reward. If the peer is not the first one, Step 6 checks if it is the second peer. If true, Step 7 assigns the peer as the Secondary Group Head (SGH), and Step 8 mints 5,000 tokens. If the peer is neither the first nor the second, Step 9 checks if it belongs to the group as another peer. If true, Step 10 assigns the peer as a Regular Member (RM), and Step 11 mints 1,000 tokens. Once the role is assigned, the peer is marked as assigned, and the process is complete. If the peer does not meet any of the conditions, the process ends without assigning a role. 

The first peer in the group is assigned as the Primary Group Head (PGH) and receives 10,000 tokens. If a PGH already exists, the second peer is assigned as the Secondary Group Head (SGH) with a reward of 5,000 tokens. Any additional peers are designated as Regular Members (RM) and receive 1,000 tokens.  

Each assigned role is recorded in the appropriate resource table, ensuring the network remains structured and scalable. This hierarchical assignment process distributes responsibilities efficiently while incentivizing participation, as seen in Figure \ref{fig:role_assignment}.

\makeatletter
\renewcommand{\ALG@name}{Algorithm} 
\renewcommand{\algorithmicindent}{1.5em} 
\makeatother

\begin{algorithm}
\caption{\scriptsize Assign Role} 
\begin{algorithmic}[1]
\scriptsize 
\Function{assignRole}{addr \_addr, str \_name, str \_type}
    \If{roles[\_addr] == Role.NONE}
        \If{isPrimaryHeadEmpty(\_type)}
            \State roles[\_addr] $\gets$ Role.PRIMARY\_GROUP\_HEAD
            \State addToGlobalResourceTable(\_addr, \_name, \_type)
            \State mint(\_addr, 1000000)
            \State emit RoleAssigned(\_addr, Role.PGH, \_type)
        \ElsIf{isSecondaryHeadEmpty(\_type)}
            \State roles[\_addr] $\gets$ Role.SECONDARY\_GROUP\_HEAD
            \State addToLocalResourceTable(\_addr, \_name, \_type)
            \State mint(\_addr, 500000)
            \State emit RoleAssigned(\_addr, Role.SGH, \_type)
        \Else
            \State roles[\_addr] $\gets$ Role.REGULAR\_MEMBER
            \State addToLocalResourceTable(\_addr, \_name, \_type)
            \State mint(\_addr, 100000)
            \State emit RoleAssigned(\_addr, Role.REGULAR\_MEMBER, \_type)
        \EndIf
    \EndIf
\EndFunction
\end{algorithmic}
\end{algorithm}

\textbf{Algorithm 2 Swap Roles Between Accounts} 
The algorithm enables smooth role adjustments based on network conditions or peer behavior, helping reassign responsibilities. It follows these steps: First, Validation ensures both accounts have assigned roles, are different, and belong to the same member type (\texttt{memberType1 = memberType2}). Second, roles are temporarily removed. Third, Role Swapping exchanges roles between accounts. Fourth, Updating Records updates role details and mappings. Finally, Event Logging emits \texttt{RoleAssigned} events to record changes. This allows the network to (1) manage its hierarchy flexibly and (2) correct role misalignments efficiently.

\makeatletter
\renewcommand{\ALG@name}{Algorithm} 
\renewcommand{\algorithmicindent}{1.5em} 
\makeatother

\begin{algorithm}
\caption{\scriptsize Swap Roles Between Accounts} 
\begin{algorithmic}[1]
\scriptsize 
\Function{swapRole}{account1, account2}
    \Require roles[account1] $\neq$ \texttt{Role.NONE} and roles[account2] $\neq$ \texttt{Role.NONE}
    \Require roles[account1] $\neq$ roles[account2]
    \Require \texttt{memberType1} = \texttt{memberType2} for both accounts

    \State role1 $\gets$ roles[account1]
    \State role2 $\gets$ roles[account2]
    \State memberType1 $\gets$ members[account1].memberType
    \State memberType2 $\gets$ members[account2].memberType

    \Comment{Remove roles from existing mappings}
    \State \texttt{removeRoleFromMappings}(account1, memberType1)
    \State \texttt{removeRoleFromMappings}(account2, memberType2)

    \Comment{Swap the roles between accounts}
    \State roles[account1] $\gets$ role2
    \State roles[account2] $\gets$ role1
    \State members[account1].role $\gets$ role2
    \State members[account2].role $\gets$ role1

    \Comment{Re-assign roles to new mappings}
    \State \texttt{assignRoleToMappings}(account1, memberType2, role2)
    \State \texttt{assignRoleToMappings}(account2, memberType1, role1)

    \State \textbf{emit} RoleAssigned(account1, role2, memberType2)
    \State \textbf{emit} RoleAssigned(account2, role1, memberType1)
\EndFunction
\end{algorithmic}
\end{algorithm}

\textbf{Algorithm 3 Revoke Role} 
This algorithm ensures network security by revoking roles from peers who are no longer suitable while preserving hierarchy and redistributing roles as needed. It follows these steps: First, it validates if the peer has an assigned role. Second, it retrieves the peer's role and member type. Third, it handles role-specific revocation by invoking \texttt{handlePGHRevocation} for PGH, promoting SGH if required, calling \texttt{handleSGHRevocation} for SGH, and removing RM from the local resource table. Fourth, it performs cleanup by setting the peer's role to \texttt{Role.NONE}, deleting their member record, and burning tokens. Finally, it emits a \texttt{RoleRevoked} event to log the revocation. This ensures clean role removal without inconsistencies, maintaining network integrity and promptly revoking access from unauthorized peers.

\makeatletter
\renewcommand{\ALG@name}{Algorithm} 
\renewcommand{\algorithmicindent}{1.5em} 
\makeatother

\begin{algorithm}
\caption{\scriptsize Revoke Role} 
\begin{algorithmic}[1]
\scriptsize 
\Function{revokeRole}{addr a}
    \Require roles[a] $\neq$ Role.NONE
    \State r $\gets$ roles[a]
    \State t $\gets$ members[a].memberType

    \If{isPrimaryHead(a, t)}
        \State handlePrimaryHeadRevocation(t, a)
    \ElsIf{roles[a] = Role.SECONDARY\_GROUP\_HEAD}
        \State handleSecondaryHeadRevocation(t, a)
    \Else
        \State removeFromLocal(t, a)
    \EndIf

    \State roles[a] $\gets$ Role.NONE
    \State delete members[a]
    \State burnTokens(a)
    \State \textbf{emit} RoleRevoked(a, r, t)
\EndFunction
\end{algorithmic}
\end{algorithm}

\subsubsection{Access Control Mechanisms}
The framework integrates static policies and dynamic threat management to ensure adaptive access control, as shown in Figure \ref{fig:communication_sequence1} and Figure \ref{fig:communication_sequence2}. Static access control enforces predefined rules via the GRT and LRT for routine operations, while dynamic mechanisms handle real-time threats by adjusting peer states (benign, suspicious, malicious) based on behavior. Smart contracts enforce policies transparently, with the JC applying penalties or rewards. This dual approach ensures robust management of both operational and threat scenarios.

\textbf{Algorithm 4 Access Control Function}

\begin{figure}[!ht]
\centering
\includegraphics[width=0.48\textwidth]{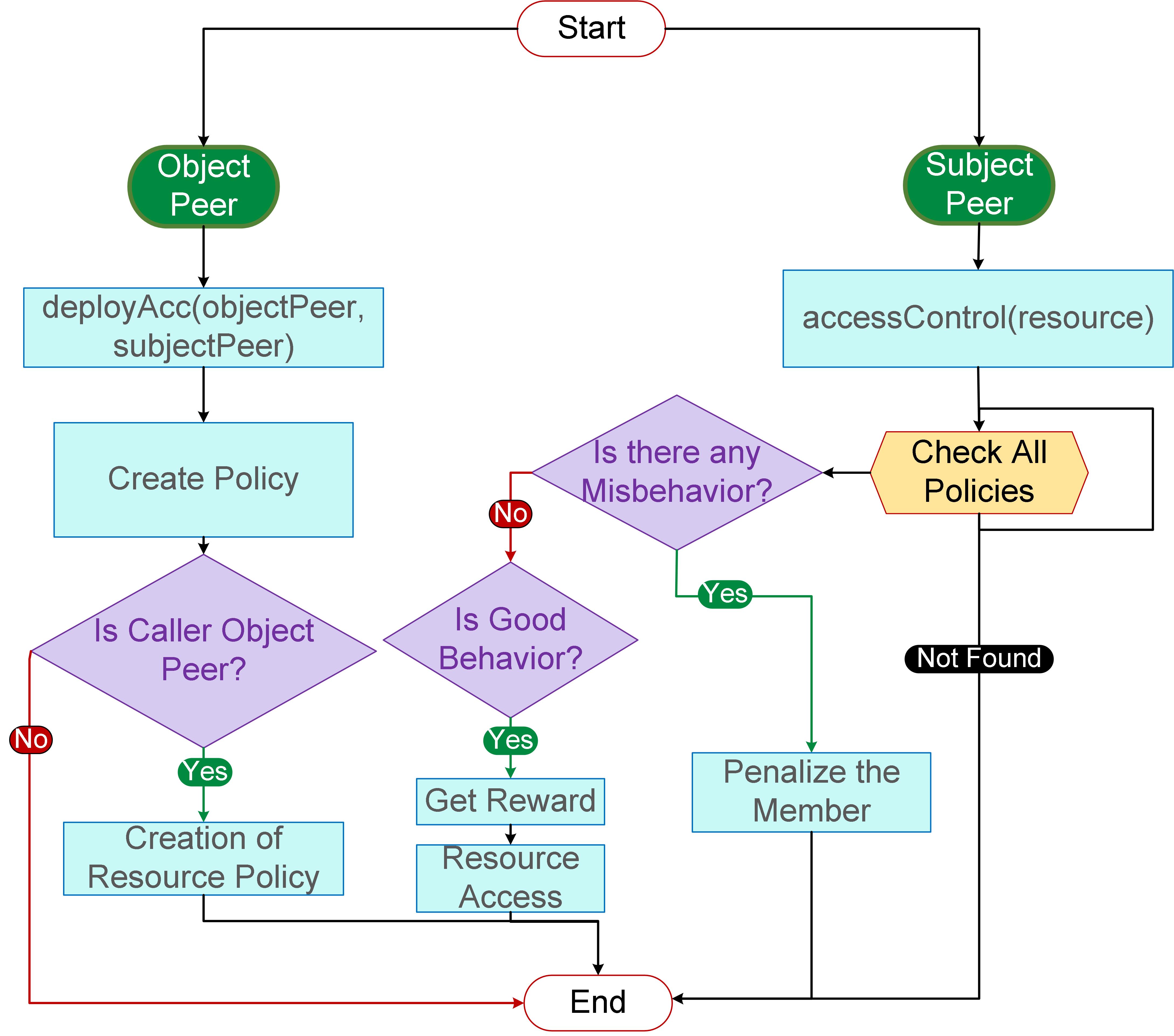} 
\caption{ Flowchart depicting the policy creation and access control workflow in the blockchain-based access control framework. Object Peers initiate by deploying an access control policy for Subject Peers. The framework verifies object-peer authority, evaluates peer behavior, and enforces reward or penalty actions accordingly. Subject Peers request resource access, prompting the system to validate permissions by checking existing policies and misbehavior records, thereby maintaining controlled and secure resource utilization. }\label{fig:flow-2}
\end{figure}

This algorithm, as illustrated in Figure \ref{fig:flow-2}, dynamically checks permissions for resource access, validates peer roles, and enforces access policies. The process begins with two types of peers: the Object Peer and the Subject Peer. The Object Peer calls \texttt{deployAcc} with the Subject Peer details to create a policy for access control. If the caller is valid, the creation of a resource policy proceeds; otherwise, the process terminates. Meanwhile, the Subject Peer initiates an \texttt{accessControl(resource)} request, triggering a policy check to determine if access is allowed. If no matching policy is found, access is denied. If policies exist, the system verifies misbehavior, penalizing violations or rewarding compliance accordingly. Finally, access is granted or denied based on the policy enforcement, and events are logged for accountability.

The implementation ensures that all access requests undergo peer validation, role verification, and frequency checks to prevent abuse. The Role-Based Access Control (RBAC) system manages role assignments and policy iterations. The Judge Contract (JC) processes misbehavior reports and compliance rewards.

\makeatletter
\renewcommand{\ALG@name}{Algorithm} 
\renewcommand{\algorithmicindent}{1.0em} 
\makeatother

\begin{algorithm}
\caption{\tiny Access Control Function} 
\begin{algorithmic}[1]
\tiny 
\Function{accessCtrl}{res: STRING, act: STRING}
    \State \textbf{ensure} active
    \State \textbf{ensure} notBlkd
    \State \textbf{ensure} subjMatch

    \State role $\gets$ roleCntr.getRole(msg.sender)

    \For{pol in pols}
        \If{pol.role $=$ role \textbf{and}
           keccak256(pol.res) $=$ keccak256(res) \textbf{and}
           keccak256(pol.act) $=$ keccak256(act)}

            \If{blk.timestamp $<$ pol.lastReq + thresh}
                \State (pen, rs, blkEnd) $\gets$ 
                judgeCntr.rptMalAct(msg.sender, ``Freq access'')
                \State \textbf{emit} MalActRep(subAddr, pen, rs, blkEnd)
                \Return FALSE
            \EndIf

            \State pol.lastReq $\gets$ blk.timestamp
            \State \textbf{emit} CtrlChk(subAddr, role, res, act, pol.perm $=$ ``allow'')

            \State isAll $\gets$ (pol.perm $=$ ``allow'')

            \If{isAll}
                \State (rew, newStat) $\gets$ 
                judgeCntr.rptNoPenMis(msg.sender)
                \State \textbf{emit} NoPenMisRep(msg.sender, rew, newStat)
            \Else
                \If{act $=$ ``edit'' \textbf{or} act $=$ ``delete''}
                    \State (pen, rs, blkEnd) $\gets$ 
                    judgeCntr.rptMalAct(msg.sender, ``Data tamper'')
                    \State \textbf{emit} MalActRep(msg.sender, pen, rs, blkEnd)
                \ElsIf{act $=$ ``view''}
                    \State (pen, rs, blkEnd) $\gets$ 
                    judgeCntr.rptMalAct(msg.sender, ``Unauth access'')
                    \State \textbf{emit} MalActRep(msg.sender, pen, rs, blkEnd)
                \EndIf
            \EndIf

            \Return isAll
        \EndIf
    \EndFor

    \If{act $=$ ``edit'' \textbf{or} act $=$ ``delete''}
        \State (pen, rs, blkEnd) $\gets$ 
        judgeCntr.rptMalAct(msg.sender, ``Data tamper'')
        \State \textbf{emit} MalActRep(msg.sender, pen, rs, blkEnd)
    \ElsIf{act $=$ ``view''}
        \State (pen, rs, blkEnd) $\gets$ 
        judgeCntr.rptMalAct(msg.sender, ``Unauth access'')
        \State \textbf{emit} MalActRep(msg.sender, pen, rs, blkEnd)
    \EndIf

    \State \textbf{emit} PolNotFndCtrl(res, act)
    \Return FALSE
\EndFunction
\end{algorithmic}
\end{algorithm}

\textbf{Algorithm 5 Create Access Request} 
This algorithm handles the creation and logging of access requests for network resources, ensuring that requests are processed accurately and efficiently while maintaining a record for auditing purposes. The implementation involves several steps: retrieving the peer's role; validating the requested action against the policy, reporting malicious activity and emitting relevant events if unauthorized; identifying whether the request is global or local and calculating the required quorum for approval; creating a new access request object and adding it to the list of access requests; reporting non-penalizable behavior to the JC for potential rewards; and emitting events to log the creation of the access request. It ensures that requests are properly vetted and that any unauthorized attempts are promptly addressed.

\makeatletter
\renewcommand{\ALG@name}{Algorithm} 
\renewcommand{\algorithmicindent}{1.5em} 
\makeatother

\begin{algorithm}
\caption{\scriptsize Create Access Request} 
\begin{algorithmic}[1]
\scriptsize 
\Function{createAccessRequest}{resource: STRING, action: STRING}
    \State role $\gets$ \text{roleBasedAccessControl.getRole(msg.sender)}
    \State allowed $\gets$ \text{checkPolicy(resource, action, role)}

    \If{not allowed}
        \If{action $=$ ``edit'' \textbf{or} action $=$ ``delete''}
            \State (penalty, reason, endTime) $\gets$ 
                \text{judgeContract.reportmaliciousActivity(msg.sender, }Tampering with data'')
            \State \textbf{emit} maliciousActivityReported(msg.sender, penalty, reason, endTime)
        \ElsIf{action $=$ ``view''}
            \State (penalty, reason, endTime) $\gets$ 
                \text{judgeContract.reportmaliciousActivity(msg.sender, }``Unauthorized access attempt'')
            \State \textbf{emit} maliciousActivityReported(msg.sender, penalty, reason, endTime)
        \EndIf
        \Return
    \EndIf

    \State isGlobal $\gets$ (resource $=$ ``GlobalResourceTable'')
    \State isLocalPrimary $\gets$ \textbf{not} isGlobal \textbf{and} (role $=$ \text{RoleToken.Role.PRIMARY\_GROUP\_HEAD})

    \If{isGlobal}
        \State total $\gets$ \text{roleBasedAccessControl.getGlobalResourceTable().length}
        \If{total $>$ 0}
            \If{role $=$ \text{RoleToken.Role.PRIMARY\_GROUP\_HEAD}}
                \State quorum $\gets$ 
                    $\lceil (total - 1) \cdot quorumPercentage / 100 \rceil$
            \Else
                \State quorum $\gets$ $(total \cdot quorumPercentage) / 100$
            \EndIf
        \Else
            \State quorum $\gets$ 0
        \EndIf
    \Else
        \State quorum $\gets$ 1
    \EndIf

    \State requestId $\gets$ \text{accessRequests.length}
    \State \text{accessRequests.push(AccessRequest(requestId, resource, action,} 
        isLocalRequestByPrimaryHead ? 1 : 0, msg.sender, isGlobal, isLocalPrimary, quorum))

    \State (reward, newStatus) $\gets$ 
        \text{judgeContract.reportNonPenalizeMisbehavior}
    \State \textbf{emit} NonPenalizeMisbehaviorReported(msg.sender, reward, newStatus)
    \State \textbf{emit} AccessRequestCreated(requestId, msg.sender, resource, 
        \text{address(this)})
\EndFunction
\end{algorithmic}
\end{algorithm}

\subsubsection{Behavioral Monitoring and Penalty Enforcement for JC}
These algorithms focus on 3 step process of tracking peer actions, enforcing penalties for violations, and rewarding compliance to encourage trustworthy behavior within the network.

\textbf{Algorithm 6 Report malicious Activity}
This algorithm ensures systematic enforcement against malicious peers, applying penalties dynamically based on the severity of reported violations. It maintains continuous monitoring by logging each instance of misbehavior into the peer's history and adjusting their status accordingly. Penalties include token deductions and time-bound blocking to discourage future infractions, ensuring accountable and secure peer interactions within the network.

\makeatletter
\renewcommand{\ALG@name}{Algorithm} 
\renewcommand{\algorithmicindent}{1.5em} 
\makeatother

\begin{algorithm}
\caption{\scriptsize Report malicious Activity} 
\begin{algorithmic}[1]
\scriptsize 
\Function{reportmaliciousActivity}{addr \_memberAddress, string reason}
    \State \textbf{Required} \text{bytes(reason).length} $>$ 0
    \State record $\gets$ misbehaviorHistory[\_memberAddress]
    \State currentTime $\gets$ \text{block.timestamp}
    
    \State penalty $\gets$ penalties[reason]
    \State penaltyDuration $\gets$ penalty.blockingDuration

    \State record.reasons.push(reason)
    \State record.lastPenaltyTime $\gets$ currentTime

    \State updateMemberStatus(\_memberAddress, penalty.level)

    \State tokenDeductionAmount $\gets$ penalty.tokenDeduction
    \Statex \hspace{2em} [roleBasedAccessContract.getRole(\_memberAddress)]
    \State penalizeMember(\_memberAddress, tokenDeductionAmount, reason)

    \If{penaltyDuration $>$ 0}
        \State blockMember(\_memberAddress, penaltyDuration, reason)
    \EndIf

    \State \Return (tokenDeductionAmount, reason, record.blockingEndTime)
\EndFunction
\end{algorithmic}
\end{algorithm}

\textbf{Algorithm 7 Compliant Behavior Reward Mechanism}
This algorithm rewards compliant behavior by peers, fostering a positive environment by updating peer statuses and providing token rewards to encourage adherence to network policies. Its purpose is to promote trustworthy behavior by recognizing and rewarding peers who consistently comply with access policies, thereby incentivizing peers to maintain good standing within the network. The implementation follows several steps: it begins by defining thresholds for benign behavior recovery based on peer status, retrieves the peer's member record and calculates the time since the last status update, and checks the peer's current status to determine if they qualify for a status change or reward. Specifically, if the peer is benign and has surpassed the threshold, the algorithm rewards them and updates their status; if the peer is suspicious or malicious and has maintained good behavior, it updates their status to benign. Finally, the algorithm outputs any rewards or status changes. The impact of this algorithm is that by reinforcing positive behavior, it encourages peers to comply with network policies and contributes to a cooperative network environment, providing a path for peers to recover from previous violations.

\makeatletter
\renewcommand{\ALG@name}{Algorithm} 
\renewcommand{\algorithmicindent}{1.5em} 
\makeatother

\begin{algorithm}
\caption{\scriptsize Compliant Behavior Reward Mechanism} 
\begin{algorithmic}[1]
\scriptsize 
\State benignThreshold $\gets$ \text{1 Day}
\State benignThresholdFormalicious $\gets$ \text{2 Days}
\Function{reportNonPenalizeMisbehavior}{addr \_memberAddress}
    \State member $\gets$ \text{roleBasedAccessContract.getMember(\_memberAddress)}
    \State timeSinceLastUpdate $\gets$ \text{block.timestamp} - member.lastStatusUpdate

    \If{keccak256(abi.encodePacked(member.status)) $=$ keccak256(abi.encodePacked(``benign''))}
        \If{timeSinceLastUpdate $>=$ benignThreshold}
            \State rewardMember(\_memberAddress)
            \State updateMemberStatus(\_memberAddress, ``nothing'')
            \State \Return (5000, ``benign'')
        \EndIf

    \ElsIf{keccak256(abi.encodePacked(member.status)) $=$ keccak256(abi.encodePacked(``suspicious''))}
        \If{timeSinceLastUpdate $>=$ benignThreshold}
            \State updateMemberStatus(\_memberAddress, ``nothing'')
            \State \Return (0, ``benign'')
        \EndIf

    \ElsIf{keccak256(abi.encodePacked(member.status)) $=$ keccak256(abi.encodePacked(``malicious''))}
        \If{timeSinceLastUpdate $>=$ benignThresholdFormalicious}
            \State updateMemberStatus(\_memberAddress, ``nothing'')
            \State \Return (0, ``benign'')
        \EndIf
    \EndIf

    \State \Return (0, ``not changed'')
\EndFunction
\end{algorithmic}
\end{algorithm}

\subsubsection{Smart Contract Deployment for RC}
Algorithms in this section handle the deployment and management of Access Control Contracts (ACCs), ensuring secure and automated enforcement of access policies.

\textbf{Algorithm 8 Deploy Access Control Contract} 
This algorithm deploys new Access Control Contracts (ACCs) tailored to specific access scenarios, ensuring that policies are enforced correctly for subject-object interactions within the network. The purpose of this algorithm is to automate policy enforcement by creating ACCs that govern access rights between subjects and objects, thereby streamlining the access control process and ensuring consistency across the network. The implementation involves several key steps: first, it validates that the subject and object are either in the same group or globally recognized; second, it creates a new ACC instance with the specified parameters; third, it configures the JC address within the ACC for misbehavior reporting; fourth, it registers the ACC in the system using the Register Contract; fifth, it emits an event to log the deployment of the ACC; and finally, it returns the address of the deployed ACC. The impact of this algorithm is that by securely deploying ACCs, it ensures that access control policies are enforced automatically and correctly.

\makeatletter
\renewcommand{\ALG@name}{Algorithm} 
\renewcommand{\algorithmicindent}{1.5em} 
\makeatother

\begin{algorithm}
\caption{\scriptsize Deploy Access Control Contract} 
\begin{algorithmic}[1]
\scriptsize 
\Function{deployAccessControlContract}{objectAddress: ADDRESS, subjectAddress: ADDRESS, accType: STRING}
    \State \textbf{require} roleBasedAccessControl.areSameOrGlobalMembers(subjectAddress, objectAddress)

    \State acc $\gets$ new AccessControlContract(subjectAddress, objectAddress, accType, address(roleBasedAccessControl), msg.sender)
    \State acc.setJudgeContract(address(judgeContract))

    \State registerContract.registerAccessControlContract(objectAddress, subjectAddress, accType, address(acc))

    \State \textbf{emit} ACCDeployed(address(acc), msg.sender)

    \Return address(acc)
\EndFunction
\end{algorithmic}
\end{algorithm}

\subsubsection{Utility Functions for Framework Operations}
Utility algorithms assist in implementing core functionality, ensuring efficiency and consistency across the framework.

\textbf{Algorithm 9 System Maintenance and Role Operations} 

This algorithm identifies, logs, and penalizes malicious behavior by peers, enforcing penalties such as token deductions and temporary blocks to deter violations and maintain network integrity. Its purpose is to ensure network security by promptly addressing violations and discouraging malicious actions, thereby maintaining a trustworthy and cooperative environment through the application of appropriate penalties. The implementation involves several steps: first, it validates that a reason for the report is provided; second, it retrieves the peer's misbehavior record and the current time; third, it determines the penalty based on the violation reason; fourth, it updates the misbehavior history with the new violation; fifth, it updates the peer's status according to the severity level; sixth, it deducts tokens from the peer's balance using the Token Contract; seventh, it blocks the peer for a specified duration if necessary; and finally, it returns details of the penalty applied. The impact of this algorithm is that by enforcing penalties consistently and proportionally, it deters malicious behavior and encourages peers to comply with network policies, thereby maintaining a secure and reliable network environment.

\makeatletter
\renewcommand{\ALG@name}{Algorithm} 
\renewcommand{\algorithmicindent}{1.5em} 
\makeatother

\begin{algorithm}
\caption{\scriptsize System Maintenance and Role Operations} 
\begin{algorithmic}[1]
\scriptsize 
\Function{addToGlobalResourceTable}{addr a, str name, str type}
    \State newHead $\gets$ Member(a, name, type)
    \State globalResourceTable.push(newHead)
\EndFunction

\Function{addToLocalResourceTable}{addr a, str name, str type}
    \State newMember $\gets$ Member(a, name, type)
    \State localResourceTable[type].push(newMember)
\EndFunction

\Function{removeFromLocal}{str type, addr a}
    \State localResourceTable[type].remove(a)
\EndFunction

\Function{promoteSecondaryHead}{str type}
    \For{each member in localResourceTable[type]}
        \If{member.role == Role.SECONDARY\_GROUP\_HEAD}
            \State roles[member.addr] $\gets$ Role.PRIMARY\_GROUP\_HEAD
            \State typePrimaryHead[type] $\gets$ member.addr
            \State \textbf{emit} RoleAssigned(member.addr, Role.PRIMARY\_GROUP\_HEAD, type)
            \State break
        \EndIf
    \EndFor
\EndFunction

\Function{burnTokens}{addr a}
    \State burn(a, balanceOf(a))
\EndFunction
\end{algorithmic}
\end{algorithm}

All tests were executed on a private Ethereum network simulated locally using the Hardhat development framework, allowing full control over test conditions and peer behavior.

\begin{table}[h!]
\centering
\caption{\large Device Specification}
\label{tab:device_spec}
\renewcommand{\arraystretch}{1.2} 
\normalsize\sffamily 
\setlength{\tabcolsep}{8pt} 
\begin{tabular}{|>{\centering\arraybackslash}m{3cm}|>{\centering\arraybackslash}m{4.0cm}|} 
\hline
\rowcolor{blue!15} 
\textbf{Component} & \textbf{Details} \\ \hline
Device & OMEN by HP, 17-an0xx \\ \hline
CPU & Intel(R) Core(TM) i7-7700HQ CPU @ 2.81 GHz, x64-based processor \\ \hline
Operating System & Microsoft Windows 10 Home (64 bit) \\ \hline
Memory & 32.0 GB RAM \\ \hline
Storage & HDD: 1 TB, SSD: 512 GB \\ \hline
\end{tabular}
\end{table}

\subsection{Hardware used}
\label{tab:device_specs}
The combination of high RAM and SSD storage, as seen in Table \ref{tab:device_spec} allowed for swift data access and processing, suitable for high-speed blockchain operations and to handle large-scale data exchanges without lag.

\subsection{Software Stack}

Our implementation uses an extensive software stack, as seen in Table \ref{tab:software_tools}. Solidity is used to write smart contracts such as RoleToken and RoleBasedAccessControl. OpenZeppelin provides pre-tested libraries for secure ERC20 token management and access control features. Truffle streamlines contract development, deployment, and testing, while Ganache offers a local blockchain environment for scenarios requiring multiple accounts. Remix IDE makes it easy to prototype and debug quickly. Testing is structured with Mocha and Chai to ensure that contracts function as intended through organized tests and clear validation. Automated deployment and interaction scripts are built in Node.js, with Web3.js enabling communication with the Ethereum blockchain.

\begin{table*}[h]
\centering
\caption{Table of Software Tools}
\normalsize\sffamily
\label{tab:software_tools}
\begin{tabular}{|m{3cm}|m{2.1cm}|m{10.5cm}|}

\rowcolor{blue!15} 
\hline
\textbf{Tool} & \textbf{Type} & \textbf{Purpose / Usage} \\
\hline
\textbf{Solidity} & Contract Language & Writing contract logic to define roles, access permissions, and token management for the system. \\
\hline
\textbf{OpenZeppelin} & Contract Library & Secure, standardized libraries such as ERC20 for token handling and access control; includes test helpers for accurate validation of contract behavior. \\
\hline
\textbf{Truffle Suite} & Development Framework & Compilation, deployment, and structured testing of contracts, integrated with Mocha for running tests and Chai for assertion handling. \\
\hline
\textbf{Ganache} & Local Blockchain & Used to overcome the limitation on the number of accounts in Truffle, creating a controlled testing environment with multiple accounts. \\
\hline
\textbf{Remix} & IDE & Browser-based IDE for quick prototyping, deploying, and debugging Solidity contracts. Useful for initial contract testing. \\
\hline
\textbf{Mocha} & Testing Framework & Structures test cases, organizes tests, and enables logical debugging for contract functionality, with concise and readable test definitions. \\
\hline
\textbf{Chai} & Assertion Library & Provides assertions to validate expected contract behaviors and outcomes, ensuring the system behaves as intended. \\
\hline
\textbf{JavaScript (Node.js)} & Scripting Language & Automated contract interaction and test execution, handling complex interactions and validating role assignments. \\
\hline
\textbf{Web3.js} & Ethereum JS API & Manages blockchain interactions and scalability strictly for testing transfers, role-based checks, and balance validation. \\
\hline
\end{tabular}
\end{table*}

\subsection{Functional Validation}

This subsection describes the testing process, summarizing eight key test cases designed to validate role assignments, state transitions, and access control policies under various scenarios. The primary goal of our testing framework is to identify operational bottlenecks in real-world scenarios and ensure that the proposed access control system performs reliably under diverse conditions. A summary table is displayed in Table \ref{tab:test_cases}. Unlike traditional single-focus frameworks that prioritize either security or performance, our approach integrates a comprehensive trade-off analysis to balance both aspects. 

Each test case includes a detailed setup, execution steps, and results and they are designed to simulate specific scenarios, such as handling excessive access requests, managing peer state transitions, or mitigating malicious activities. 

Our framework identifies and addresses trade-offs between security and performance effectively. For instance, while robust security measures might introduce minor latency, the framework ensures that such trade-offs do not compromise system usability. The following subsections describe the various test cases and their outcomes in detail.

\section{Evaluation Methodology} \label{Sec:testcases}
Our test plan highlights the unique strengths and state-of-the-art features of our access control architecture. \textit{Functional Testing} ensures our system's role assignments, token rewards, penalties, and state transitions work correctly. \textit{Performance Testing} verifies the responsiveness and stability of our framework under high load, proving its suitability for blockchain applications. \textit{Security Testing} confirms that our architecture effectively handles blockchain-specific threats, such as Sybil attacks and smart contract vulnerabilities. \textit{Scalability Testing} demonstrates our system’s capability to efficiently support a growing number of peers and transactions. Lastly, \textit{Integration Testing} shows that smart contracts, external libraries, and user interfaces function together smoothly. Together, these tests validate that our architecture offers a scalable, secure, and integrated solution for decentralized access control.

\begin{table*}[h]
\centering
\caption{Test Case Summary}
\label{tab:test_cases}
\normalsize\sffamily 

\resizebox{\textwidth}{!}{
\begin{tabular}{|m{0.9cm}|m{3.5cm}|m{2.5cm}|m{2.5cm}|m{4.5cm}|m{3.5cm}|}
\hline
\rowcolor{blue!15} 
\textbf{Test Cases} & \textbf{Objective} & \textbf{Prerequisites} & \textbf{Contracts Affected} & \textbf{Test Steps} & \textbf{Results} \\ \hline
T1 & Verify role assignment, token initialization, resource table population & Contracts deployed. Accounts initialized. & RoleBased-AccessControl-Contract, TokenContract & 1. Assign roles to peers with initial tokens. 2. Retrieve resource tables. & Roles and tokens assigned correctly. Resource tables accurate. peers in ``benign'' state. \\ \hline
T2 & Verify token rewards for successful actions, peer remains in ``benign'' state. & Access-ControlContract deployed for secondaryGroupHead1. & AccessControl-Contract, JudgeContract, TokenContract & 1. Add policy allowing SGH to ``view'' ``test.jpg''. 2. SGH accesses ``test.jpg''. 3. Check token balance. & Access granted. Token balance increased by +5. State remains ``benign''. \\ \hline
T3 & Unauthorized attempt triggers ``benign'' $\rightarrow$ ``suspicious'' state transition with token deduction. & RegularMember1 in ``benign'' state with +30 tokens. & AccessControl-Contract, JudgeContract, TokenContract & 1. RegularMember1 attempts unauthorized ``edit'' on ``test.jpg''. 2. Observe response. & Access denied. Token balance decreased by -10. State is ``suspicious''. \\ \hline
T4 & Severe misbehavior (data tampering) leads to ``suspicious'' $\rightarrow$ ``malicious'' transition with severe penalties. & RegularMember1 in ``suspicious'' state. & AccessControl-Contract, JudgeContract, TokenContract & 1. RegularMember1 attempts ``delete'' action on ``test.jpg''. 2. Observe response. 3. Check state and tokens. & Action denied. Token balance decreased by -20. State is ``malicious''. \\ \hline
T5 & Verify recovery from ``malicious'' to ``benign'' after 2 days with no violations. & RegularMember1 in ``malicious'' state. & JudgeContract, RoleBasedAccessControlContract & 1. Wait 2 days without violations. 2. Check state transition. & State transitioned to ``benign''. All access restored. Tokens unchanged. \\ \hline
T6 & Ensure peers in ``suspicious'' state can return to ``benign'' after successful actions. & RegularMember1 in ``suspicious'' state. & AccessControl-Contract, JudgeContract, TokenContract & 1. RegularMember1 performs authorized actions. 2. Check state transition. 3. Verify full access and tokens. & State transitioned to ``benign''. Full access restored. Tokens increased accordingly. \\ \hline
T7 & Excessive access attempts lead to ``benign'' $\rightarrow$ ``suspicious'' transition with token deduction. & RegularMember2 in ``benign'' state. & AccessControl-Contract, JudgeContract, TokenContract & 1. RegularMember2 makes excessive access attempts. 2. Observe response. 3. Check state and tokens. & Excessive attempts detected. Token balance decreased by -5. State is ``suspicious''. \\ \hline
T8 & Service disruption attempts result in immediate ``malicious'' state and severe penalties. & Peer Attempts to Disrupt Service. & AccessControl-Contract, JudgeContract, TokenContract & 1. peer simulates service disruption. 2. Observe detection. 3. Check state and tokens. & Service disruption detected. Token balance decreased by -15. State is ``malicious''. \\ \hline
\end{tabular}
}
\end{table*}

It is important to note that, we purposefully include the full output of each test case to ensure complete traceability. This decision is motivated by two key reasons. First, it allows readers to follow the exact sequence of state transitions, event emissions, and token operations as they unfold during execution. Second, it improves reproducibility, addressing a common limitation we observed in several related works where implementation details were either omitted or only partially disclosed.


\subsection{T1: Verify Role Assignment}

\begin{figure*}[htbp]
\centering
\includegraphics[width=0.98\textwidth]{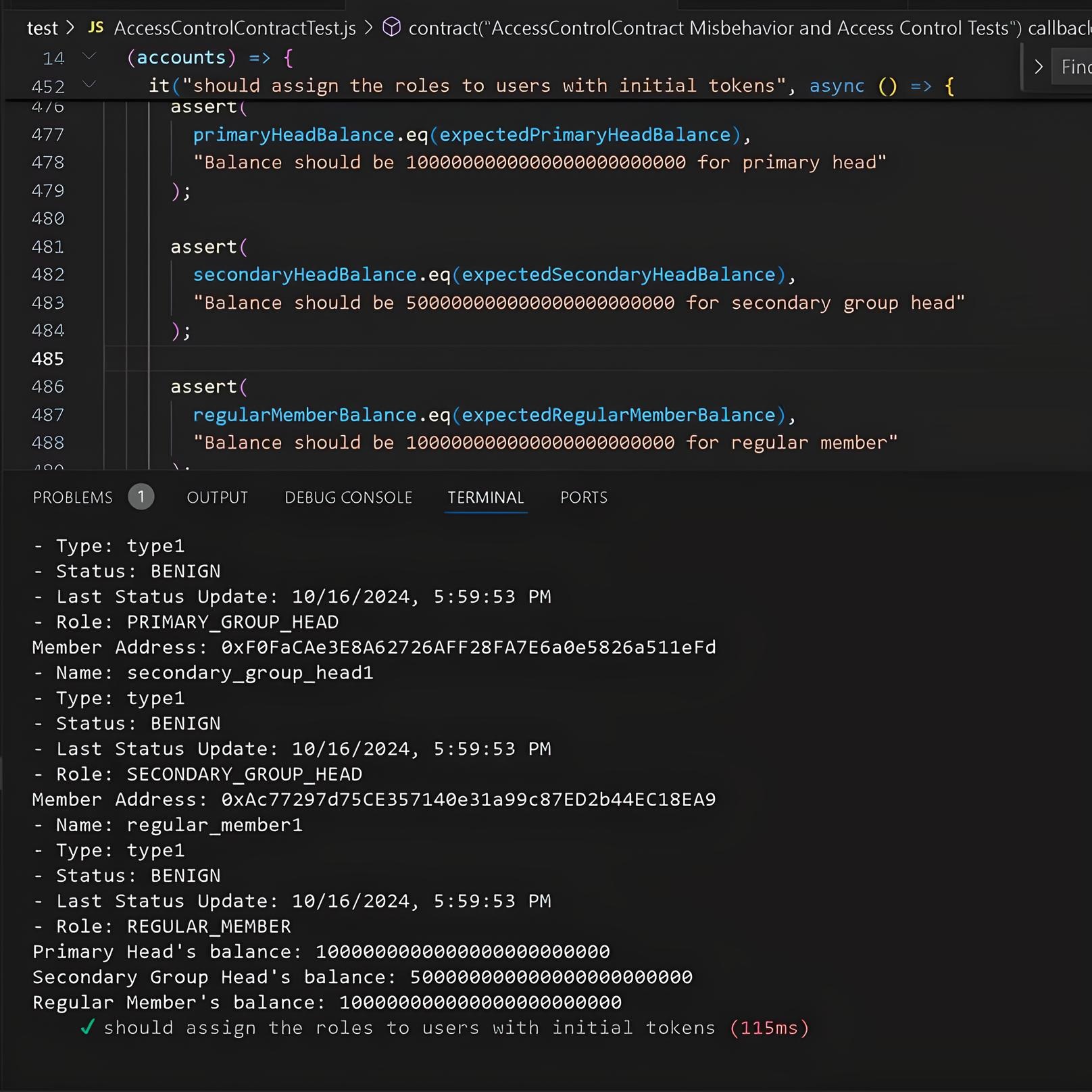}
\caption{Role Assignment and Initialization (Test Case T1)}
\label{fig:role_assignment}
\end{figure*}

This test case, as seen in Figure \ref{fig:role_assignment} verifies the system's functionality in assigning roles to peers and initializing their token balances accurately. It focuses on three hierarchical roles: ``PGH'', ``SGH'', and ``RM''. Each role is assigned a specific initial token balance to reflect its network responsibilities: 1,000,000 tokens for the ``PGH'', 500,000 tokens for the ``SGH'', and 1,000,000 tokens for the ``RM''. The test uses assertions to ensure that the actual balances match the expected values and confirms that all peers remain in the ``benign'' state upon initialization. Additionally, the system logs and validates the role assignments and balances in real-time, completing the operation within an efficient 115 ms. This test is essential for establishing the core functionality of role assignment and resource allocation, ensuring the network's structure and initial configuration operate as intended. Figure \ref{fig:role_assignment} presents the corresponding code snippet, showing how authorized roles are verified before granting permissions. 

\subsection{T2: Verify Token Rewards for Successful Actions}

\begin{figure*}[htbp]
\centering
\includegraphics[width=0.98\textwidth]{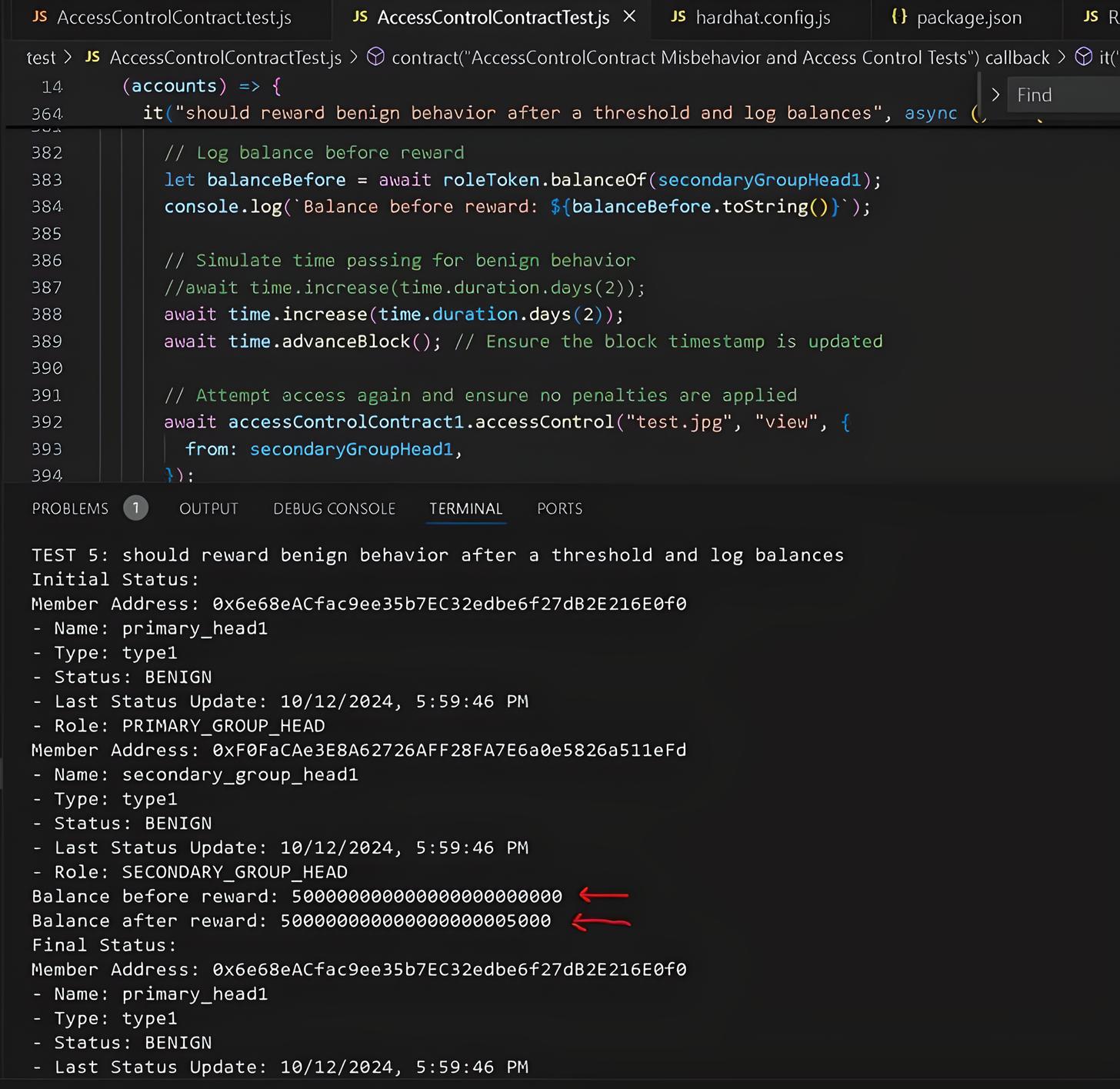}
\caption{Successful Access and Token Reward (Test Case T2)}
\label{fig:test_case_t2}
\end{figure*}

This test case, as seen in Figure \ref{fig:test_case_t2} evaluates the system's ability to reward ``benign'' behavior after a predefined threshold and validate the logging of token balances before and after the reward. The test involves continuous compliant for extended period (i.e., 2 days) by a ``SGH'' role, during which no violations are committed. The system then processes an access request for a resource, ensuring no penalties are applied. The token balance is logged before and after the reward, with an increase observed, indicating the reward has been successfully granted. This test verifies the system's responsiveness to positive peer actions and its ability to enforce rewards as part of its token-based incentive mechanism. A representative code snippet implementing this test case is provided in Figure \ref{fig:test_case_t2}. 

\subsection{T3: Unauthorized Attempt Triggers State Transition}
\begin{figure*}[htbp]
\centering
\includegraphics[width=0.98\textwidth]{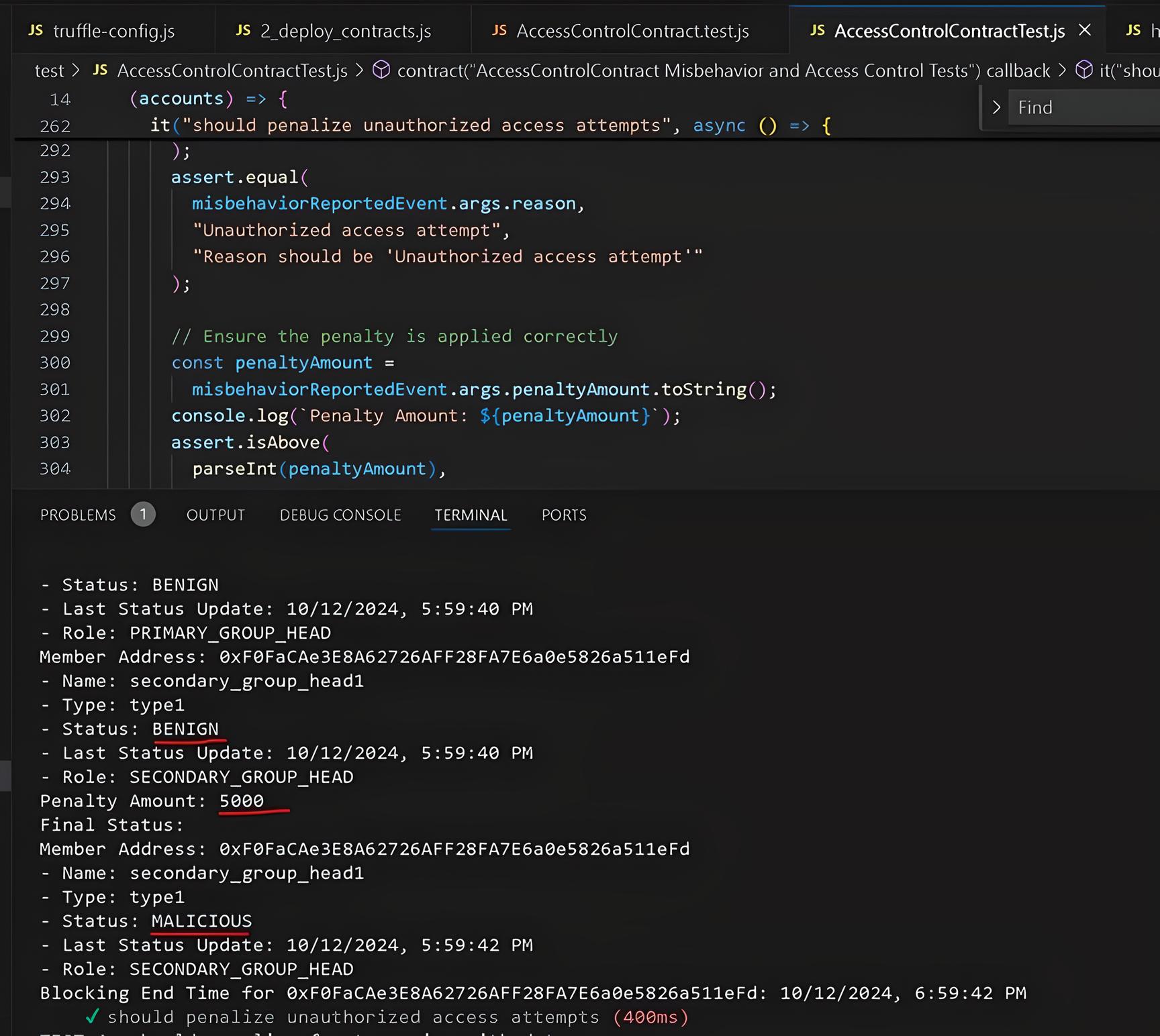}
\caption{Unauthorized Access Attempt and Penalty (Test Case T3)}
\label{fig:test_case_t3}
\end{figure*}

This test case, as seen in Figure \ref{fig:test_case_t3} assesses the system's ability to detect and penalize unauthorized access attempts appropriately. In the scenario, a ``SGH'' peer attempts to access a resource without proper authorization. The system identifies the violation and logs it under the reason ``Unauthorized access attempt.'' A penalty of 5,000 tokens is applied to the violating peer, and their status transitions from ``benign'' to ``malicious''. Additionally, the peer is temporarily blocked until the violation is resolved or the penalty conditions are met. This test verifies the correctness of the penalty application and ensures the system maintains the integrity of its access control policies by discouraging unauthorized actions. The corresponding implementation for this scenario is shown in the code snippet in Figure \ref{fig:test_case_t3}.

\subsection{T4: Severe Misbehavior Leads to Malicious State}

\begin{figure*}[htbp]
\centering
\includegraphics[width=0.98\textwidth]{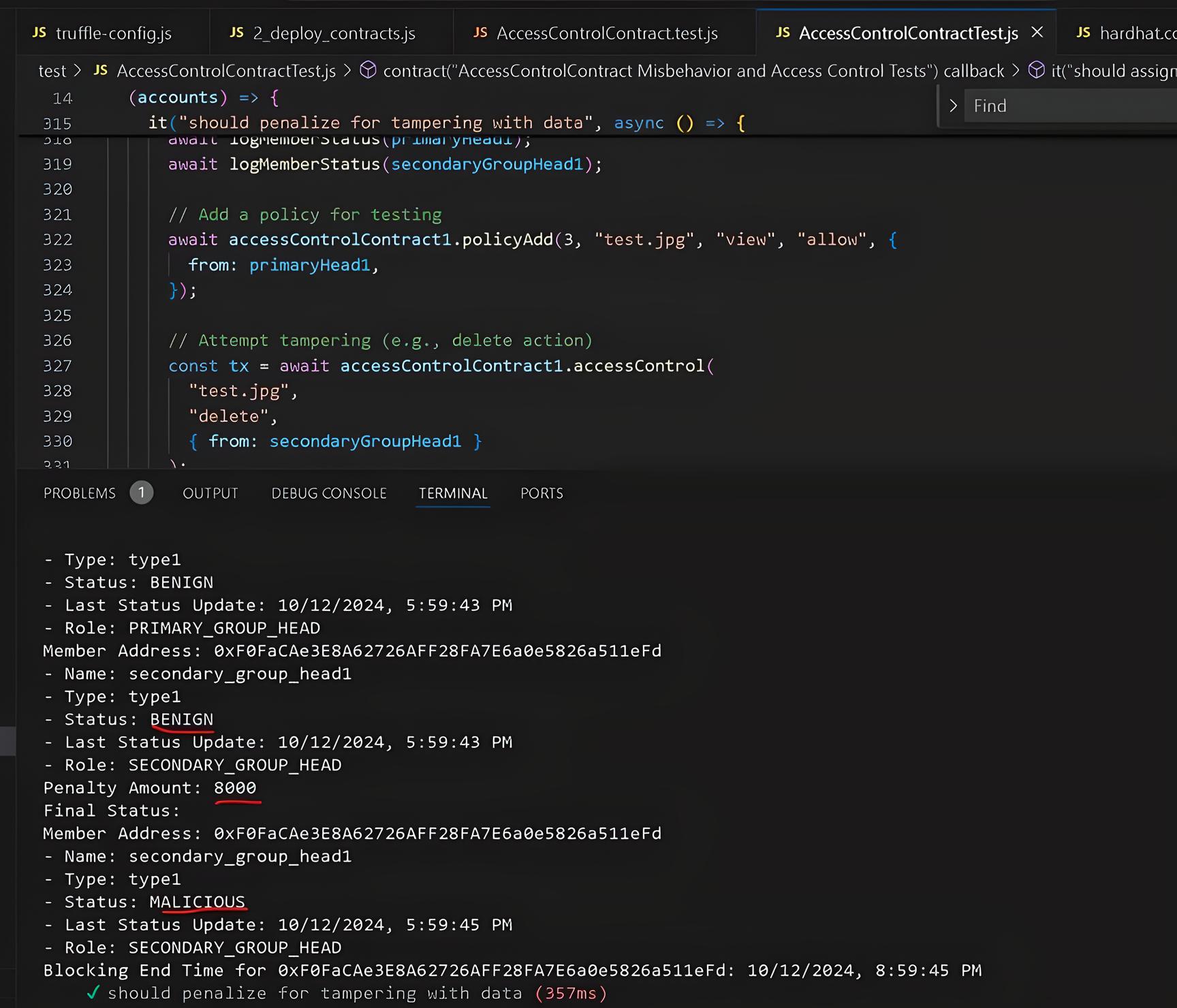}
\caption{Severe Misbehavior and State Transition (Test Case T4)}
\label{fig:test_case_t4}
\end{figure*}

This test case, as seen in Figure \ref{fig:test_case_t4} evaluates the system's ability to penalize attempts at tampering with data, ensuring compliance with the access control policies. A ``SGH'' peer attempts an unauthorized ``delete'' action on a resource (``test.jpg'') after a legitimate policy for ``view'' access is established by the ``PGH''. The system detects the tampering action and imposes a penalty of 8,000 tokens on the violator. As a result, the violator’s status is changed from ``benign'' to ``malicious'', and they are temporarily blocked from further actions until the penalty or blocking conditions are resolved. This test confirms the sturdiness of the system's tamper detection mechanism and its adherence to access control rules by penalizing violators effectively. The code implementation of this test scenario is illustrated in Figure \ref{fig:test_case_t4}.

\subsection{T5: Verify Recovery from malicious to benign}

\begin{figure*}[htbp]
\centering
\includegraphics[width=0.98\textwidth]{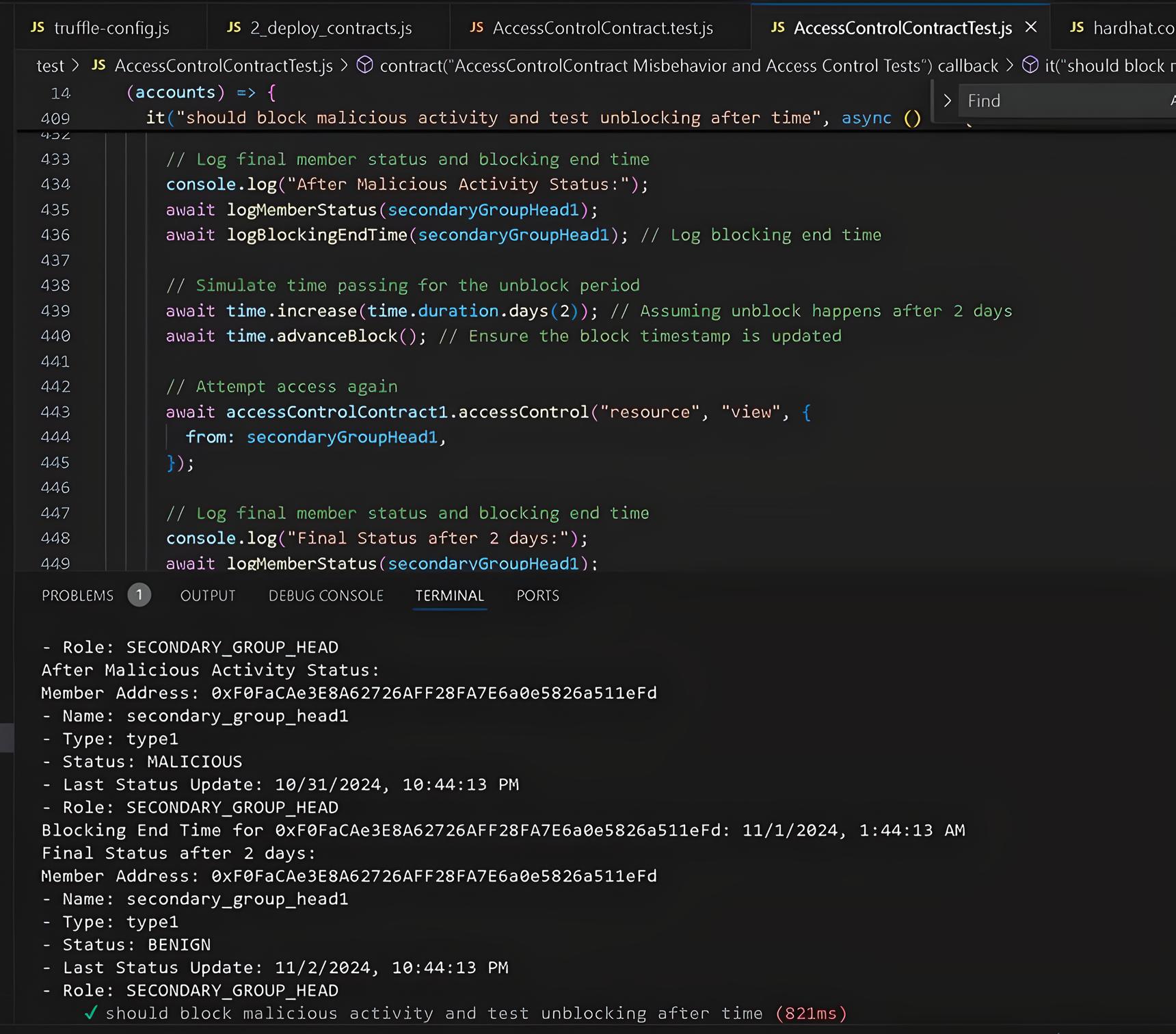}
\caption{Recovery from Malicious State (Test Case T5)}
\label{fig:test_case_t5}
\end{figure*}

This test case, as seen in Figure \ref{fig:test_case_t5} validates the system's ability to handle malicious activity by temporarily blocking the violator and subsequently unblocking them after the designated penalty period. A ``SGH'' engages in unauthorized activity, triggering a status change from ``benign'' to ``malicious'' and a blocking period. The system logs the blocking end time, indicating the precise duration of the penalty. After simulating the passage of the blocking period (2 days in this case), the peer attempts to regain again. The system processes the attempt and restores the peer’s status to ``benign'', confirming that the penalty mechanism is correctly time-bound and functional. This test ensures the system enforces access control policies effectively while providing a recovery path for penalized peers after compliance with the penalty conditions. The code corresponding to this recovery scenario is provided in Figure \ref{fig:test_case_t5}.

\subsection{T6: Recovery from suspicious to benign}
\begin{figure*}[htbp]
\centering
\includegraphics[width=0.98\textwidth]{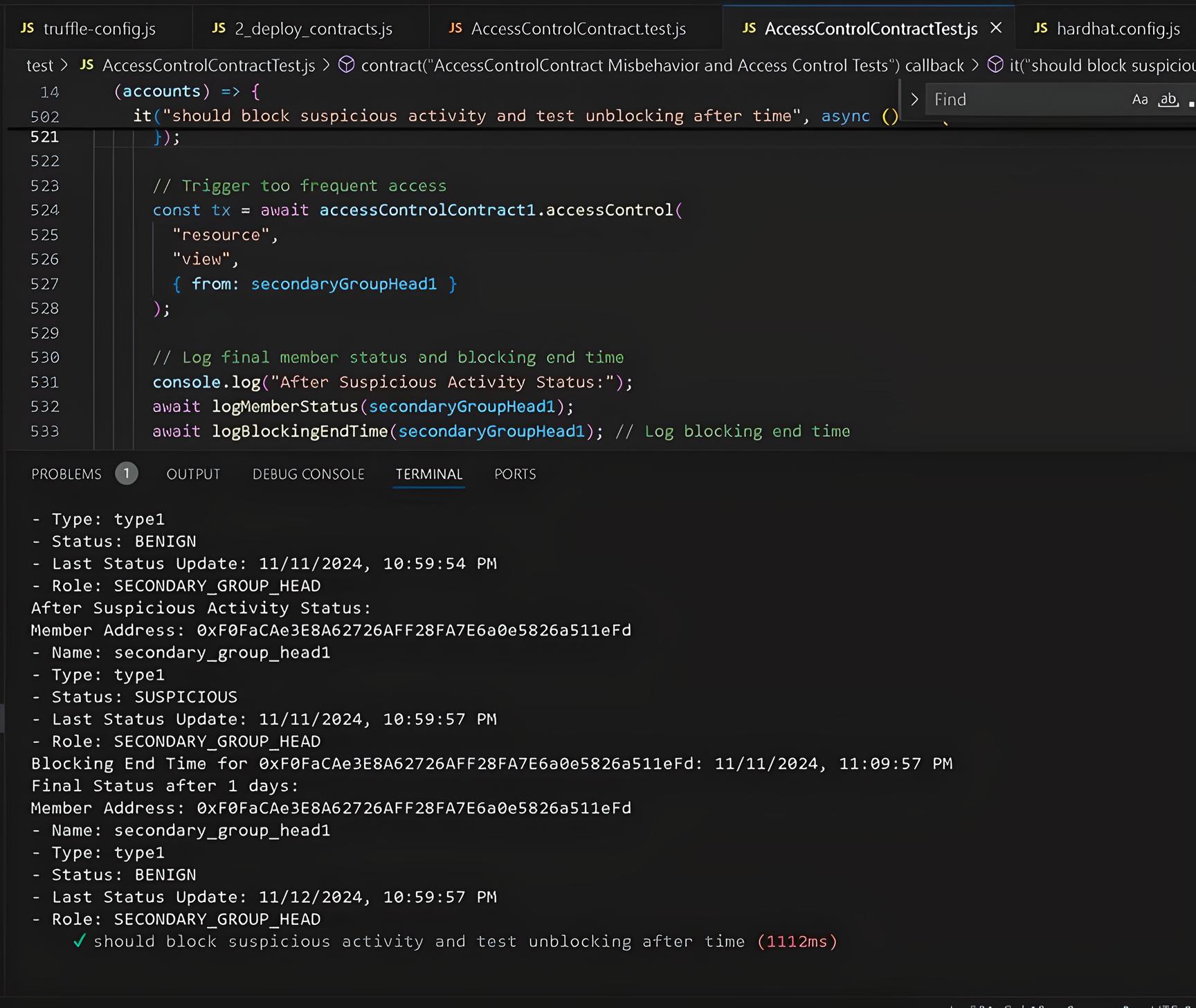}
\caption{Recovery from Suspicious State (Test Case T6)}
\label{fig:test_case_t6}
\end{figure*}

This test case, as seen in Figure \ref{fig:test_case_t6}, demonstrates the system's capability to identify and respond to suspicious activity, applying a temporary block as a preventative measure. This recovery test case T6 is different and more complex than the previous test case T5. In T5, recovery was from ``malicious'' to ``benign'' after two days without any violations. But in T6, the recovery is from ``suspicious'' to ``benign'', and involves both detection of suspicious behavior and shorter block duration. A ``SGH'' triggers the system by performing frequent unauthorized access attempts, causing their status to change from ``benign'' to ``suspicious''. The system logs the blocking end time, reflecting a 1-day blocking duration. After simulating the lapse of the blocking period, the peer's status is reverted to ``benign'', which eventually ensures they regain access upon serving the block. The implementation details of this recovery scenario are shown in the code snippet in Figure \ref{fig:test_case_t6}.

\subsection{T7: Excessive Access Attempts Lead to State Transition}
\begin{figure*}[htbp]
\centering
\includegraphics[width=0.98\textwidth]{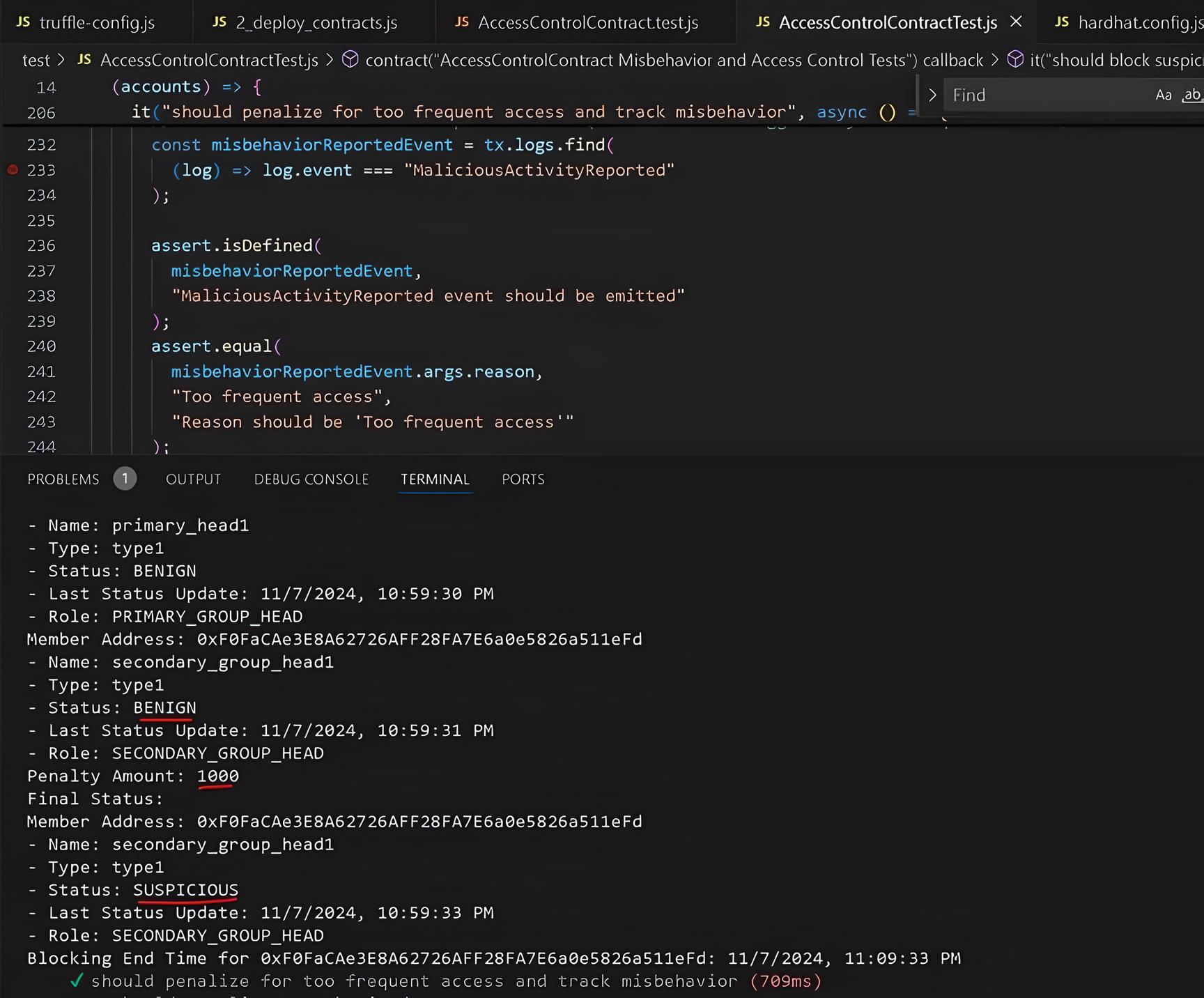}
\caption{Excessive Access Attempts (Test Case T7)}
\label{fig:test_case_t7}
\end{figure*}

This test case, as seen in Figure \ref{fig:test_case_t7}, validates the system's ability to penalize and track misbehavior caused by too frequent access attempts. In the system, normal behavior refers to actions that comply with predefined access policies, such as adhering to frequency limits for resource requests. peers exhibiting normal behavior are considered ``benign'' and maintain their current status without penalties. In contrast, suspicious behavior arises when a peer exceeds acceptable thresholds, such as access frequency limits, signaling potential misuse but not yet escalating to malicious intent. In this scenario, a ``SGH'' triggers the system by exceeding acceptable access frequency, resulting in a penalty of ``1000'' tokens and the emission of a ``maliciousActivityReported'' event. The system logs the reason as ``Too frequent access,'' and the peer's status transitions from ``benign'' to ``suspicious''. Additionally, the system logs the blocking period to prevent further activity during this state. This test ensures the system can handle access violations effectively by imposing penalties, updating peer status, and tracking misbehavior in the event logs. The corresponding code snippet for this excessive access scenario is presented in Figure \ref{fig:test_case_t7}.

\subsection{T8: Service Disruption Attempts Result in Penalties}

\begin{figure*}[htbp]
\centering
\includegraphics[width=0.98\textwidth]{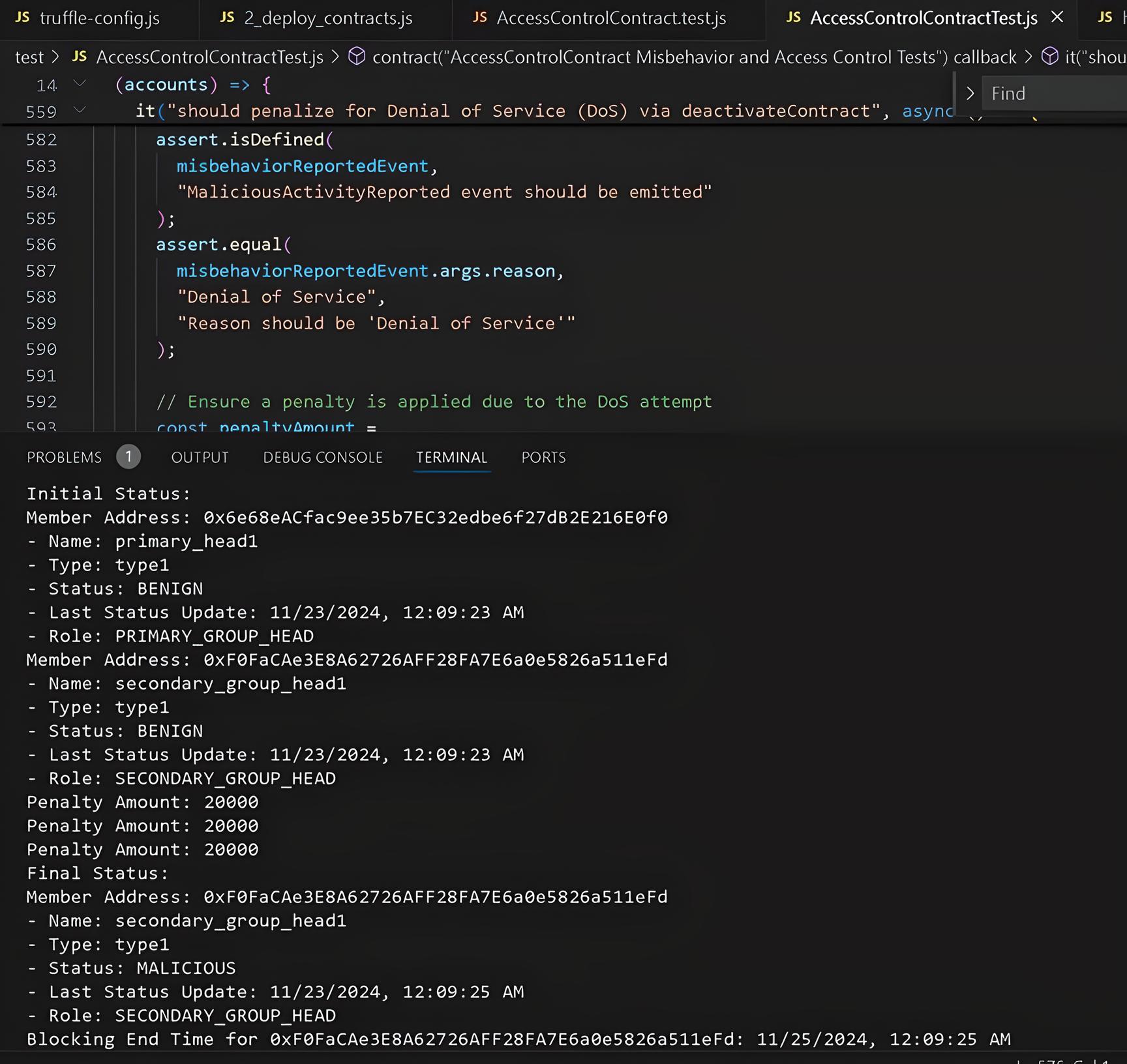}
\caption{Service Disruption Attempt (Test Case T8)}
\label{fig:test_case_t8}
\end{figure*}

This test case, as seen in Figure \ref{fig:test_case_t8} demonstrates the system's ability to handle a Denial of Service (DoS) attempt initiated by a ``SGH'' through a ``deactivateContract'' action. The system successfully detects the malicious behavior, emitting a ``maliciousActivityReported'' event with the reason ``Denial of Service.'' A penalty of ``20000'' tokens is imposed on the offending peer, whose status is subsequently updated from ``benign'' to ``malicious''. To mitigate further harm, the system enforces a blocking period, logging the blocking end time. Furthermore, this framework is designed with flexibility to identify and respond to newer zero-day attack scenarios. Future test scenarios will explore such novel threats by extending existing monitoring capabilities and anomaly detection mechanisms. This test validates the system's sturdiness in identifying, penalizing, and mitigating attacks.

\textit{Code Snippet for T8 can be found in Figure \ref{fig:test_case_t8}}.

Overall, our framework is supported by 9 algorithms that govern key functions. This section also presented 8 distinct test cases to support it's claim. In addition, this section assess the system's capability to withstand operational demands through trade-off analyses.

\section{Performance Evaluation} \label{Sec:perf_eval}

This section presents the comprehensive evaluation framework used to measure \textbf{four} different categories of metrics over \textbf{eight} comprehensive tests. Efficiency metrics, such as latency, execution time, and throughput, measure the system’s responsiveness and operational speed. Efficacy metrics, including transaction success rates, assess the system’s reliability in processing requests. Cost metrics, such as blockchain gas costs and transaction fees, quantify the economic impact of system operations. Standard blockchain and network metrics, including gas consumption, latency, and execution types, provide a broader view of resource utilization and system behavior. Together, these metrics offer an all-inclusive analysis of the framework’s performance under various conditions.

\subsection{Measurement Framework} \label{Sec:meas_framework}
Our systematic approach provides a complete understanding of the system’s performance, and resource efficiency under varying conditions. We present a comprehensive measurement framework that sets a new standard in performance evaluation for blockchain-based systems. We predefined eight comprehensive test cases covering various operational and threat scenarios, measuring five distinct performance metrics consistently across all tests. This structured and homogeneous evaluation approach sets a new state-of-the-art benchmark for blockchain-based access control systems.

We follow industry-standard metrics and best practices while leveraging advanced tools to simulate real-world scenarios. Using a comprehensive software stack, detailed in Table \ref{tab:software_tools}, and automated scripts within Truffle, we simulated diverse peer actions such as access requests, policy updates, and token-based transitions. Each action was precisely logged through smart contracts, capturing critical data such as timestamps, token balances, and role-specific operations. To ensure accuracy, Ganache’s in-memory blockchain was employed to monitor gas usage and execution times in real-time. 

Detailed descriptions of each metric, along with their associated Tools, Techniques, and Processes (TTP), are presented systematically in the subsequent sections.


\subsection{Result Visualization}

This section presents a detailed description of our architecture's comprehensive evaluation metrics. These metrics expand beyond conventional computational approaches. A key strength of our contribution lies in the careful development and refinement of these metrics. Additionally, we present a sophisticated instrumentation of Tools, Techniques, and Processes (TTP) tailored specifically for blockchain environments. This instrumentation can help various network functions, enabling better security and performance.

Our results showcase how the framework stands out due to its ability to optimize gas consumption through a refined smart contract design, ensuring cost-effective operations. Rapid execution times and minimal response delays are achieved, even under high-load or stress scenarios, thanks to dynamic resource and penalty management. The hierarchical architecture distributes workloads efficiently to maintain high performance as the network grows. Moreover, the framework demonstrates exceptional resilience against malicious activities, including threats such as privilege escalation and DDoS attacks, through adaptive state management and automated penalties.

These findings highlight the versatility of our framework, positioning it as a significant advancement over state-of-art models. 

\subsubsection{Gas Usage Analysis}
It represents the computational cost required to execute smart contract operations on the Ethereum blockchain. Optimizing this matric is critical for ensuring the cost-efficiency of blockchain-based frameworks, particularly when deployed in large peer-to-peer networks. For our analysis, we focused on specific events that represent key policy deployment points and critical operations where the core functionalities of our framework are exercised. These include role assignment, access control enforcement, and table access management. They are central to maintaining system security, governance, and resource integrity.

\textit{Tools, Techniques, and Processes (TTP):}
We used Truffle Suite and Ganache in combination with Web3.js scripts that automated the execution of critical operations, such as \texttt{assignRole()}, \texttt{createAccessRequest()}, and \texttt{\\swapRolesBetweenAccounts()}. Each function execution logged gas costs in real-time, ensuring accurate insights into the system's performance under various scenarios. The operations chosen for measurement are role assignment and anomaly handling, which were deliberately selected. These events align with key policy enforcement points where our framework's primary mechanisms, such as token-based incentives, state transitions, and access control decisions, are triggered.

We drew inspiration from existing research, which highlights the importance of analyzing similar policy-critical events in decentralized systems; however, our choice of these specific events showcases our novel contributions, such as dynamic state management, decentralized governance, and efficient resource allocation.

\textit{Implications:} Lower gas usage is essential for designing scalable blockchain-based frameworks, as it reduces operational costs while supporting critical functionalities. By measuring gas consumption at policy deployment points, such as role assignments and access control checks, we identified resource-intensive processes that may create bottlenecks in large-scale networks.

Focusing on key functional events—where our framework actively enforces roles, policies, and state transitions—provided a clear understanding of our system's performance. Based on our findings, optimizing these policy-critical points significantly improved cost efficiency, and reduced gas overhead. Future network designers can similarly identify and optimize such crucial points in their frameworks to maintain cost efficiency.

\vspace{10pt}

\begin{figure}[htbp]
    \centering
\begin{tikzpicture}[scale=0.45]
    \definecolor{myblue}{HTML}{1F7ED2}
    \begin{axis}[
        title={\LARGE Gas Used for Different Operations},
        ymin=0,
        ymax=400000,
        ytick={0, 50000, 100000, 150000, 200000, 250000, 300000, 350000, 400000},
        yticklabels={0, 50000, 100000, 150000, 200000, 250000, 300000, 350000, 400000}, 
        enlarge x limits=0.15,
        bar width=30pt,
        xticklabel style={rotate=45, anchor=north east},
        symbolic x coords={
            Policy Add, 
            Access Control without Misbehavior, 
            Access Control with Misbehavior, 
            Role Assignment, 
            Create Table Access Request, 
            Create Table Access Request with Misbehavior, 
            View Local Resource Table, 
            View Global Resource Table
        },
        xtick=data,
        ybar,
        width=1\textwidth, 
        ymajorgrids=true,
        nodes near coords, 
        every node near coord/.append style={font=\small, anchor=south}, 
        yticklabel style={/pgf/number format/fixed, /pgf/number format/precision=0}, 
        xticklabel style={font=\LARGE},
        ylabel={Gas Used},
        yticklabel style={font=\LARGE},
        ylabel style={yshift=10pt, font=\LARGE},
        every node near coord/.append style={/pgf/number format/fixed, /pgf/number format/precision=0, font=\large}
    ]
        \addplot[draw=none, fill=myblue] coordinates {
            (Policy Add, 155292)
            (Access Control without Misbehavior, 131359)
            (Access Control with Misbehavior, 216165)
            (Role Assignment, 361664)
            (Create Table Access Request, 282930)
            (Create Table Access Request with Misbehavior, 243640)
            (View Local Resource Table, 224943)
            (View Global Resource Table, 197166)
        };
    \end{axis}
\end{tikzpicture}
\caption{Gas Consumption Comparison for Various Operations}
\label{fig:gas_used_operations}
\end{figure}
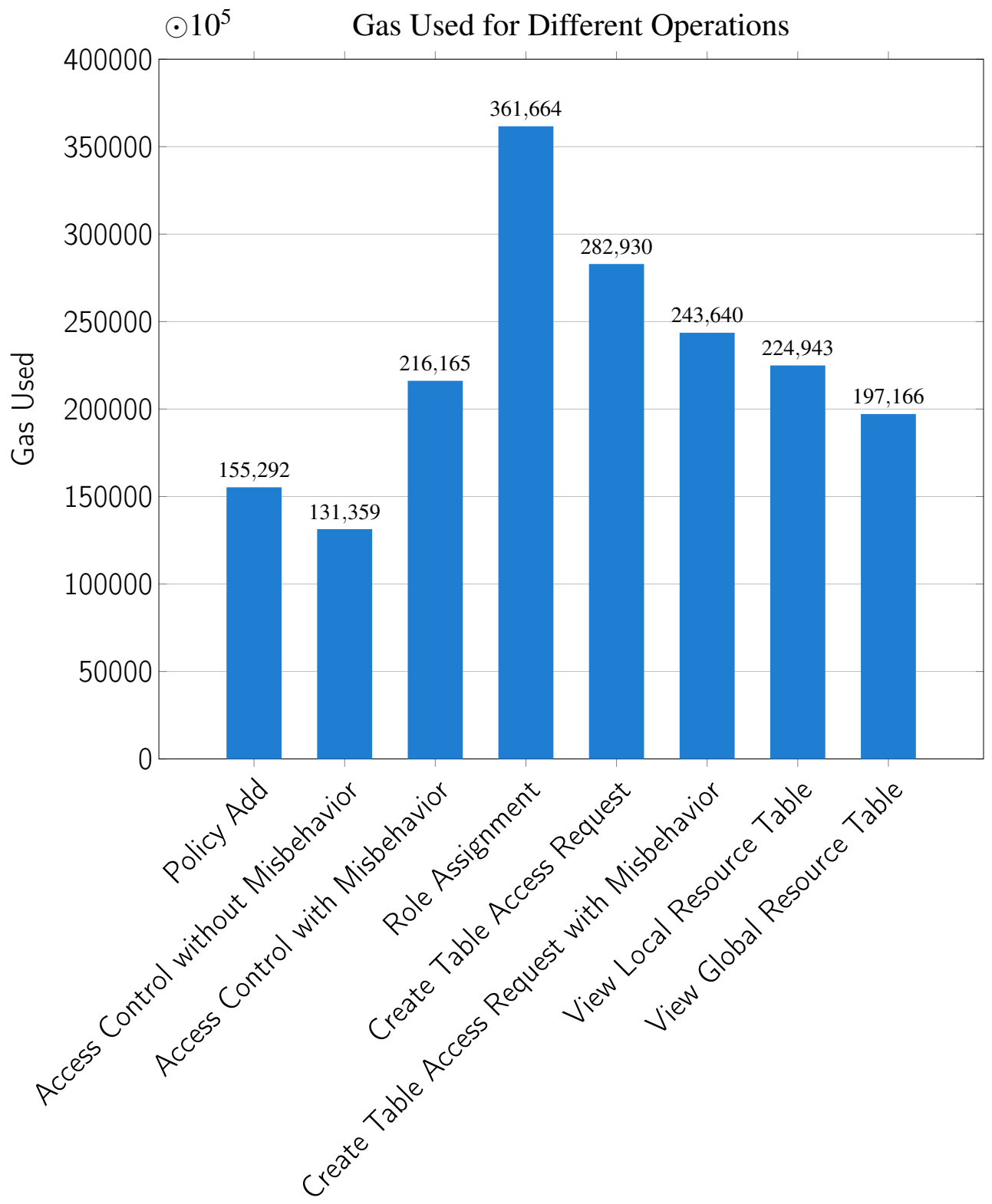

Figure \ref{fig:gas_used_operations} showcases the gas consumption for various system operations, emphasizing the resource demands of each function. Complex operations such as \texttt{Role Assignment} and \texttt{Create Table Access Request} exhibit higher gas usage due to their multifaceted processes, including state transitions, token minting, and data updates. In contrast, simpler tasks such as Access Control without Misbehavior and Viewing the GRT demonstrate lower gas consumption, reflecting their streamlined computational requirements. Operations addressing misbehavior, such as \texttt{Access Control with Misbehavior} and \texttt{Create Table Access Request with Misbehavior}, incur additional gas costs due to anomaly detection, penalty enforcement, and state updates. This analysis highlights resource-intensive areas, guiding optimization efforts to improve efficiency and reduce operational costs.

\begin{figure}[htbp]
    \centering
\begin{tikzpicture}[scale=0.42]
\begin{axis}[
    title={\huge Gas Costs for malicious Attacks (During and After Attack)},
    title style={
            at={(0.5,1.12)}, 
            anchor=north,   
        },
    xlabel={Type of Attack},
    xlabel style={font=\huge}, 
    ylabel={Gas Cost ($\times 10^5$)},
    ylabel style={font=\huge}, 
    ymin=0,
    ymax=3.5,
    ytick={0, 0.5, 1, 1.5, 2, 2.5, 3.0, 3.5},
    yticklabels={0, 0.5, 1, 1.5, 2, 2.5, 3, 3.5},
    ybar,
    bar width=0.9cm,
    enlargelimits=0.15,
    legend style={at={(0.5,1)},
        anchor=north,legend columns=-1},
    xtick={1,2,3,4,5},
    xticklabels={
        Privilege \\ Escalation,
        Unauthorized \\ Access,
        Tampering \\ with Data,
        Too Frequent \\ Access,
        Denial \\ of Service
    },
    xticklabel style={align=center, font=\LARGE}, 
    yticklabel style={font=\LARGE}, 
    width=1.1\textwidth, 
    ymajorgrids=true,
    nodes near coords, 
    every node near coord/.append style={font=\large, anchor=south}, 
    yticklabel style={/pgf/number format/fixed, /pgf/number format/precision=1},
    every node near coord/.append style={/pgf/number format/fixed, /pgf/number format/precision=1, font=\large},
    legend style={
        font=\Huge 
    }
]

    \definecolor{duringColor}{RGB}{255,10,0}   
    \definecolor{afterColor}{RGB}{0,128,0}       

\addplot[fill=duringColor] coordinates {
    (1, 2.47936)
    (2, 2.33356)
    (3, 2.10821)
    (4, 1.64925)
    (5, 1.42534)
};
\addlegendentry{During Attack}

\addplot[fill=afterColor] coordinates {
    (1, 3.07377)
    (2, 1.15239)
    (3, 1.37655)
    (4, 1.14787)
    (5, 0.36762)
};
\addlegendentry{After Attack}

\end{axis}
\end{tikzpicture}
\caption{Cost analysis for malicious attacks (both during and post-attack) incorporates STRIDE threat modeling}
\label{fig:gas_costs_malicious_attacks}
\end{figure}

Figure \ref{fig:gas_costs_malicious_attacks} highlights the gas costs associated with various types of malicious attacks during and after their occurrence. Notably, ``Privilege Escalation'' and ``Unauthorized Access'' incur high gas costs during the attack phase, with ``Privilege Escalation'' peaking at 307,377 gas. These elevated costs stem from intensive operations such as validation checks and countermeasures to prevent attacks. Post-attack gas costs are generally lower for actions such as ``Unauthorized Access'' and ``Too Frequent Access'', as recovery mechanisms such as logging and isolating malicious peers are less resource-intensive compared to active prevention.

Conversely, attacks such as ``Denial of Service'' and ``Tampering with Data'' exhibit a marked decrease in gas usage post-attack. This decrease such asly happens because resource-heavy processes, such as access control or data validation, are bypassed during recovery. However, this reduction in gas costs might introduce vulnerabilities or incomplete resolutions. Possible improvements include strategies such as adopting stronger intra-group isolation or stricter inter-group access policies. Additionally, implementing more aggressive measures, such as completely boycotting malicious peers temporarily, could balance cost savings with security. Regular monitoring instead of time-bound sanctions might also offer a cost-effective recovery, ensuring continuous assessment of peer trustworthiness without incurring high gas expenses. These insights underscore the importance of carefully designing response mechanisms to balance effective prevention and efficient mitigation.


\subsubsection{Execution Time and Responsiveness}
Execution time measures the duration taken to complete specific operations within the smart contract. Responsiveness is vital in access control frameworks to ensure seamless peer experiences, particularly under time-sensitive conditions such as resource access requests or state transitions. Lower execution times improve system performance and operational efficiency.

\textit{Tools, Techniques, and Processes (TTP): }

We measured execution times accurately using high-resolution timers in Node.js scripts. These timers tracked each operation precisely from start to finish. We monitored important events such as role assignments, table access requests, and access control checks involving anomaly handling. These events represent different functionalities and complexities in our system.

We also used cycle accounting to track computational effort across various network events. This method recorded detailed execution cycles for tasks such as state transitions, resource validations, and misbehavior handling. By checking timings at each step, we found performance bottlenecks and chances for improvement.

\textit{Implications:} Lower execution times mean better system responsiveness. Good responsiveness ensures a smooth peer experience and reliable system operation under different conditions. Cycle accounting helped us find specific steps, such as policy validation or state updates, causing most delays. This information is very helpful for improving operations that happen often or at critical points, such as access control enforcement.

Cycle accounting is a reliable way to analyze timings across different operation stages. It helps us understand system performance fully. Future designers can use this approach to improve not only single operations but the complete workflow. This will make the system faster and more efficient.

\begin{figure}[htbp]
    \centering
\begin{tikzpicture}[scale=0.55]
\begin{axis}[
    title={\Large Execution Time of Operations},
    height=\textwidth * 0.73,
    width=\textwidth * 0.63,
    xlabel={\Large Execution Time (ms)},
    xmax=700,
    yticklabels={
        Policy Add, 
            Access Control without \\Misbehavior, 
            Access Control with \\Misbehavior, 
            Role Assignment, 
            Create Table Request, 
            Create Table Request \\with Misbehavior, 
            View LRT,
        View GRT
    },
    xbar,  
    bar width=0.85cm,
    yticklabel style={font=\Large},
    xticklabel style={font=\Large},
    enlargelimits=0.15,
    x tick label style={/pgf/number format/.cd,fixed,precision=2},
    yticklabel style={align=left, xshift=-0.5em},
    ytick={0,1,2,3,4,5,6,7} 
]
\addplot[fill=cyan!50] coordinates {
    (78.12,0) (230.11,1) (292.13,2) (178.84,3) (361.04,4) (393.13,5) (546.27,6) (301.33,7)
};

\node at (axis cs:78.12,0) [anchor=west, font=\large] {78.12 ms};
\node at (axis cs:230.11,1) [anchor=west, font=\large] {230.11 ms};
\node at (axis cs:292.13,2) [anchor=west, font=\large] {292.13 ms};
\node at (axis cs:178.84,3) [anchor=west, font=\large] {178.84 ms};
\node at (axis cs:361.04,4) [anchor=west, font=\large] {361.04 ms};
\node at (axis cs:393.13,5) [anchor=west, font=\large] {393.13 ms};
\node at (axis cs:546.27,6) [anchor=west, font=\large] {546.27 ms};
\node at (axis cs:301.33,7) [anchor=west, font=\large] {301.33 ms};

\end{axis}
\end{tikzpicture}
\caption{Execution Time of Operations}
\label{fig:execution_time_operations}
\end{figure}
 
 Figure \ref{fig:execution_time_operations} displays the execution times of various operations within the system, measured in milliseconds (ms). The operation View Local Resource Table is the most time-consuming, taking 546.27 ms to execute, possibly due to extensive data processing or validation steps. Other high-duration operations include \texttt{Create Table Access Request} with Misbehavior at 393.13 ms and View GRT at 301.33 ms, which may reflect the additional handling required for access control and misbehavior monitoring. In contrast, Policy Add is the fastest operation at 78.12 ms, indicating its relative simplicity. The analysis reveals that functions involving table access and misbehavior handling generally take longer to execute, highlighting areas that may benefit from optimization to improve overall system efficiency.

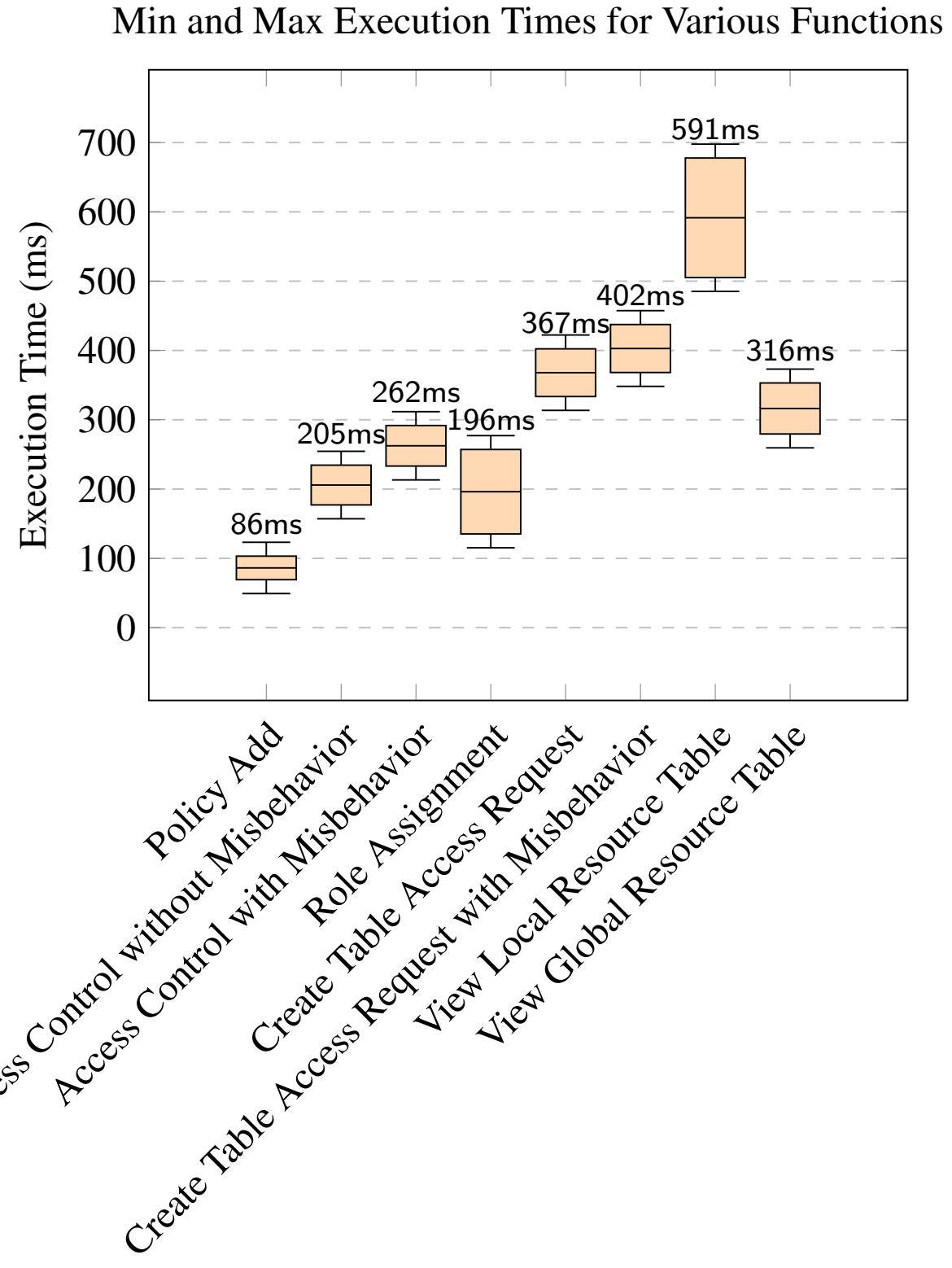
\begin{figure}[htbp]
\centering
\begin{tikzpicture}[scale=0.9]
    \begin{axis}[
        boxplot/draw direction=y,
        ylabel={Execution Time (ms)},
        xtick={1,2,3,4,5,6,7,8},
        xticklabels={Policy Add, 
            Access Control without Misbehavior, 
            Access Control with Misbehavior, 
            Role Assignment, 
            Create Table Access Request, 
            Create Table Access Request with Misbehavior, 
            View Local Resource Table, 
            View Global Resource Table},
        ymajorgrids = true,
        enlargelimits=0.15,
        ytick={0, 100, 200, 300, 400, 500, 600, 700},  
        ymin=0,  
        ymax=700,  
        grid style=dashed,
        title={Min and Max Execution Times for Various Functions},
        xticklabel style={rotate=45, anchor=north east}, 
    ]


        \addplot+[
            boxplot prepared={
                lower whisker=49.23, 
                lower quartile=69.23,
                median=86.23, 
                upper quartile=103.22,
                upper whisker=123.22, 
            },
            fill=orange!30,
            draw=black
        ] coordinates {(1,69.23) (1,103.22)};
        \node[font=\footnotesize] at (axis cs:1, 150) {86ms};

        \addplot+[
            boxplot prepared={
                lower whisker=157.13, 
                lower quartile=177.13,
                median=205.82, 
                upper quartile=234.50,
                upper whisker=254.50, 
            },
            fill=orange!30,
            draw=black
        ] coordinates {(2,177.13) (2,234.50)};
        \node[font=\footnotesize] at (axis cs:2, 280) {205ms};

        \addplot+[
            boxplot prepared={
                lower whisker=213.20, 
                lower quartile=233.20,
                median=262.44, 
                upper quartile=291.68,
                upper whisker=311.68, 
            },
            fill=orange!30,
            draw=black
        ] coordinates {(3,233.20) (3,291.68)};
        \node[font=\footnotesize] at (axis cs:3, 340) {262ms};

        \addplot+[
            boxplot prepared={
                lower whisker=115.25, 
                lower quartile=135.25,
                median=196.19, 
                upper quartile=257.12,
                upper whisker=277.12, 
            },
            fill=orange!30,
            draw=black
        ] coordinates {(4,135.25) (4,257.12)};
        \node[font=\footnotesize] at (axis cs:4, 300) {196ms};

        \addplot+[
            boxplot prepared={
                lower whisker=313.52, 
                lower quartile=333.52,
                median=367.93, 
                upper quartile=402.33,
                upper whisker=422.33, 
            },
            fill=orange!30,
            draw=black
        ] coordinates {(5,333.52) (5,402.33)};
        \node[font=\footnotesize] at (axis cs:5, 440) {367ms};

        \addplot+[
            boxplot prepared={
                lower whisker=348.19, 
                lower quartile=368.19,
                median=402.81, 
                upper quartile=437.43,
                upper whisker=457.43, 
            },
            fill=orange!30,
            draw=black
        ] coordinates {(6,368.19) (6,437.43)};
        \node[font=\footnotesize] at (axis cs:6, 480) {402ms};

        \addplot+[
            boxplot prepared={
                lower whisker=485.21, 
                lower quartile=505.21,
                median=591.46, 
                upper quartile=677.71,
                upper whisker=697.71, 
            },
            fill=orange!30,
            draw=black
        ] coordinates {(7,505.21) (7,677.71)};
        \node[font=\footnotesize] at (axis cs:7, 720) {591ms};

        \addplot+[
            boxplot prepared={
                lower whisker=259.48, 
                lower quartile=279.48,
                median=316.29, 
                upper quartile=353.10,
                upper whisker=373.10, 
            },
            fill=orange!30,
            draw=black
        ] coordinates {(8,279.48) (8,353.10)};
        \node[font=\footnotesize] at (axis cs:8, 400) {316ms};

    \end{axis}
\end{tikzpicture}
\caption{Min and Max Execution Times for Various Functions}
\label{fig:min_max_execution_times}
\end{figure}

This boxplot, shown in Figure \ref{fig:min_max_execution_times}, presents the minimum, median, and maximum execution times (in milliseconds) for various system operations, with average times labeled above each whisker. The results indicate that View Local Resource Table has the highest average execution time at 591.46 ms, attributed to intensive data processing and resource validation, followed by \texttt{Create Table Access Request} with Misbehavior at 402.81 ms, reflecting the added computational overhead of handling misbehavior. In contrast, Policy Add demonstrates the lowest average execution time at 86.23 ms, showcasing its efficiency. This analysis highlights the most resource-intensive operations, emphasizing the impact of misbehavior handling on execution time.

\subsubsection{Scalability with Simultaneous Requests}

Scalability refers to the framework's ability to handle increasing numbers of simultaneous requests without significant performance degradation. This metric is crucial for peer-to-peer networks, where multiple peers interact concurrently, especially during high-demand periods. Measuring scalability allows us to evaluate the framework's capacity to manage diverse workloads effectively while ensuring responsiveness and reliability.

\textit{Tools, Techniques, and Processes (TTP):} To evaluate scalability, we simulated 1, 3, and 5 concurrent requests using automated Web3.js scripts. The operations included critical tasks such as \texttt{viewResourceTable()}, \texttt{revokeRole()}, and \texttt{swapRolesBetweenAccounts()}. Gas costs and execution times were logged for each scenario. The X-axis in the results represents the concurrency level, while the Y-axis measures gas consumption and execution delays. 

This approach is easy to expand further. For example, higher numbers of concurrent requests, such as 10 or 20, can be tested. Different types of operations and their combinations can also be added. This expansion will provide deeper insights into how our system handles larger-scale scenarios.

\textit{Implications:} 
The results demonstrated that the framework effectively managed up to five simultaneous requests with only a proportional increase in gas usage and execution time. This indicates the design's robustness and its ability to maintain consistent performance under moderate concurrency. While the current measurements were conducted using available hardware, the results also highlight opportunities for future scaling. With improved hardware resources, such as improved computational power and memory, the framework can seamlessly handle larger request volumes.

The key takeaways are that the framework’s current design focuses on establishing a measurement-ready environment capable of capturing scalability metrics under diverse conditions. Future prospects for the system include supporting higher concurrency levels, as the architecture is adaptable to better hardware configurations. These insights are valuable for network designers aiming to build systems that can grow in size and complexity while maintaining stability and performance.

\begin{figure}[htbp]
\centering
\begin{tikzpicture}[scale=0.47]
\begin{axis}[
    width=\textwidth * 1.0,  
    title={\LARGE Gas Cost for Functions with Simultaneous Requests on Global Resource Table},
    title style={
            at={(0.5,1.1)}, 
            anchor=north,   
        },
    ylabel={\Huge Gas Cost (units)},
    ylabel style={font=\LARGE},
    ybar,
    bar width=1.1cm,  
    enlargelimits=0.45,
    legend style={at={(0.5,1)},
        anchor=north,legend columns=-1},
    xtick={1,2,3},
    title style={yshift=0.3cm},
    xticklabels={
        View GRT,
        Revoke Role on GRT,
        Swap Role on GRT
    },
    xticklabel style={rotate=45, anchor=north east, font=\LARGE},
    yticklabel style={font=\LARGE},
    nodes near coords,
    nodes near coords align={vertical},
    every node near coord/.append style={/pgf/number format/fixed, /pgf/number format/precision=0, font=\large},
    legend style={
        font=\LARGE 
    }
]

\addplot[fill=cyan!50] coordinates {
    (1, 315243)
    (2, 317483)
    (3, 306457)
};
\addlegendentry{1 req Gas Cost}

\addplot[fill=orange!50] coordinates {
    (1, 927795)
    (2, 927795)
    (3, 907329)
};
\addlegendentry{3 req Gas Cost}

\addplot[fill=green!50] coordinates {
    (1, 1362641)
    (2, 1360401)
    (3, 1412252)
};
\addlegendentry{5 req Gas Cost}

\end{axis}
\end{tikzpicture}
\caption{Gas Cost for Functions with Simultaneous Requests on Global Resource Table}
\label{fig:gas_cost_simultaneous_requests}
\end{figure}
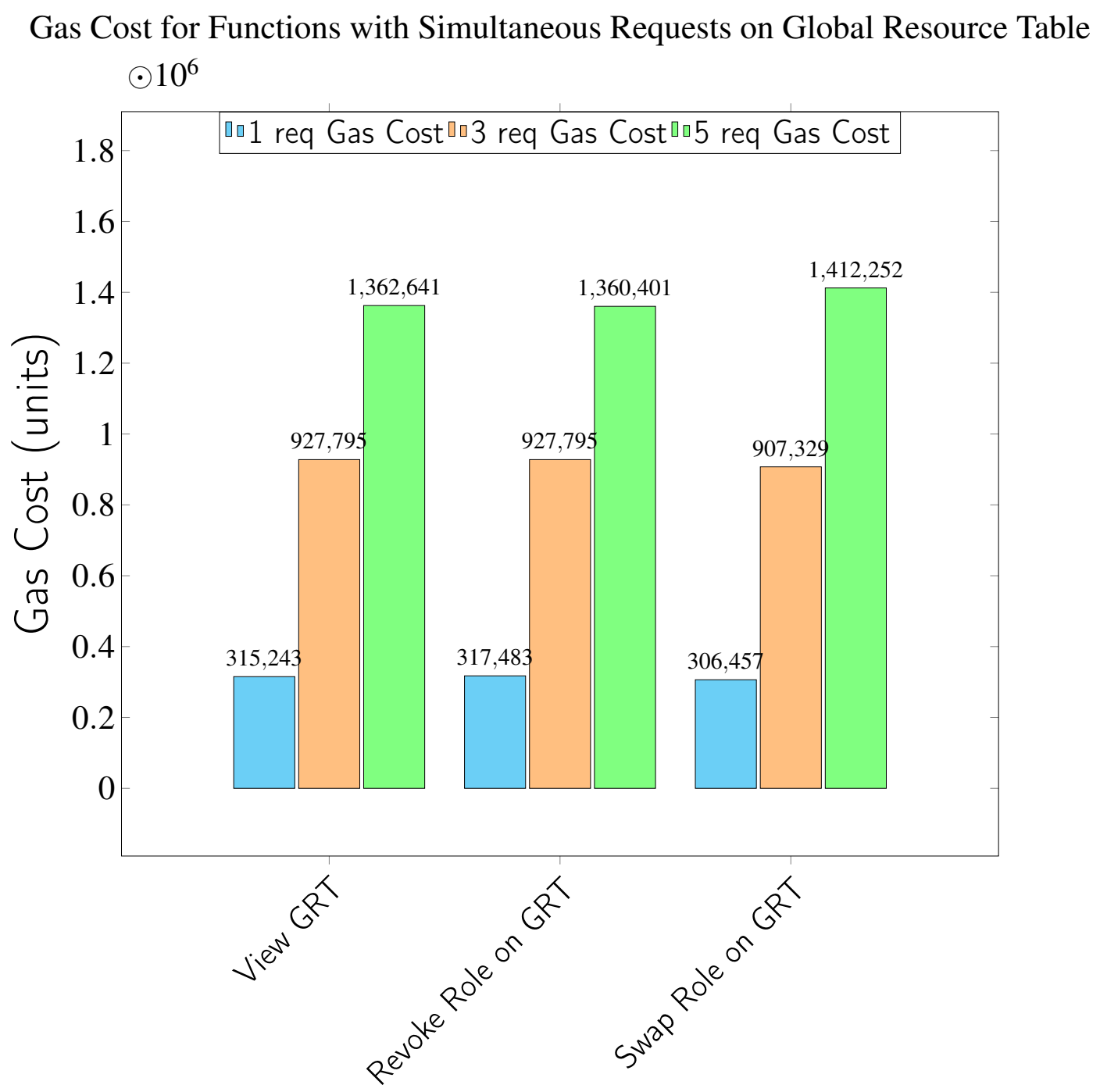

This bar graph, as illustrated in Figure \ref{fig:gas_cost_simultaneous_requests}, presents the gas costs (in units) for three operations on the Global Resource Table (GRT) — View, Revoke Role, and Swap Role — under different levels of simultaneous requests (1, 3, and 5). As expected, gas costs increase with the number of concurrent requests. Notably, the Swap Role operation with five requests shows the highest gas cost at 1,412,252 units, such asly due to the complexity of role reassignments. The View Global Resource Table operation, though generally less intensive, still reaches 1,362,641 units for five requests, indicating that simultaneous access can substantially impact resource consumption across all operations.

This analysis focuses on GRT because it represents the overall network state and is accessed by multiple groups, making it more susceptible to congestion under high concurrency. In contrast, the Local Resource Table (LRT) operates within specific groups, affecting only localized interactions. While LRT operations may have lower gas costs due to limited scope, analyzing GRT highlights the scalability challenges of managing global access control in a decentralized network. 

Additionally, our framework can easily accommodate larger or more realistic problem sizes by extending the number of concurrent requests, increasing the number of nodes, or introducing more complex access scenarios. This flexibility ensures the system remains scalable and adaptable to real-world network conditions.

\begin{figure*}[htbp]
\centering
\begin{tikzpicture}[scale=0.8]
    \definecolor{1Color}{rgb}{0.4, 0.7, 1.0}
    \definecolor{3Color}{rgb}{1.0, 0.6, 0.4}
    \definecolor{5Color}{rgb}{0.6, 1.0, 0.4}

    \begin{axis}[
        xshift=-1cm, 
        width=6cm,
        boxplot/draw direction=y,
        title={View Global Resource Table},
        ylabel={Latency (ms)},
        ymajorgrids = true,
        grid style=dashed,
        xtick={1, 2, 3},
        xticklabels={1 Request, 3 Requests, 5 Requests},
        xticklabel style={rotate=45, anchor=north east},
    ]

    \addplot+[
        boxplot prepared={
            lower whisker=353.75,
            lower quartile=553.75,
            median=562.13,
            upper quartile=581.62,
            upper whisker=821.62,
        },
        fill=1Color,
        draw=black,
    ] coordinates {};
    \node[font=\footnotesize] at (axis cs:1, 1100) {562.13ms};
    
    \addplot+[
        boxplot prepared={
            lower whisker=1162.19,
            lower quartile=1382.19,
            median=1463.76,
            upper quartile=1582.26,
            upper whisker=1802.26,
        },
        fill=3Color,
        draw=black,
    ] coordinates {};
    \node[font=\footnotesize] at (axis cs:2, 2100) {1463.76ms};
    
    \addplot+[
        boxplot prepared={
            lower whisker=2080.23,
            lower quartile=2300.23,
            median=2448.89,
            upper quartile=2845.69,
            upper whisker=3065.69,
        },
        fill=5Color,
        draw=black,
    ] coordinates {};
    \node[font=\footnotesize] at (axis cs:2.1, 3065) {2448.89ms};

    \end{axis}

    \begin{axis}[
        at={(5cm,0)},
        width=6cm,
        boxplot/draw direction=y,
        title={Revoke Role on Global Resource Table},
        ymajorgrids = true,
        grid style=dashed,
        xtick={1, 2, 3},
        xticklabels={1 Request, 3 Requests, 5 Requests},
        xticklabel style={rotate=45, anchor=north east},
        yticklabel pos=left
    ]

    \addplot+[
        boxplot prepared={
            lower whisker=723.65,
            lower quartile=923.65,
            median=961.87,
            upper quartile=1046.93,
            upper whisker=1246.93,
        },
        fill=1Color,
        draw=black,
    ] coordinates {};
    \node[font=\footnotesize] at (axis cs:1, 1500) {961.87ms};
    
    \addplot+[
        boxplot prepared={
            lower whisker=1872.30,
            lower quartile=2092.30,
            median=2151.27,
            upper quartile=2208.57,
            upper whisker=2428.57,
        },
        fill=3Color,
        draw=black,
    ] coordinates {};
    \node[font=\footnotesize] at (axis cs:2, 2700) {2151.27ms};
    
    \addplot+[
        boxplot prepared={
            lower whisker=2687.27,
            lower quartile=2907.27,
            median=2998.12,
            upper quartile=3107.04,
            upper whisker=3327.04,
        },
        fill=5Color,
        draw=black,
    ] coordinates {};
    \node[font=\footnotesize] at (axis cs:2.1, 3327) {2998.12ms};

    \end{axis}

    \begin{axis}[
        at={(11cm,0)},
        ymin=2000,
        ymax=9750,
        ytick={2000, 3500, 7000, 9750},
        yticklabels={2000, 3500, 7000, 9750},
        width=6cm,
        boxplot/draw direction=y,
        title={Swap Role on Global Resource Table},
        ymajorgrids = true,
        grid style=dashed,
        xtick={1, 2, 3},
        xticklabels={1 Request, 3 Requests, 5 Requests},
        xticklabel style={rotate=45, anchor=north east},
        yticklabel pos=left
    ]

    \addplot+[
        boxplot prepared={
            lower whisker=3699.21,
            lower quartile=3919.21,
            median=4037.18,
            upper quartile=4307.61,
            upper whisker=4527.61,
        },
        fill=1Color,
        draw=black,
    ] coordinates {};
    \node[font=\footnotesize] at (axis cs:1, 4800) {4037.18ms};

    \addplot+[
        boxplot prepared={
            lower whisker=6259.16,
            lower quartile=6479.16,
            median=6665.71,
            upper quartile=6912.82,
            upper whisker=7132.82,
        },
        fill=3Color,
        draw=black,
    ] coordinates {};
    \node[font=\footnotesize] at (axis cs:2, 7400) {6665.71ms};
    
    \addplot+[
        boxplot prepared={
            lower whisker=8611.91,
            lower quartile=8831.91,
            median=9017.95,
            upper quartile=9331.49,
            upper whisker=9551.49,
        },
        fill=5Color,
        draw=black,
    ] coordinates {};
    \node[font=\footnotesize] at (axis cs:2.1, 9017) {9017.95ms};

    \end{axis}

\end{tikzpicture}
\caption{Delays by Number of Simultaneous Requests}
\label{fig:delays_simultaneous_requests}
\end{figure*}
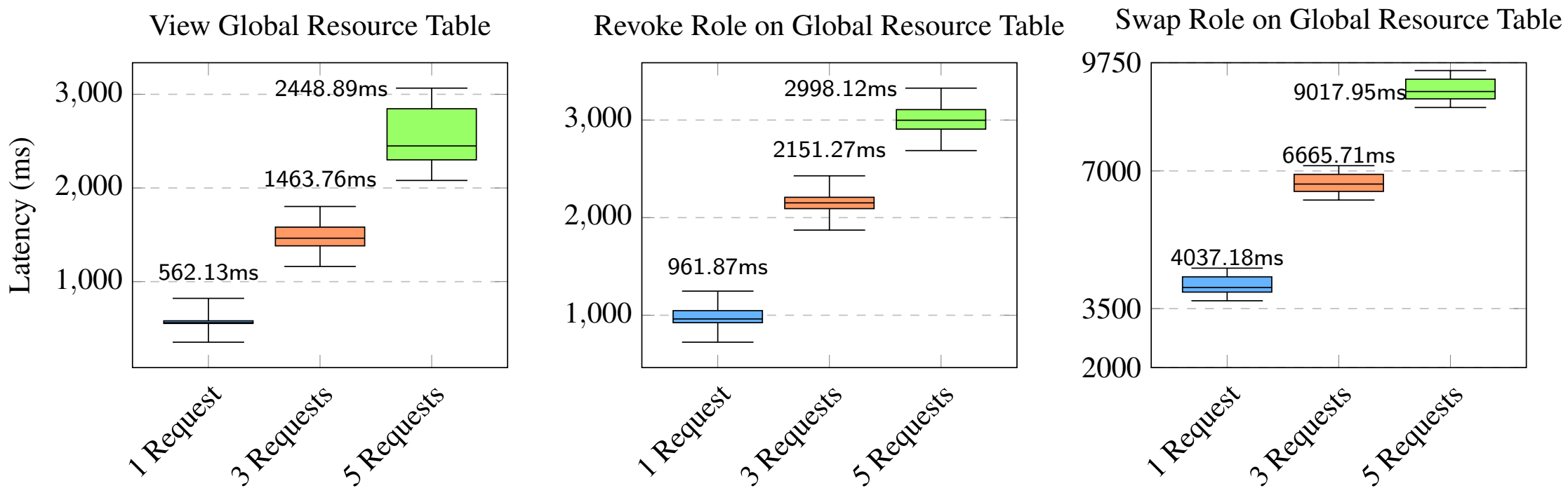

The boxplot in Figure \ref{fig:delays_simultaneous_requests} presents the execution times (in milliseconds) for three operations on the GRT—View, Revoke Role, and Swap Role—under 1, 3, and 5 simultaneous requests. Average execution times are displayed above the whiskers for clarity. Swap Role consistently exhibits the longest delays, peaking at an average of 9,017.95 ms for five simultaneous requests, highlighting the operational complexity of role-swapping tasks. In comparison, the View GRT operation demonstrates shorter delays, emphasizing its efficiency even under higher request loads. This analysis underscores how request concurrency impacts the response times of different operations.  

Beyond performance analysis, these findings have larger implications for securing planned response mechanisms in decentralized systems. High execution delays in critical tasks such as role reassignment may lead to bottlenecks in security enforcement, especially during high-traffic periods or attack scenarios. Ensuring that key operations execute within acceptable timeframes is essential for maintaining system integrity. Future optimizations, such as parallel processing or load-balancing strategies, can further improve response efficiency while upholding security guarantees.


\subsubsection{Attack Impact on System Delays}
This section evaluates impact of malicious activities on system performance and delay metrics during three key phases: normal operation, active attack scenarios, and post-attack recovery. The key takeaway is that this comprehensive testing framework goes beyond the limited scope of prior research. This approach also highlights the resilience of our framework and its ability to mitigate security threats such as privilege escalation, unauthorized access, and DoS.  

To gauge the maximum possible damage, we intentionally focus on high-impact threats that pose the greatest risk to system stability and security. By analyzing worst-case scenarios, such as large-scale privilege escalation attempts or prolonged DoS attacks, we assess how effectively the framework detects, mitigates, and recovers from severe disruptions. Understanding these dynamics is crucial for designing robust networks capable of handling security challenges without significant performance degradation.

\textit{Tools, Techniques, and Processes (TTP):} To analyze attack impact, we conducted a comprehensive set of tests using eight smart contracts that implement the attack-defense-recovery cycle. These contracts simulate malicious activities such as data tampering, excessive access attempts, and service disruptions, and they apply penalties or defensive actions in response.  

A detailed monitoring and instrumentation setup was used to track system behavior at each stage of the attack cycle. Function call logging captured critical events, allowing us to trace execution paths, measure response times, and analyze computational overhead. Gas costs, execution times, and response delays were continuously monitored through embedded metrics.

\textit{Implications:} Unlike previous works that focused on limited scenarios, such as a system under attack vs. during normal operations, our framework provides a more holistic view by integrating performance data across all three stages. This broader analysis enables a clear understanding of how the system reacts during an attack and how quickly it stabilizes after implementing defenses. For example, while denial-of-service attacks caused the most significant delays during active attacks, the system demonstrated notable recovery efficiency by minimizing post-attack disruptions. These results underscore the effectiveness of our penalty-based mechanisms in discouraging malicious behaviors and maintaining operational integrity.  

By running numerous tests across a variety of contracts and attack scenarios, we developed a structured measurement approach to evaluate system resilience. Future network designers can leverage these findings to identify critical bottlenecks and optimize recovery mechanisms for large-scale networks, ensuring security without compromising performance. 

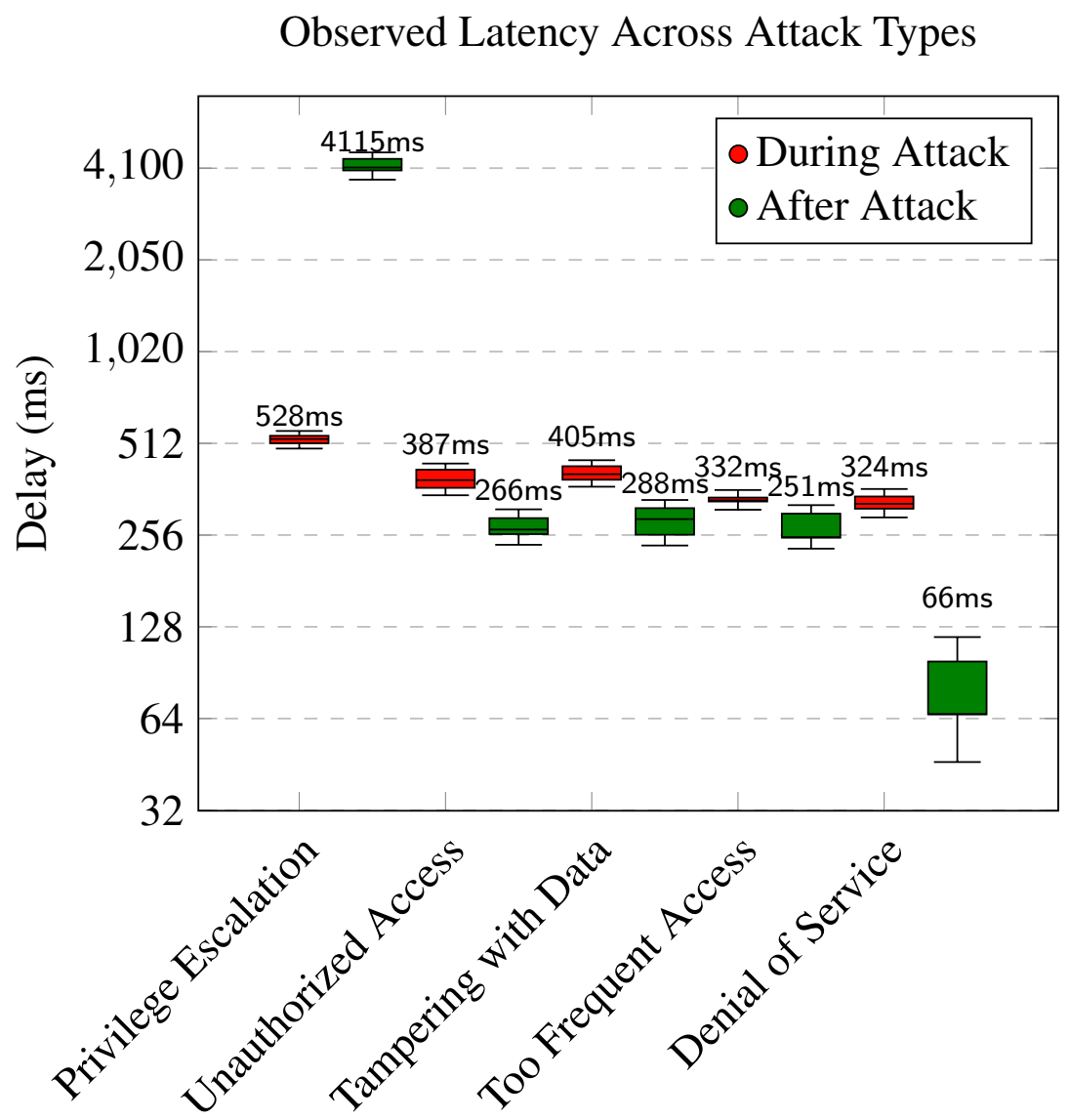
\begin{figure}[htbp]
\centering
\begin{tikzpicture}[scale=0.85]
    \begin{semilogyaxis}[
        boxplot/draw direction=y,
        title={Observed Latency Across Attack Types},
        ylabel={Delay (ms)},
        xtick={1, 3, 5, 7, 9},
        enlargelimits=0.10,
        xticklabels={
            Privilege Escalation, 
            Unauthorized Access, 
            Tampering with Data, 
            Too Frequent Access, 
            Denial of Service
        },
        ytick={2,4,8,16,32,64,128,256,512,1024,2048,4096},
        ymajorgrids=true,
        yminorticks=false,
        log ticks with fixed point,
        grid style=dashed,
        xticklabel style={rotate=45, anchor=north east},
        legend style={at={(1.05,1)}, anchor=north west},
        legend cell align={left},
        legend entries={During Attack, After Attack},
        legend image post style={mark=*},
        ymin=50, ymax=4500,
        legend pos=north east,
    ]

    \definecolor{duringColor}{RGB}{255,10,0}   
    \definecolor{afterColor}{RGB}{0,128,0}       

    \addlegendimage{only marks, mark=*, fill=duringColor}
    \addlegendimage{only marks, mark=*, fill=afterColor}

    \addplot+[
        boxplot prepared={
            lower whisker=492.20,
            lower quartile=512.20,
            median=528.82,
            upper quartile=542.11,
            upper whisker=562.11,
        },
        fill=duringColor,
        draw=black,
    ] coordinates {(1,528.82)};
    \node[font=\scriptsize] at (axis cs:1, 632) {528ms};
    
    \addplot+[
        boxplot prepared={
            lower whisker=3754.06,
            lower quartile=4025.92,
            median=4115.86,
            upper quartile=4393.00,
            upper whisker=4613.00,
        },
        fill=afterColor,
        draw=black,
    ] coordinates {(2,4115.86)};
    \node[font=\scriptsize] at (axis cs:2, 5003) {4115ms};
    
    \addplot+[
        boxplot prepared={
            lower whisker=345.85,
            lower quartile=365.85,
            median=387.84,
            upper quartile=419.56,
            upper whisker=439.56,
        },
        fill=duringColor,
        draw=black,
    ] coordinates {(3,387.84)};
    \node[font=\scriptsize] at (axis cs:3, 510) {387ms};
    
    \addplot+[
        boxplot prepared={
            lower whisker=237.87,
            lower quartile=257.87,
            median=266.94,
            upper quartile=290.95,
            upper whisker=310.95,
        },
        fill=afterColor,
        draw=black,
    ] coordinates {(4,266.94)};
    \node[font=\scriptsize] at (axis cs:4, 360) {266ms};
    
    \addplot+[
        boxplot prepared={
            lower whisker=368.96,
            lower quartile=388.96,
            median=405.50,
            upper quartile=430.71,
            upper whisker=450.71,
        },
        fill=duringColor,
        draw=black,
    ] coordinates {(5,405.50)};
    \node[font=\scriptsize] at (axis cs:5, 540) {405ms};
    
    \addplot+[
        boxplot prepared={
            lower whisker=236.59,
            lower quartile=256.59,
            median=288.81,
            upper quartile=313.75,
            upper whisker=333.75,
        },
        fill=afterColor,
        draw=black,
    ] coordinates {(6,288.81)};
    \node[font=\scriptsize] at (axis cs:6, 380) {288ms};
    
    \addplot+[
        boxplot prepared={
            lower whisker=309.96,
            lower quartile=329.96,
            median=332.83,
            upper quartile=339.88,
            upper whisker=359.88,
        },
        fill=duringColor,
        draw=black,
    ] coordinates {(7,332.83)};
    \node[font=\scriptsize] at (axis cs:7, 430) {332ms};
    
    \addplot+[
        boxplot prepared={
            lower whisker=231.03,
            lower quartile=251.03,
            median=251.84,
            upper quartile=300.99,
            upper whisker=320.99,
        },
        fill=afterColor,
        draw=black,
    ] coordinates {(8,251.84)};
    \node[font=\scriptsize] at (axis cs:8, 370) {251ms};
    
    \addplot+[
        boxplot prepared={
            lower whisker=292.46,
            lower quartile=312.46,
            median=324.14,
            upper quartile=342.66,
            upper whisker=362.66,
        },
        fill=duringColor,
        draw=black,
    ] coordinates {(9,324.14)};
    \node[font=\scriptsize] at (axis cs:9, 430) {324ms};
    
    \addplot+[
        boxplot prepared={
            lower whisker=46.07,
            lower quartile=66.07,
            median=66.18,
            upper quartile=98.45,
            upper whisker=118.45,
        },
        fill=afterColor,
        draw=black,
    ] coordinates {(10,66.18)};
    \node[font=\scriptsize] at (axis cs:10, 160) {66ms};

    \end{semilogyaxis}
\end{tikzpicture}
\caption{Latency measured (in milliseconds) for five attack scenarios}
\label{fig:delays_before_after_attacks}
\end{figure}

Figure \ref{fig:delays_before_after_attacks} presents the observed latency in milliseconds, capturing the total response time from request initiation to policy decision enforcement, under various attack conditions. The Y-axis reflects latency on a log scale to accommodate the spike in delays observed during certain attack types such as privilege escalation. Here, privilege escalation exhibits a substantial increase in median delay after mitigation, rising from 528.82 ms to 4115.86 ms, such as due to added security measures. In contrast, operations such as Unauthorized Access and Denial of Service demonstrate reduced delays post-attack, reflecting improved system resilience. Average delays are indicated above the whiskers for each category, highlighting the varying impacts of each attack on system response time.

\subsubsection{Role-Based Delay Analysis}
Role-based delay analysis examines the implication of hierarchical roles (PGH, SGH, and RM) affect system performance during various operations. This analysis introduces a novel metric for evaluating fairness and efficiency in role-based structures, particularly in large networks with two-layer hierarchies.

\textit{Tools, Techniques, and Processes (TTP):} 
Using timing measurements from high-precision Node.js scripts and smart contract logs, we tracked execution times for key operations such as Role Assignment, Access Control, and Table Access Requests. These measurements were conducted across PGH, SGH, and RM roles under identical conditions to ensure consistency.

\textit{Implications:} The results show that despite the large size and complexity of the network, the two-layer role structure maintains efficient performance. PGHs exhibit slightly faster execution times due to their optimized responsibilities, while RMs and SGHs achieve near-comparable results. This demonstrates that our hierarchical design effectively distributes workloads without introducing significant delays.

The key takeaways are that our scalable and well-organized role structure ensures fairness across roles and avoids performance bottlenecks, even as network size grows. These results reinforce that the framework’s design is both robust and efficient, delivering performance comparable to smaller networks while accommodating larger, decentralized systems.

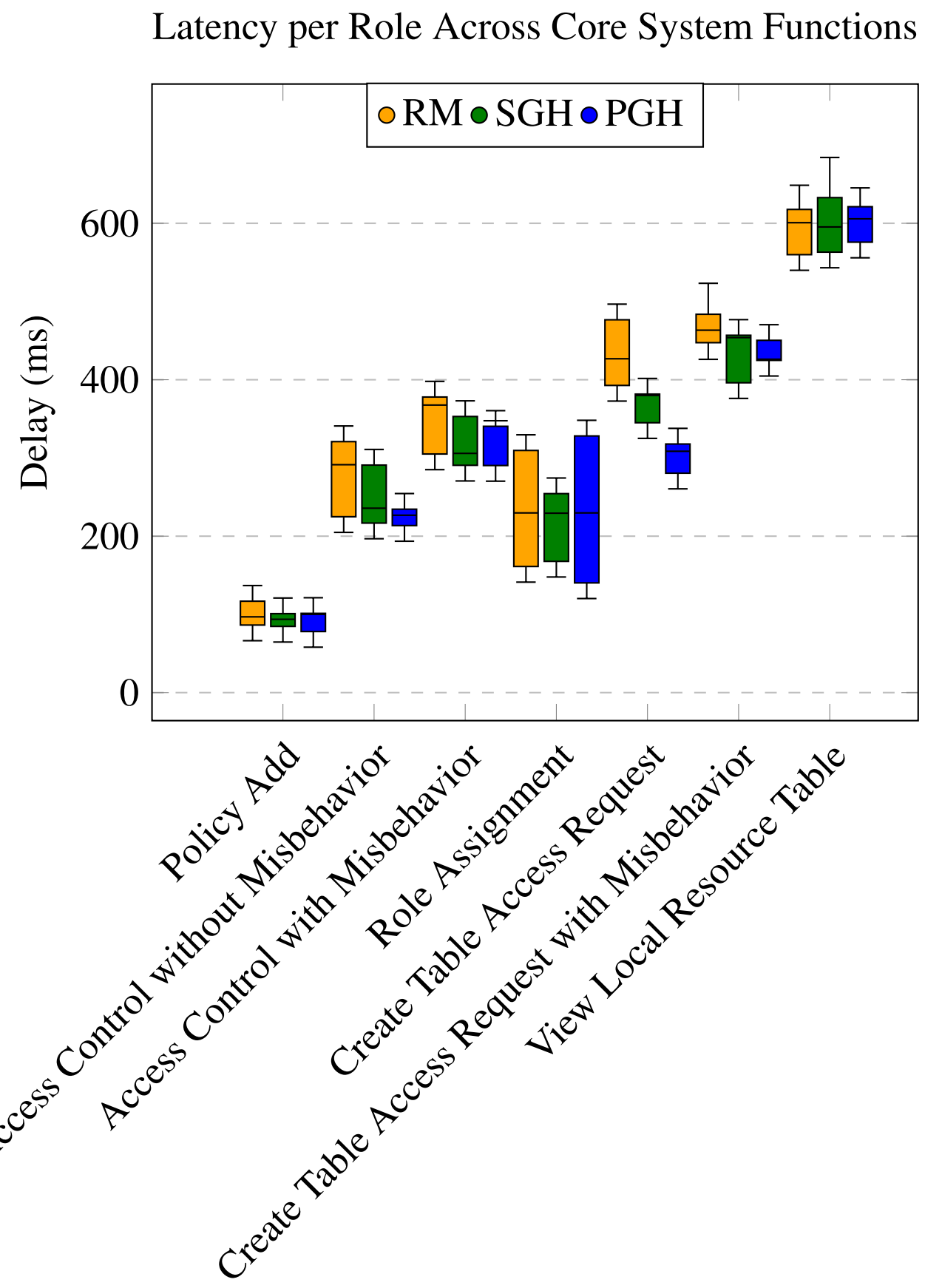
\begin{figure}[htbp]
\centering
\begin{tikzpicture}[scale=1.0]
        \begin{axis}[
            boxplot/draw direction=y,
            xmax=20,
            ylabel={Delay (ms)},
            title={Latency per Role Across Core System Functions},
            xtick={2, 5, 8, 11, 14, 17, 20},
            enlargelimits=0.15,
            xticklabels={
                Policy Add, 
            Access Control without Misbehavior, 
            Access Control with Misbehavior, 
            Role Assignment, 
            Create Table Access Request, 
            Create Table Access Request with Misbehavior, 
            View Local Resource Table
            },
            ymajorgrids=true,
            grid style=dashed,
            xticklabel style={rotate=45, anchor=north east},
            legend style={at={(0.5,0.90)}, anchor=south, legend columns=-1},
            legend cell align={left},
            legend entries={RM, SGH, PGH},
            legend image post style={mark=*}
    ]

    \definecolor{rmColor}{RGB}{255,165,0}    
    \definecolor{shColor}{RGB}{0,128,0}      
    \definecolor{phColor}{RGB}{0,0,255}      

    \addlegendimage{only marks, mark=*, fill=rmColor}
    \addlegendimage{only marks, mark=*, fill=shColor}
    \addlegendimage{only marks, mark=*, fill=phColor}

        \addplot+[
            boxplot prepared={
                lower whisker=66.38,
                lower quartile=86.38,
                median=96.87,
                upper quartile=116.87,
                upper whisker=136.87,
            },
            fill=rmColor,
            draw=black,
        ] coordinates {(1,90.08)};

        \addplot+[
            boxplot prepared={
                lower whisker=64.67,
                lower quartile=84.67,
                median=93.73,
                upper quartile=100.94,
                upper whisker=120.94,
            },
            fill=shColor,
            draw=black,
        ] coordinates {(2,90.44)};

        \addplot+[
            boxplot prepared={
                lower whisker=58.09,
                lower quartile=78.09,
                median=100.08,
                upper quartile=101.31,
                upper whisker=121.31,
            },
            fill=phColor,
            draw=black,
        ] coordinates {(3,91.52)};

        \addplot+[
            boxplot prepared={
                lower whisker=204.85,
                lower quartile=224.85,
                median=291.37,
                upper quartile=320.88,
                upper whisker=340.88,
            },
            fill=rmColor,
            draw=black,
        ] coordinates {(4,254.23)};

        \addplot+[
            boxplot prepared={
                lower whisker=196.73,
                lower quartile=216.73,
                median=235.76,
                upper quartile=290.87,
                upper whisker=310.87,
            },
            fill=shColor,
            draw=black,
        ] coordinates {(5,236.49)};

        \addplot+[
            boxplot prepared={
                lower whisker=193.54,
                lower quartile=213.54,
                median=226.78,
                upper quartile=234.49,
                upper whisker=254.49,
            },
            fill=phColor,
            draw=black,
        ] coordinates {(6,234.13)};

        \addplot+[
            boxplot prepared={
                lower whisker=285.04,
                lower quartile=305.04,
                median=367.61,
                upper quartile=377.88,
                upper whisker=397.88,
            },
            fill=rmColor,
            draw=black,
        ] coordinates {(7,338.97)};

        \addplot+[
            boxplot prepared={
                lower whisker=270.60,
                lower quartile=290.60,
                median=305.84,
                upper quartile=353.02,
                upper whisker=373.02,
            },
            fill=shColor,
            draw=black,
        ] coordinates {(8,322.83)};

        \addplot+[
            boxplot prepared={
                lower whisker=270.32,
                lower quartile=290.32,
                median=347.56,
                upper quartile=340.45,
                upper whisker=360.45,
            },
            fill=phColor,
            draw=black,
        ] coordinates {(9,344.63)};

        \addplot+[
            boxplot prepared={
                lower whisker=141.25,
                lower quartile=161.25,
                median=229.75,
                upper quartile=309.59,
                upper whisker=329.59,
            },
            fill=rmColor,
            draw=black,
        ] coordinates {(10,181.16)};

        \addplot+[
            boxplot prepared={
                lower whisker=147.81,
                lower quartile=167.81,
                median=229.40,
                upper quartile=254.37,
                upper whisker=274.37,
            },
            fill=shColor,
            draw=black,
        ] coordinates {(11,198.43)};

        \addplot+[
            boxplot prepared={
                lower whisker=120.27,
                lower quartile=140.27,
                median=229.75,
                upper quartile=328.12,
                upper whisker=348.12,
            },
            fill=phColor,
            draw=black,
        ] coordinates {(12,165.29)};

        \addplot+[
            boxplot prepared={
                lower whisker=372.70,
                lower quartile=392.70,
                median=426.81,
                upper quartile=476.62,
                upper whisker=496.62,
            },
            fill=rmColor,
            draw=black,
        ] coordinates {(13,421.91)};

        \addplot+[
            boxplot prepared={
                lower whisker=325.01,
                lower quartile=345.01,
                median=379.95,
                upper quartile=381.63,
                upper whisker=401.63,
            },
            fill=shColor,
            draw=black,
        ] coordinates {(14,353.30)};

        \addplot+[
            boxplot prepared={
                lower whisker=260.49,
                lower quartile=280.49,
                median=308.54,
                upper quartile=317.82,
                upper whisker=337.82,
            },
            fill=phColor,
            draw=black,
        ] coordinates {(15,283.31)};

    \addplot+[
        boxplot prepared={
            lower whisker=426.06,
            lower quartile=447.37,
            median=463.34,
            upper quartile=483.69,
            upper whisker=523.27,
        },
        fill=rmColor,
        draw=black,
    ] coordinates {(17,463.34)};

    \addplot+[
        boxplot prepared={
            lower whisker=376.10,
            lower quartile=396.10,
            median=453.84,
            upper quartile=456.81,
            upper whisker=476.81,
        },
        fill=shColor,
        draw=black,
    ] coordinates {(17,453.84)};

    \addplot+[
        boxplot prepared={
            lower whisker=404.75,
            lower quartile=424.75,
            median=426.21,
            upper quartile=450.36,
            upper whisker=470.36,
        },
        fill=phColor,
        draw=black,
    ] coordinates {(17,426.21)};

    \addplot+[
        boxplot prepared={
            lower whisker=539.87,
            lower quartile=559.87,
            median=600.75,
            upper quartile=617.71,
            upper whisker=648.59,
        },
        fill=rmColor,
        draw=black,
    ] coordinates {(20,600.75)};

    \addplot+[
        boxplot prepared={
            lower whisker=543.12,
            lower quartile=563.12,
            median=595.29,
            upper quartile=632.81,
            upper whisker=684.12,
        },
        fill=shColor,
        draw=black,
    ] coordinates {(20,595.29)};

    \addplot+[
        boxplot prepared={
            lower whisker=555.87,
            lower quartile=575.87,
            median=605.70,
            upper quartile=621.19,
            upper whisker=645.24,
        },
        fill=phColor,
        draw=black,
    ] coordinates {(20,605.70)};

        \end{axis}
\end{tikzpicture}
\caption{
Measured latency (in milliseconds) for each system function, categorized by user roles: Regular Member (RM), Secondary Group Head (SGH), and Primary Group Head (PGH). 
} 
\label{fig:delays_roles_functions}
\end{figure}

Figure~\ref{fig:delays_roles_functions} presents the latency results for seven key access control and policy management operations, segmented by user role: Regular Member (RM), Secondary Group Head (SGH), and Primary Group Head (PGH). The measured latency reflects end-to-end execution time from request issuance to policy enforcement. While all roles follow similar latency trends across operations, RM users consistently show slightly higher latency in tasks such as Access Control with Misbehavior and Create Table Access Request, due to additional permission verification steps and lack of elevated privileges. In contrast, PGH users often demonstrate improved response times, particularly in high-privilege operations, suggesting that hierarchical access shortens decision paths within the policy enforcement mechanism. This analysis confirms that the system efficiently respects role-based performance variance while maintaining overall consistency in response time across roles and functions.

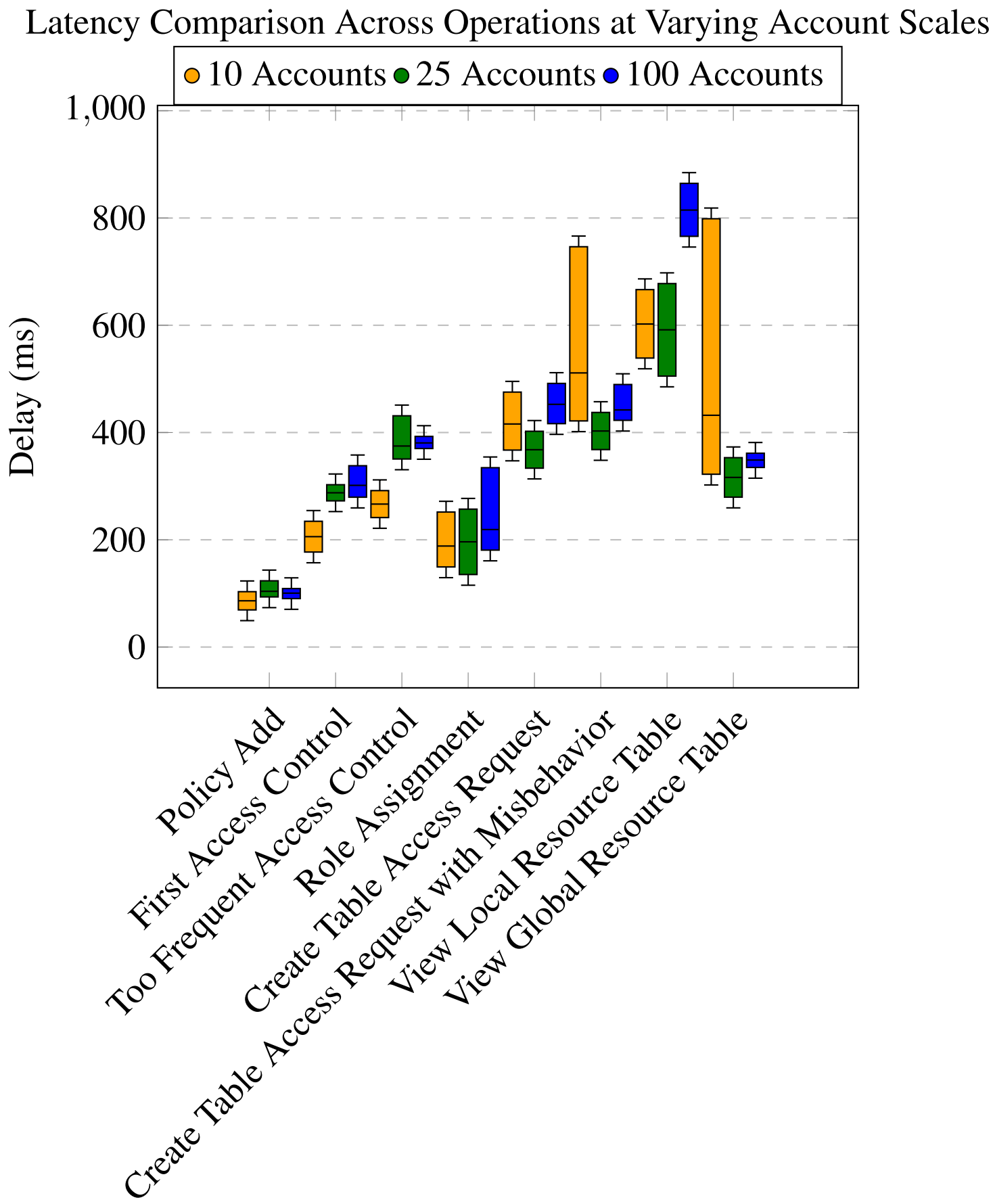
\begin{figure}[htbp]
\centering
\begin{tikzpicture}[scale=1.0]
    \begin{axis}[
        boxplot/draw direction=y,
        xmax=25,
        title={Latency Comparison Across Operations at Varying Account Scales},
        title style={yshift=0.3cm},
        ylabel={Delay (ms)},
        xtick={2, 5, 8, 11, 14, 17, 20, 23},
        enlargelimits=0.15,
        xticklabels={
            Policy Add, 
            First Access Control, 
            Too Frequent Access Control, 
            Role Assignment, 
            Create Table Access Request, 
            Create Table Access Request with Misbehavior, 
            View Local Resource Table,
            View Global Resource Table
        },
        ymajorgrids=true,
        grid style=dashed,
        xticklabel style={rotate=45, anchor=north east},
        legend style={at={(0.5,1.00)}, anchor=south, legend columns=-1},
        legend cell align={left},
        legend entries={10 Accounts, 25 Accounts, 100 Accounts},
        legend image post style={mark=*}
    ]

    \definecolor{tenColor}{RGB}{255,165,0}       
    \definecolor{twentyFiveColor}{RGB}{0,128,0}  
    \definecolor{hundredColor}{RGB}{0,0,255}     

    \addlegendimage{only marks, mark=*, fill=tenColor}
    \addlegendimage{only marks, mark=*, fill=twentyFiveColor}
    \addlegendimage{only marks, mark=*, fill=hundredColor}

    \addplot+[
        boxplot prepared={
            lower whisker=49.23,
            lower quartile=69.23,
            median=86.225,
            upper quartile=103.22,
            upper whisker=123.22,
        },
        fill=tenColor,
        draw=black,
    ] coordinates {(2,86.225)};
    \addplot+[
        boxplot prepared={
            lower whisker=73.45,
            lower quartile=93.45,
            median=103.88,
            upper quartile=123.48,
            upper whisker=143.48,
        },
        fill=twentyFiveColor,
        draw=black,
    ] coordinates {(2,103.88)};
    \addplot+[
        boxplot prepared={
            lower whisker=70.33,
            lower quartile=90.33,
            median=100.49,
            upper quartile=109.05,
            upper whisker=129.05,
        },
        fill=hundredColor,
        draw=black,
    ] coordinates {(2,100.49)};

    \addplot+[
        boxplot prepared={
            lower whisker=157.13,
            lower quartile=177.13,
            median=205.815,
            upper quartile=234.50,
            upper whisker=254.50,
        },
        fill=tenColor,
        draw=black,
    ] coordinates {(5,205.815)};
    \addplot+[
        boxplot prepared={
            lower whisker=252.56,
            lower quartile=272.56,
            median=287.61,
            upper quartile=302.66,
            upper whisker=322.66,
        },
        fill=twentyFiveColor,
        draw=black,
    ] coordinates {(5,287.61)};
    \addplot+[
        boxplot prepared={
            lower whisker=259.41,
            lower quartile=279.41,
            median=301.49,
            upper quartile=338.06,
            upper whisker=358.06,
        },
        fill=hundredColor,
        draw=black,
    ] coordinates {(5,301.49)};

    \addplot+[
        boxplot prepared={
            lower whisker=221.42,
            lower quartile=241.42,
            median=266.55,
            upper quartile=291.68,
            upper whisker=311.68,
        },
        fill=tenColor,
        draw=black,
    ] coordinates {(8,266.55)};
    \addplot+[
        boxplot prepared={
            lower whisker=330.60,
            lower quartile=350.60,
            median=374.82,
            upper quartile=431.17,
            upper whisker=451.17,
        },
        fill=twentyFiveColor,
        draw=black,
    ] coordinates {(8,374.82)};
    \addplot+[
        boxplot prepared={
            lower whisker=350.18,
            lower quartile=370.18,
            median=380.58,
            upper quartile=392.73,
            upper whisker=412.73,
        },
        fill=hundredColor,
        draw=black,
    ] coordinates {(8,380.58)};

    \addplot+[
        boxplot prepared={
            lower whisker=129.36,
            lower quartile=149.36,
            median=188.33,
            upper quartile=251.74,
            upper whisker=271.74,
        },
        fill=tenColor,
        draw=black,
    ] coordinates {(11,188.33)};
    \addplot+[
        boxplot prepared={
            lower whisker=115.25,
            lower quartile=135.25,
            median=196.19,
            upper quartile=257.12,
            upper whisker=277.12,
        },
        fill=twentyFiveColor,
        draw=black,
    ] coordinates {(11,196.19)};
    \addplot+[
        boxplot prepared={
            lower whisker=160.74,
            lower quartile=180.74,
            median=219.06,
            upper quartile=334.39,
            upper whisker=354.39,
        },
        fill=hundredColor,
        draw=black,
    ] coordinates {(11,219.06)};

    \addplot+[
        boxplot prepared={
            lower whisker=347.22,
            lower quartile=367.22,
            median=415.88,
            upper quartile=475.36,
            upper whisker=495.36,
        },
        fill=tenColor,
        draw=black,
    ] coordinates {(10,415.88)};
    \addplot+[
        boxplot prepared={
            lower whisker=313.52,
            lower quartile=333.52,
            median=367.93,
            upper quartile=402.33,
            upper whisker=422.33,
        },
        fill=twentyFiveColor,
        draw=black,
    ] coordinates {(25,367.93)};
    \addplot+[
        boxplot prepared={
            lower whisker=396.55,
            lower quartile=416.55,
            median=452.57,
            upper quartile=491.57,
            upper whisker=511.57,
        },
        fill=hundredColor,
        draw=black,
    ] coordinates {(100,452.57)};

    \addplot+[
        boxplot prepared={
            lower whisker=401.63,
            lower quartile=421.63,
            median=511.13,
            upper quartile=746.40,
            upper whisker=766.40,
        },
        fill=tenColor,
        draw=black,
    ] coordinates {(10,511.13)};
    \addplot+[
        boxplot prepared={
            lower whisker=348.19,
            lower quartile=368.19,
            median=402.81,
            upper quartile=437.43,
            upper whisker=457.43,
        },
        fill=twentyFiveColor,
        draw=black,
    ] coordinates {(25,402.81)};
    \addplot+[
        boxplot prepared={
            lower whisker=402.74,
            lower quartile=422.74,
            median=442.14,
            upper quartile=489.50,
            upper whisker=509.50,
        },
        fill=hundredColor,
        draw=black,
    ] coordinates {(100,442.14)};

    \addplot+[
        boxplot prepared={
            lower whisker=518.83,
            lower quartile=538.83,
            median=602.30,
            upper quartile=666.46,
            upper whisker=686.46,
        },
        fill=tenColor,
        draw=black,
    ] coordinates {(10,602.30)};
    \addplot+[
        boxplot prepared={
            lower whisker=485.21,
            lower quartile=505.21,
            median=591.46,
            upper quartile=677.71,
            upper whisker=697.71,
        },
        fill=twentyFiveColor,
        draw=black,
    ] coordinates {(25,591.46)};
    \addplot+[
        boxplot prepared={
            lower whisker=745.85,
            lower quartile=765.85,
            median=814.77,
            upper quartile=864.59,
            upper whisker=884.59,
        },
        fill=hundredColor,
        draw=black,
    ] coordinates {(100,814.77)};

    \addplot+[
        boxplot prepared={
            lower whisker=302.21,
            lower quartile=322.21,
            median=432.15,
            upper quartile=798.59,
            upper whisker=818.59,
        },
        fill=tenColor,
        draw=black,
    ] coordinates {(10,432.15)};
    \addplot+[
        boxplot prepared={
            lower whisker=259.48,
            lower quartile=279.48,
            median=316.29,
            upper quartile=353.10,
            upper whisker=373.10,
        },
        fill=twentyFiveColor,
        draw=black,
    ] coordinates {(25,316.29)};
    \addplot+[
        boxplot prepared={
            lower whisker=314.87,
            lower quartile=334.87,
            median=348.67,
            upper quartile=361.40,
            upper whisker=381.40,
        },
        fill=hundredColor,
        draw=black,
    ] coordinates {(100,348.67)};

    \end{axis}
\end{tikzpicture}
\caption{
Measured latency (in milliseconds) for core operations under three account group sizes (10, 25, and 100 accounts).
}
\label{fig:delays_by_account_numbers}
\end{figure}

Figure~\ref{fig:delays_by_account_numbers} illustrates the measured latency (end-to-end response delay) for eight key system operations across three different peer group sizes: 10, 25, and 100 accounts. These operations span core functionalities such as policy addition, role assignment, creation of table access requests (both normal and with misbehavior), and reading from local/global resource tables. As expected, most operations exhibit a modest increase in latency with scale. Notably, operations involving interaction with shared resources—such as View Local Resource Table and Create Table Access Request with Misbehavior—show a more noticeable latency growth, especially at 100-account scale. The View Global Resource Table operation, in contrast, maintains consistent latency due to distributed data lookup and caching optimizations. These results affirm that the proposed access control framework maintains reliable and scalable performance across increasing network sizes while ensuring real-time responsiveness.

\subsubsection{Token Economics and Security Actions}
Token economics evaluates how token-based incentives and penalties promote compliant behavior and discourage malicious activities. This metric is crucial for maintaining accountability, trust, and long-term system integrity in decentralized environments.

\textit{Tools, Techniques, and Processes (TTP):} We tracked token balances using smart contract logs from RoleToken operations, such as \texttt{grantAccess()}, \texttt{reportMaliciousActivity()}, and \texttt{applyPenalty()}. Sampling was conducted across benign actions (e.g., successful access requests) and malicious events (e.g., unauthorized access attempts). The X-axis represents peer actions, while the Y-axis shows changes in token balances.

\textit{Implications:} Benign actions lead to token rewards or stable balances, while malicious activities result in token deductions and state changes. Severe violations, such as data tampering, incur significant penalties, reducing a peer’s reputation and privileges. This approach ensures a dynamic and self-regulating network.

The key takeaways are that well-calibrated token incentives and penalties effectively align peer behavior with system security goals. Future network designers can adopt similar token-based mechanisms to foster trust and accountability in decentralized systems.

\begin{figure}[htbp]
\centering
\begin{tikzpicture}[scale=0.85]
    \begin{axis}[
        xlabel style={yshift=-1cm},
        ylabel=Token Balance,
        xtick={0,1,2,3,4,5,6,7,8,9,10},
        xticklabels={
            Initial Balance,
            Successful Access to test1.jpg,
            Successful Access to test2.jpg,
            Successful Access to video1.mp4,
            Successful Local Table Access Request,
            Unauthorized Access,
            Tampering Data,
            Too Frequent Access,
            Denial of Service,
            Successful Request Approval,
            Successful Global Table Access Request
        },
        x tick label style={rotate=40, anchor=east},
        grid=major,
        grid style={dashed, gray!30},
        legend style={at={(0.95,0.95)}, anchor=north east},
        ymin=4.5e23,
        ymax=5.8e23,
        ytick={4.5e23, 4.6e23, 4.7e23, 4.8e23, 4.9e23, 5.0e23, 5.1e23, 5.2e23, 5.3e23, 5.4e23, 5.5e23, 5.6e23, 5.7e23, 5.8e23},
        nodes near coords,
        nodes near coords style={font=\scriptsize, color=black}
    ]

    \addplot[color=green, mark=*, mark options={solid}] coordinates {
        (0, 5e23)
        (1, 5.1e23)
        (2, 5.2e23)
        (3, 5.3e23)
        (4, 5.4e23)
        (5, 5.2e23)  
        (6, 5.0e23)  
        (7, 4.8e23)
        (8, 4.6e23)
        (9, 4.7e23)  
        (10, 4.8e23)
    };

    \addplot[color=red, mark=*, mark options={solid}] coordinates {
        (4, 5.4e23)
        (5, 5.2e23)
        (6, 5.0e23)
        (7, 4.8e23)
        (8, 4.6e23)
    };

    \legend{benign Activities, malicious Activities}
    \end{axis}
\end{tikzpicture}
\caption{Token Balance Change Over Time}
\label{fig:token_balance_over_time}
\end{figure}
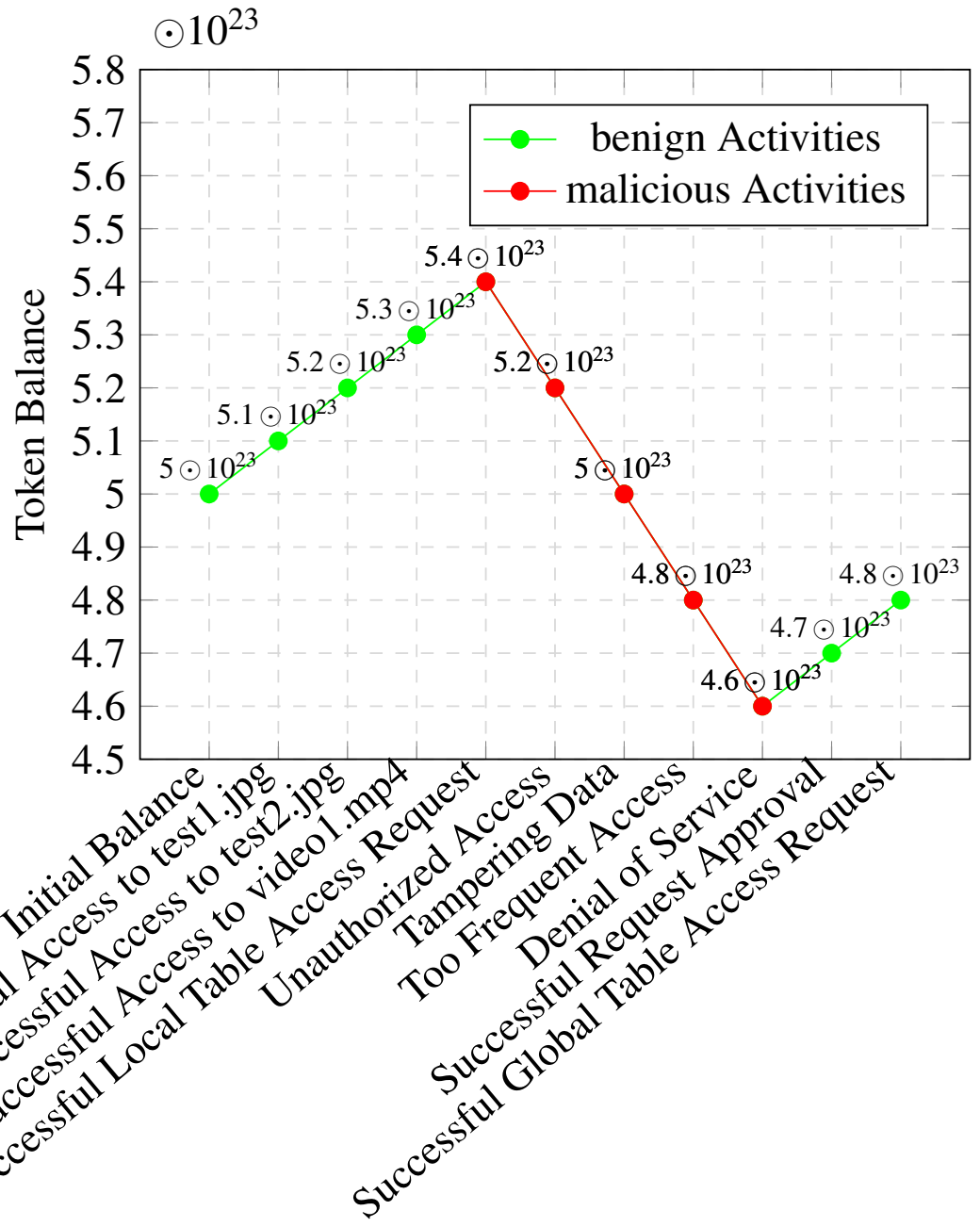

The line graph seen in Figure \ref{fig:token_balance_over_time}, illustrates token balance changes across various actions. It compares benign activities (green line) and malicious activities (red dots). The x-axis represents different actions, from initial balance to successful accesses, unauthorized attempts, and request approvals. The y-axis shows token balance fluctuations.  

Benign actions, such as accessing files or submitting requests, cause minor increases or stable balances. In contrast, malicious actions, such as unauthorized access and data tampering, result in sharp token balance drops. This indicates their negative impact on the system. The graph clearly shows the difference between benign and malicious activities in maintaining system integrity. The analysis demonstrates systematic accountability enforcement. It rewards compliant behavior and penalizes violations. The steep decline in token balance for malicious activities highlights the system’s ability to deter harmful actions. By maintaining this balance-driven approach, the framework ensures long-term stability and encourages peers to follow security policies.


\subsection{Security of Smart Contracts}

\begingroup

Smart contracts are susceptible to several types of vulnerabilities, including reentrancy attacks, integer overflows, underflows, and timestamp manipulations. These flaws may allow attackers to execute unintended behaviors such as unauthorized access or token balance manipulation. In our framework, these issues are mitigated through a combination of techniques. We use safe math libraries (such as OpenZeppelin’s SafeMath) to prevent overflow and underflow during arithmetic operations. Additionally, reentrancy guards are applied to critical functions particularly for govern role assignment and token transfers—to prevent recursive calls that could disrupt the contract state.

The ERC-20 token is vulnerable to token inflation, minting exploits, and price manipulation~\cite{goswami2021tokencheck}. These attacks may allow adversaries to circumvent control mechanisms and gain unfair advantages. Our framework addresses these risks by applying a \textit{series of guard rail mechanisms} as the crux of our framework. First, token minting restrictions are enforced through strict access controls (permissioned blockchain) that limits to authorized entities governed by Judge Contract. This security hardening is extends over key function such as role assignment to prevent privilege escalation attack. Finally, our dynamic penalty system scales the response based on the severity and frequency of the offense, thereby deterring low-impact exploits.


Alongside targeted protections, the system implements broader security best practices to maintain resilience in a decentralized environment. The framework supports security patches and upgrade paths, ensuring that smart contracts can be updated without compromising integrity. Most importantly, the design encourages incentive-driven security by rewarding nodes who report valid vulnerabilities, creating an additional line of defense through community participation. Although no system is hundred percent fool-proof, we present our case as implementation-ready robust and scalable network design.

\endgroup

\subsection{Qualitative Comparison}
This section provides a qualitative comparison of different blockchain-based access control architectures. Table \ref{tab:comparison} focuses on critical aspects such as access control design, system capabilities and incentivization for trust management that showcases our contributions. Previous works such as Fabric-IoT~\cite{ning2014aggregated} and SGX-Blockchain~\cite{wang2021} rely on RBAC with static policies. However, our system introduces dynamic behavior-driven policies enabling real-time adaptability with STRIDE threat modeling. Unlike Dynamic IoT~\cite{wang2021} that supports trust management, but lacks structured state transitions (viz., benign, suspicious, malicious). Furthermore, we implement a recovery mechanism for penalized peers after a re-evaluation phase, thereby mitigating permanent expulsion. Although DBACP-IoTSG~\cite{bera2021} employs contextual policies, it does not provide such a rehabilitative path. Unlike SGX-Blockchain that incurs high gas costs due to frequent on-chain cryptographic operations and storage overhead, our framework reduces gas consumption by refining smart contract execution, optimizing storage operations, and minimizing redundant computations. 
Our hierarchical design efficiently distributes workloads, preventing cost spikes as the network scales and provides adaptive cost control.

\begin{table*}[ht]
\centering
\caption{Comparison of Blockchain-Based Access Control Architectures}
\label{tab:comparison}
\small
\resizebox{\textwidth}{!}{%
\sffamily
\begin{tabular}{|c|c|c|c|c|c|}
\hline
\rowcolor{blue!15}
\multicolumn{1}{|c|}{\textbf{Feature}} & 
\shortstack{\textbf{Fabric-IoT} \\ \cite{ning2014aggregated}} & 
\shortstack{\textbf{Dynamic IoT} \\ \cite{wang2021}} & 
\shortstack{\textbf{SGX-Blockchain} \\ \cite{han2022}} & 
\shortstack{\textbf{DBACP-IoTSG} \\ \cite{bera2021}} & 
\multicolumn{1}{c|}{\textbf{Our System}} \\
\hline

\rowcolor{gray!15}
\multicolumn{6}{|c|}{\textbf{Access Control Design}} \\
\hline
Access Control Logic & RBAC & ABAC + Trust Mgmt & RBAC & Contextual & RBAC + Hierarchical \\
\arrayrulecolor{gray!30}\hline\arrayrulecolor{black}
Dynamic State-Based Policies & \xmark & \cmark & \xmark & \xmark & \cmark \\
\arrayrulecolor{gray!30}\hline\arrayrulecolor{black}
Recovery from Violations & \xmark & \xmark & \xmark & \xmark & \cmark \\
\hline

\rowcolor{gray!15}
\multicolumn{6}{|c|}{\textbf{System Capabilities and Performance}} \\
\hline
Decentralized Enforcement & \cmark & \cmark & \cmark & \xmark & \cmark \\
\arrayrulecolor{gray!30}\hline\arrayrulecolor{black}
Gas Efficiency (Ethereum-Compatible) & \xmark & \cmark & \xmark & \xmark & \cmark \\
\arrayrulecolor{gray!30}\hline\arrayrulecolor{black}
Lightweight Implementation & \cmark & \cmark & \xmark & \cmark & \cmark \\
\hline

\rowcolor{gray!15}
\multicolumn{6}{|c|}{\textbf{Trust and Incentive Mechanisms}} \\
\hline
Trust Management Layer & \xmark & \cmark & \xmark & \xmark & \cmark \\
\arrayrulecolor{gray!30}\hline\arrayrulecolor{black}
Token Incentive Mechanism & \xmark & \xmark & \xmark & \xmark & \cmark \\
\hline
\end{tabular}
}
\end{table*}

\subsection{Discussion}  
Conceptually, our framework showcases the decentralized access control by prioritizing controllability and observability. The \textit{controllability} lies in its ability to enforce variable levels of control, ranging from lenient permissions for low-risk interactions to stringent restrictions for high-risk, sensitive operations. This adaptability ensures that the framework can scale its security measures dynamically, guided by real-time evaluations of peer behavior and threat severity. \textit{Observability} is achieved through a robust monitoring mechanism embedded within the system that ensures that operations are fully traceable even under subpar operating conditions.  

The system remains efficient even as network size and complexity increases also represented as \textit{scalability}. This is made possible by designing a unique mapping relationship between the network roles (viz., PGH, SGH, RM) and dynamic state management (viz., benign, suspicious, malicious). Handling normal and threat situations using incentivization mechanism ensures \textit{fine-grained control} and thereby fosters node compliance. These design constraints also lend themselves into \textit{lightweight design} as they limit the communication scope as shown in Table~\ref{tab:access_control}. Lastly, three-level smart contract mechanism explained in Figure~\ref{fig:double_column_architecture} simplifies the abstraction and shows the realistic deployement.
By integrating blockchain's immutable audit logs and a finite state machine, the system achieves both reliability and transparency, making it suitable for a wide range of applications in dynamic peer-to-peer environments.
Thus, the system efficiently manages complex relationships between peers, access permissions, and different peer states (benign, suspicious, and malicious).

\begingroup 

\textit{Limitations:} The current implementation is deployed on web simulation platform and lacks hardware setup. The design and evaluation of specialized hardware nodes to carry out compute-heavy token calculation on resource constrained peer nodes is part future hardware extension. Analytics engine for ML-based behavioral predictions is another known limitation. Although we have implemented several optimization techniques, lightweight ML deployed on live peer nodes remains an open area for ongoing development.

\endgroup

\section{Conclusion} \label{Sec:conclusion}

Trust enforcement in lightweight devices with scalable realistic implementation, rigorous security evaluations and reproducible code is lacking in the state-of-art. Our blockchain-based operational framework goes beyond the conceptual state-of-art architecture to showcase realistic mapping between resources, peer roles and standard threats in P2P network. It features token-based incentives to enforce compliance and security situations with multi-level abstractions in terms of dynamic state transitions (benign, suspicious, malicious) that is mapped to realistic STRIDE threat model. We are using hierarchical role assignments (Primary Group Heads, Secondary Group Heads, and Regular Members) for network scalability.

Our deployment showcases three types of Ethereum smart contracts ranging from policy monitoring to low-level access control using Truffle and Ganache. To ensure controllability and observability, we extensively tested these contracts using eight diverse and realistic test cases using Mocha and Chai. We have presented easily reproducible code for all benign and malicious threat situations. Our detailed evaluation uses comprehensive metrics: gas efficiency, execution time, response delays, scalability with simultaneous requests to evaluate resilience against variety of threat severity. 

The experimental results shows low gas costs (upto 1.4 million gas) and fast execution times (80–550 ms) with maximum of 100 nodes, confirming the design efficiency. The framework handles up to five concurrent requests with minimal performance impact, demonstrating strong scalability across roles and threat conditions. Our operational smart contract framework highlights the  strong system controllability with fine-grained, adaptive policy enforcement and observability in terms of clear monitoring and traceability of network activities.
In future, we want to integrate our work with ML-based analytics engines running on hardware testbeds with IoT devices serving as peer nodes.

\bibliographystyle{cas-model2-names}
\bibliography{references_test} 

\begin{thebibliography}{59}
\expandafter\ifx\csname natexlab\endcsname\relax\def\natexlab#1{#1}\fi
\providecommand{\url}[1]{\texttt{#1}}
\providecommand{\href}[2]{#2}
\providecommand{\path}[1]{#1}
\providecommand{\DOIprefix}{doi:}
\providecommand{\ArXivprefix}{arXiv:}
\providecommand{\URLprefix}{URL: }
\providecommand{\Pubmedprefix}{pmid:}
\providecommand{\doi}[1]{\href{http://dx.doi.org/#1}{\path{#1}}}
\providecommand{\Pubmed}[1]{\href{pmid:#1}{\path{#1}}}
\providecommand{\bibinfo}[2]{#2}
\ifx\xfnm\relax \def\xfnm[#1]{\unskip,\space#1}\fi
\bibitem[{Abdulrahman et~al.(2021)Abdulrahman, Alshehri and
  Cherif}]{abdulrahman2021survey}
\bibinfo{author}{Abdulrahman, E.}, \bibinfo{author}{Alshehri, S.},
  \bibinfo{author}{Cherif, A.}, \bibinfo{year}{2021}.
\newblock \bibinfo{title}{Blockchain-based access control for the internet of
  things: A survey}, in: \bibinfo{booktitle}{2021 IEEE Asia-Pacific Conference
  on Computer Science and Data Engineering (CSDE)}, pp. \bibinfo{pages}{1--8}.
\newblock \DOIprefix\doi{10.1109/CSDE53843.2021.9718468}.
\bibitem[{Ahsan and Pathan(2025)}]{ahsan2025comprehensive}
\bibinfo{author}{Ahsan, M.S.}, \bibinfo{author}{Pathan, A.S.K.},
  \bibinfo{year}{2025}.
\newblock \bibinfo{title}{A comprehensive survey on the requirements,
  applications, and future challenges for access control models in {I}o{T}: The
  state of the art}.
\newblock \bibinfo{journal}{IoT} \bibinfo{volume}{6}, \bibinfo{pages}{9}.
\bibitem[{Ali et~al.(2020)Ali, Ahmad, Cao, Khan, Cruickshank, Qazi and
  Ali}]{ali2020xdbauth}
\bibinfo{author}{Ali, G.}, \bibinfo{author}{Ahmad, N.}, \bibinfo{author}{Cao,
  Y.}, \bibinfo{author}{Khan, S.}, \bibinfo{author}{Cruickshank, H.},
  \bibinfo{author}{Qazi, E.A.}, \bibinfo{author}{Ali, A.},
  \bibinfo{year}{2020}.
\newblock \bibinfo{title}{xdbauth: Blockchain-based cross domain authentication
  and authorization framework for internet of things}.
\newblock \bibinfo{journal}{IEEE Access} \bibinfo{volume}{8},
  \bibinfo{pages}{58800--58816}.
\bibitem[{Ameer et~al.(2022)Ameer, Benson and Sandhu}]{hybrid_abac_rbac_2024}
\bibinfo{author}{Ameer, S.}, \bibinfo{author}{Benson, J.},
  \bibinfo{author}{Sandhu, R.}, \bibinfo{year}{2022}.
\newblock \bibinfo{title}{Hybrid approaches ({ABAC} and {RBAC}) toward secure
  access control in smart home {I}o{T}}.
\newblock \bibinfo{journal}{IEEE Transactions on Dependable and Secure
  Computing} \bibinfo{volume}{20}, \bibinfo{pages}{4032--4051}.
\bibitem[{Bakir et~al.(2021)Bakir, Krintz and Wolski}]{bakir2021caplets}
\bibinfo{author}{Bakir, F.}, \bibinfo{author}{Krintz, C.},
  \bibinfo{author}{Wolski, R.}, \bibinfo{year}{2021}.
\newblock \bibinfo{title}{Caplets: Resource aware, capability-based access
  control for {I}o{T}}, in: \bibinfo{booktitle}{2021 IEEE/ACM Symposium on Edge
  Computing (SEC)}, \bibinfo{organization}{IEEE}. pp.
  \bibinfo{pages}{106--120}.
\bibitem[{Bera et~al.(2020)Bera, Saha, Das and Vasilakos}]{bera2021}
\bibinfo{author}{Bera, B.}, \bibinfo{author}{Saha, S.}, \bibinfo{author}{Das,
  A.K.}, \bibinfo{author}{Vasilakos, A.V.}, \bibinfo{year}{2020}.
\newblock \bibinfo{title}{Designing blockchain-based access control protocol in
  {I}o{T}-enabled smart-grid system}.
\newblock \bibinfo{journal}{IEEE Internet of Things Journal}
  \bibinfo{volume}{8}, \bibinfo{pages}{5744--5761}.
\bibitem[{Bertino and Islam(2017)}]{bertino2017botnets}
\bibinfo{author}{Bertino, E.}, \bibinfo{author}{Islam, N.},
  \bibinfo{year}{2017}.
\newblock \bibinfo{title}{Botnets and internet of things security}.
\newblock \bibinfo{journal}{Computer} \bibinfo{volume}{50},
  \bibinfo{pages}{76--79}.
\bibitem[{Choksy et~al.(2023)Choksy, Chaurasia, Rao and
  Kumar}]{choksy2023attribute}
\bibinfo{author}{Choksy, P.}, \bibinfo{author}{Chaurasia, A.},
  \bibinfo{author}{Rao, U.P.}, \bibinfo{author}{Kumar, S.},
  \bibinfo{year}{2023}.
\newblock \bibinfo{title}{Attribute based access control (abac) scheme with a
  fully flexible delegation mechanism for {I}o{T} healthcare}.
\newblock \bibinfo{journal}{Peer-to-Peer Networking and Applications}
  \bibinfo{volume}{16}, \bibinfo{pages}{1445--1467}.
\bibitem[{Chopra et~al.(2019)Chopra, Gupta and Lambora}]{Chopra2019}
\bibinfo{author}{Chopra, K.}, \bibinfo{author}{Gupta, K.},
  \bibinfo{author}{Lambora, A.}, \bibinfo{year}{2019}.
\newblock \bibinfo{title}{Future internet: The internet of things-a literature
  review}, in: \bibinfo{booktitle}{2019 International Conference on Machine
  Learning, Big Data, Cloud and Parallel Computing (COMITCon)},
  \bibinfo{organization}{IEEE}. pp. \bibinfo{pages}{135--139}.
\bibitem[{Clark(2022)}]{clark2022enhancing}
\bibinfo{author}{Clark, N.K.}, \bibinfo{year}{2022}.
\newblock \bibinfo{title}{Enhancing an information-centric network of things at
  the internet edge with trust-based access control}, in:
  \bibinfo{booktitle}{2022 IEEE 8th World Forum on Internet of Things
  (WF-{I}o{T})}, \bibinfo{organization}{IEEE}. pp. \bibinfo{pages}{1--6}.
\bibitem[{Cruz et~al.(2018)Cruz, Kaji and Yanai}]{Cruz2018}
\bibinfo{author}{Cruz, J.P.}, \bibinfo{author}{Kaji, Y.},
  \bibinfo{author}{Yanai, N.}, \bibinfo{year}{2018}.
\newblock \bibinfo{title}{Rbac-{S}{C}: Role-based access control using smart
  contract}.
\newblock \bibinfo{journal}{IEEE Access} \bibinfo{volume}{6},
  \bibinfo{pages}{12240--12251}.
\bibitem[{Da~Xu et~al.(2014)Da~Xu, He and Li}]{xu2014iotindustries}
\bibinfo{author}{Da~Xu, L.}, \bibinfo{author}{He, W.}, \bibinfo{author}{Li,
  S.}, \bibinfo{year}{2014}.
\newblock \bibinfo{title}{Internet of things in industries: A survey}.
\newblock \bibinfo{journal}{IEEE Transactions on Industrial Informatics}
  \bibinfo{volume}{10}, \bibinfo{pages}{2233--2243}.
\bibitem[{Da~Xu et~al.(2021)Da~Xu, Lu and Li}]{Xu2021}
\bibinfo{author}{Da~Xu, L.}, \bibinfo{author}{Lu, Y.}, \bibinfo{author}{Li,
  L.}, \bibinfo{year}{2021}.
\newblock \bibinfo{title}{Embedding blockchain technology into {I}o{T} for
  security: A survey}.
\newblock \bibinfo{journal}{IEEE Internet of Things Journal}
  \bibinfo{volume}{8}, \bibinfo{pages}{10452--10473}.
\bibitem[{Ding et~al.(2019)Ding, Cao, Li, Fan and Li}]{ding2019novelabac}
\bibinfo{author}{Ding, S.}, \bibinfo{author}{Cao, J.}, \bibinfo{author}{Li,
  C.}, \bibinfo{author}{Fan, K.}, \bibinfo{author}{Li, H.},
  \bibinfo{year}{2019}.
\newblock \bibinfo{title}{A novel attribute-based access control scheme using
  blockchain for {I}o{T}}.
\newblock \bibinfo{journal}{IEEE Access} \bibinfo{volume}{7},
  \bibinfo{pages}{38431--38441}.
\bibitem[{Dorri et~al.(2017)Dorri, Kanhere, Jurdak and Gauravaram}]{Dorri2017}
\bibinfo{author}{Dorri, A.}, \bibinfo{author}{Kanhere, S.S.},
  \bibinfo{author}{Jurdak, R.}, \bibinfo{author}{Gauravaram, P.},
  \bibinfo{year}{2017}.
\newblock \bibinfo{title}{Blockchain for {I}o{T} security and privacy: The case
  study of a smart home}, in: \bibinfo{booktitle}{2017 IEEE international
  conference on pervasive computing and communications workshops (PerCom
  workshops)}, \bibinfo{organization}{IEEE}. pp. \bibinfo{pages}{618--623}.
\bibitem[{Dram\'{e}-Maign\'{e} et~al.(2021)Dram\'{e}-Maign\'{e}, Laurent,
  Castillo and Ganem}]{Drame2021ACMSurv}
\bibinfo{author}{Dram\'{e}-Maign\'{e}, S.}, \bibinfo{author}{Laurent, M.},
  \bibinfo{author}{Castillo, L.}, \bibinfo{author}{Ganem, H.},
  \bibinfo{year}{2021}.
\newblock \bibinfo{title}{Centralized, distributed, and everything in between:
  Reviewing access control solutions for the {I}o{T}}.
\newblock \bibinfo{journal}{ACM Comput. Surv.} \bibinfo{volume}{54}.
\newblock \URLprefix \url{https://doi.org/10.1145/3465170},
  \DOIprefix\doi{10.1145/3465170}.
\bibitem[{D’Orazio et~al.(2016)D’Orazio, Choo and
  Yang}]{darazio2017dataexfiltration}
\bibinfo{author}{D’Orazio, C.J.}, \bibinfo{author}{Choo, K.K.R.},
  \bibinfo{author}{Yang, L.T.}, \bibinfo{year}{2016}.
\newblock \bibinfo{title}{Data exfiltration from internet of things devices:
  ios devices as case studies}.
\newblock \bibinfo{journal}{IEEE Internet of Things Journal}
  \bibinfo{volume}{4}, \bibinfo{pages}{524--535}.
\bibitem[{Fotiou et~al.(2022)Fotiou, Siris, Polyzos, Kortesniemi and
  Lagutin}]{fotiou2022capabilities}
\bibinfo{author}{Fotiou, N.}, \bibinfo{author}{Siris, V.A.},
  \bibinfo{author}{Polyzos, G.C.}, \bibinfo{author}{Kortesniemi, Y.},
  \bibinfo{author}{Lagutin, D.}, \bibinfo{year}{2022}.
\newblock \bibinfo{title}{Capabilities-based access control for {I}o{T} devices
  using verifiable credentials}, in: \bibinfo{booktitle}{2022 IEEE Security and
  Privacy Workshops (SPW)}, \bibinfo{organization}{IEEE}. pp.
  \bibinfo{pages}{222--228}.
\bibitem[{Ghosh et~al.(2024)Ghosh, Mitra, Roy and Gupta}]{ghosh2024securing}
\bibinfo{author}{Ghosh, S.}, \bibinfo{author}{Mitra, R.}, \bibinfo{author}{Roy,
  I.}, \bibinfo{author}{Gupta, B.}, \bibinfo{year}{2024}.
\newblock \bibinfo{title}{Securing rc based p2p networks: A blockchain-based
  access control framework utilizing ethereum smart contracts for {I}o{T} and
  web 3.0}, in: \bibinfo{booktitle}{International Conference on Computer
  Applications in Industry and Engineering}, \bibinfo{organization}{Springer}.
  pp. \bibinfo{pages}{73--85}.
\bibitem[{Goswami et~al.(2021)Goswami, Singh, Saikia, Bora and
  Sharma}]{goswami2021tokencheck}
\bibinfo{author}{Goswami, S.}, \bibinfo{author}{Singh, R.},
  \bibinfo{author}{Saikia, N.}, \bibinfo{author}{Bora, K.K.},
  \bibinfo{author}{Sharma, U.}, \bibinfo{year}{2021}.
\newblock \bibinfo{title}{Tokencheck: towards deep learning based security
  vulnerability detection in erc-20 tokens}, in: \bibinfo{booktitle}{2021 IEEE
  region 10 symposium (TENSYMP)}, \bibinfo{organization}{IEEE}. pp.
  \bibinfo{pages}{1--8}.
\bibitem[{Gusmeroli et~al.(2013)Gusmeroli, Piccione and
  Rotondi}]{gusmeroli2013capbac}
\bibinfo{author}{Gusmeroli, S.}, \bibinfo{author}{Piccione, S.},
  \bibinfo{author}{Rotondi, D.}, \bibinfo{year}{2013}.
\newblock \bibinfo{title}{A capability-based security approach to manage access
  control in the internet of things}.
\newblock \bibinfo{journal}{Mathematical and Computer Modelling}
  \bibinfo{volume}{58}, \bibinfo{pages}{1189--1205}.
\bibitem[{Han et~al.(2022)Han, Zhang, Liu, Li, Xian, Wang, Mao and
  Chen}]{han2022}
\bibinfo{author}{Han, J.}, \bibinfo{author}{Zhang, Y.}, \bibinfo{author}{Liu,
  J.}, \bibinfo{author}{Li, Z.}, \bibinfo{author}{Xian, M.},
  \bibinfo{author}{Wang, H.}, \bibinfo{author}{Mao, F.}, \bibinfo{author}{Chen,
  Y.}, \bibinfo{year}{2022}.
\newblock \bibinfo{title}{A blockchain-based and sgx-enabled access control
  framework for {I}o{T}}.
\newblock \bibinfo{journal}{Electronics} \bibinfo{volume}{11},
  \bibinfo{pages}{2710}.
\bibitem[{Hu et~al.(2006)Hu, Ferraiolo and Kuhn}]{hu2015abac}
\bibinfo{author}{Hu, V.C.}, \bibinfo{author}{Ferraiolo, D.F.},
  \bibinfo{author}{Kuhn, D.R.}, \bibinfo{year}{2006}.
\newblock \bibinfo{title}{Assessment of Access Control Systems}.
\newblock \bibinfo{type}{Technical Report} \bibinfo{number}{NIST IR 7316}.
  National Institute of Standards and Technology.
\newblock \DOIprefix\doi{10.6028/NIST.IR.7316}.
\bibitem[{Hu et~al.(2018)Hu, Liyanage, Mansoor, Thilakarathna, Jourjon and
  Seneviratne}]{Hu2018BlockchainSC}
\bibinfo{author}{Hu, Y.}, \bibinfo{author}{Liyanage, M.},
  \bibinfo{author}{Mansoor, A.}, \bibinfo{author}{Thilakarathna, K.},
  \bibinfo{author}{Jourjon, G.}, \bibinfo{author}{Seneviratne, A.},
  \bibinfo{year}{2018}.
\newblock \bibinfo{title}{Blockchain-based smart contracts-applications and
  challenges}.
\newblock \bibinfo{journal}{arXiv preprint arXiv:1810.04699} .
\bibitem[{Kumar and Tripathi(2021)}]{kumar2021eblp}
\bibinfo{author}{Kumar, R.}, \bibinfo{author}{Tripathi, R.},
  \bibinfo{year}{2021}.
\newblock \bibinfo{title}{Scalable and secure access control policy for
  healthcare system using blockchain and enhanced bell–lapadula model}.
\newblock \bibinfo{journal}{Journal of Ambient Intelligence and Humanized
  Computing} \bibinfo{volume}{12}, \bibinfo{pages}{2321--2338}.
\bibitem[{Liang et~al.(2019)Liang, Tang, Long, Peng, Xu and
  Li}]{liang2019secure}
\bibinfo{author}{Liang, W.}, \bibinfo{author}{Tang, M.}, \bibinfo{author}{Long,
  J.}, \bibinfo{author}{Peng, X.}, \bibinfo{author}{Xu, J.},
  \bibinfo{author}{Li, K.C.}, \bibinfo{year}{2019}.
\newblock \bibinfo{title}{A secure fabric blockchain-based data transmission
  technique for industrial internet-of-things}.
\newblock \bibinfo{journal}{IEEE Transactions on Industrial Informatics}
  \bibinfo{volume}{15}, \bibinfo{pages}{3582--3592}.
\bibitem[{Lin et~al.(2017)Lin, Yu, Zhang, Yang, Zhang and Zhao}]{Lin2017}
\bibinfo{author}{Lin, J.}, \bibinfo{author}{Yu, W.}, \bibinfo{author}{Zhang,
  N.}, \bibinfo{author}{Yang, X.}, \bibinfo{author}{Zhang, H.},
  \bibinfo{author}{Zhao, W.}, \bibinfo{year}{2017}.
\newblock \bibinfo{title}{A survey on internet of things: Architecture,
  enabling technologies, security and privacy, and applications}.
\newblock \bibinfo{journal}{IEEE Internet of Things Journal}
  \bibinfo{volume}{4}, \bibinfo{pages}{1125--1142}.
\bibitem[{Liu et~al.(2020a)Liu, Han and Li}]{liu2020fabricoT}
\bibinfo{author}{Liu, H.}, \bibinfo{author}{Han, D.}, \bibinfo{author}{Li, D.},
  \bibinfo{year}{2020}a.
\newblock \bibinfo{title}{Fabric-{I}o{T}: A blockchain-based access control
  system in {I}o{T}}.
\newblock \bibinfo{journal}{IEEE Access} \bibinfo{volume}{8},
  \bibinfo{pages}{18207--18218}.
\bibitem[{Liu et~al.(2020b)Liu, Lu, Chen, Qu, O’Connor, Choo and
  Zhang}]{liu2020capbac}
\bibinfo{author}{Liu, Y.}, \bibinfo{author}{Lu, Q.}, \bibinfo{author}{Chen,
  S.}, \bibinfo{author}{Qu, Q.}, \bibinfo{author}{O’Connor, H.},
  \bibinfo{author}{Choo, K.K.R.}, \bibinfo{author}{Zhang, H.},
  \bibinfo{year}{2020}b.
\newblock \bibinfo{title}{Capability-based {I}o{T} access control using
  blockchain}.
\newblock \bibinfo{journal}{Digital Communications and Networks}
  \bibinfo{volume}{6}, \bibinfo{pages}{453--461}.
\bibitem[{Lu et~al.(2021)Lu, Fu, Jiang and Lio}]{lu2021fine}
\bibinfo{author}{Lu, X.}, \bibinfo{author}{Fu, S.}, \bibinfo{author}{Jiang,
  C.}, \bibinfo{author}{Lio, P.}, \bibinfo{year}{2021}.
\newblock \bibinfo{title}{A fine-grained {I}o{T} data access control scheme
  combining attribute-based encryption and blockchain}.
\newblock \bibinfo{journal}{Security and communication networks}
  \bibinfo{volume}{2021}, \bibinfo{pages}{5308206}.
\bibitem[{Ma et~al.(2019)Ma, Shi and Li}]{Ma2019}
\bibinfo{author}{Ma, M.}, \bibinfo{author}{Shi, G.}, \bibinfo{author}{Li, F.},
  \bibinfo{year}{2019}.
\newblock \bibinfo{title}{Privacy-oriented blockchain-based distributed key
  management architecture for hierarchical access control in the {I}o{T}
  scenario}.
\newblock \bibinfo{journal}{IEEE Access} \bibinfo{volume}{7},
  \bibinfo{pages}{34045--34059}.
\bibitem[{Maesa et~al.(2019)Maesa, Mori and Ricci}]{maesa2019blockchain}
\bibinfo{author}{Maesa, D.D.F.}, \bibinfo{author}{Mori, P.},
  \bibinfo{author}{Ricci, L.}, \bibinfo{year}{2019}.
\newblock \bibinfo{title}{A blockchain based approach for the definition of
  auditable access control systems}.
\newblock \bibinfo{journal}{Computers \& Security} \bibinfo{volume}{84},
  \bibinfo{pages}{93--119}.
\bibitem[{Nakamura et~al.(2020)Nakamura, Enokido and
  Takizawa}]{capbac_flow_control_2020}
\bibinfo{author}{Nakamura, S.}, \bibinfo{author}{Enokido, T.},
  \bibinfo{author}{Takizawa, M.}, \bibinfo{year}{2020}.
\newblock \bibinfo{title}{Information flow control based on the capbac
  (capability-based access control) model in the {I}o{T}}.
\newblock \bibinfo{journal}{International Journal of Information Security and
  Privacy} \bibinfo{volume}{14}, \bibinfo{pages}{1--15}.
\bibitem[{Nakamura et~al.(2019)Nakamura, Zhang, Sasabe and
  Kasahara}]{nakamura2019capability}
\bibinfo{author}{Nakamura, Y.}, \bibinfo{author}{Zhang, Y.},
  \bibinfo{author}{Sasabe, M.}, \bibinfo{author}{Kasahara, S.},
  \bibinfo{year}{2019}.
\newblock \bibinfo{title}{Capability-based access control for the internet of
  things: An ethereum blockchain-based scheme}, in: \bibinfo{booktitle}{2019
  IEEE Global Communications Conference (GLOBECOM)},
  \bibinfo{organization}{IEEE}. pp. \bibinfo{pages}{1--6}.
\bibitem[{Ning et~al.(2014)Ning, Liu and Yang}]{ning2014aggregated}
\bibinfo{author}{Ning, H.}, \bibinfo{author}{Liu, H.}, \bibinfo{author}{Yang,
  L.T.}, \bibinfo{year}{2014}.
\newblock \bibinfo{title}{Aggregated-proof based hierarchical authentication
  scheme for the internet of things}.
\newblock \bibinfo{journal}{IEEE Transactions on Parallel and Distributed
  Systems} \bibinfo{volume}{26}, \bibinfo{pages}{657--667}.
\bibitem[{Novo(2018)}]{Novo2018}
\bibinfo{author}{Novo, O.}, \bibinfo{year}{2018}.
\newblock \bibinfo{title}{Blockchain meets {I}o{T}: An architecture for
  scalable access management in {I}o{T}}.
\newblock \bibinfo{journal}{IEEE Internet of Things Journal}
  \bibinfo{volume}{5}, \bibinfo{pages}{1184--1195}.
\bibitem[{Oktian and Lee(2020)}]{okian2020borderchain}
\bibinfo{author}{Oktian, Y.E.}, \bibinfo{author}{Lee, S.G.},
  \bibinfo{year}{2020}.
\newblock \bibinfo{title}{Borderchain: Blockchain-based access control
  framework for the internet of things endpoint}.
\newblock \bibinfo{journal}{IEEE Access} \bibinfo{volume}{8},
  \bibinfo{pages}{177759--177774}.
\bibitem[{Ouaddah et~al.(2016)Ouaddah, Abou~Elkalam and
  Ait~Ouahman}]{ouaddah2016fairaccess}
\bibinfo{author}{Ouaddah, A.}, \bibinfo{author}{Abou~Elkalam, A.},
  \bibinfo{author}{Ait~Ouahman, A.}, \bibinfo{year}{2016}.
\newblock \bibinfo{title}{Fairaccess: a new blockchain-based access control
  framework for the internet of things}.
\newblock \bibinfo{journal}{Security and communication networks}
  \bibinfo{volume}{9}, \bibinfo{pages}{5943--5964}.
\bibitem[{Pal et~al.(2019)Pal, Rabehaja, Hill, Hitchens and
  Varadharajan}]{pal2019delegation}
\bibinfo{author}{Pal, S.}, \bibinfo{author}{Rabehaja, T.},
  \bibinfo{author}{Hill, A.}, \bibinfo{author}{Hitchens, M.},
  \bibinfo{author}{Varadharajan, V.}, \bibinfo{year}{2019}.
\newblock \bibinfo{title}{On the integration of blockchain to the internet of
  things for enabling access right delegation}.
\newblock \bibinfo{journal}{IEEE Internet of Things Journal}
  \bibinfo{volume}{7}, \bibinfo{pages}{2630--2639}.
\bibitem[{Palattella et~al.(2016)Palattella, Dohler, Grieco, Rizzo, Torsner,
  Engel and Ladid}]{palattella2016iot5g}
\bibinfo{author}{Palattella, M.R.}, \bibinfo{author}{Dohler, M.},
  \bibinfo{author}{Grieco, A.}, \bibinfo{author}{Rizzo, G.},
  \bibinfo{author}{Torsner, J.}, \bibinfo{author}{Engel, T.},
  \bibinfo{author}{Ladid, L.}, \bibinfo{year}{2016}.
\newblock \bibinfo{title}{Internet of things in the 5g era: Enablers,
  architecture, and business models}.
\newblock \bibinfo{journal}{IEEE Journal on Selected Areas in Communications}
  \bibinfo{volume}{34}, \bibinfo{pages}{510--527}.
\bibitem[{Putra et~al.(2021)Putra, Dedeoglu, Kanhere, Jurdak and
  Ignjatovic}]{putra2021trust}
\bibinfo{author}{Putra, G.D.}, \bibinfo{author}{Dedeoglu, V.},
  \bibinfo{author}{Kanhere, S.S.}, \bibinfo{author}{Jurdak, R.},
  \bibinfo{author}{Ignjatovic, A.}, \bibinfo{year}{2021}.
\newblock \bibinfo{title}{Trust-based blockchain authorization for {I}o{T}}.
\newblock \bibinfo{journal}{IEEE Transactions on Network and Service
  Management} \bibinfo{volume}{18}, \bibinfo{pages}{1646--1658}.
\bibitem[{Qin et~al.(2021)Qin, Huang, Yang and Li}]{qin2021lbac}
\bibinfo{author}{Qin, X.}, \bibinfo{author}{Huang, Y.}, \bibinfo{author}{Yang,
  Z.}, \bibinfo{author}{Li, X.}, \bibinfo{year}{2021}.
\newblock \bibinfo{title}{Lbac: A lightweight blockchain-based access control
  scheme for the internet of things}.
\newblock \bibinfo{journal}{Information Sciences} \bibinfo{volume}{554},
  \bibinfo{pages}{222--235}.
\bibitem[{Ragothaman et~al.(2023)Ragothaman, Wang, Rimal and
  Lawrence}]{ragothaman2023access}
\bibinfo{author}{Ragothaman, K.}, \bibinfo{author}{Wang, Y.},
  \bibinfo{author}{Rimal, B.}, \bibinfo{author}{Lawrence, M.},
  \bibinfo{year}{2023}.
\newblock \bibinfo{title}{Access control for {I}o{T}: A survey of existing
  research, dynamic policies and future directions}.
\newblock \bibinfo{journal}{Sensors} \bibinfo{volume}{23},
  \bibinfo{pages}{1805}.
\bibitem[{Reyna et~al.(2018)Reyna, Mart{\'\i}n, Chen, Soler and
  D{\'\i}az}]{Reyna2018}
\bibinfo{author}{Reyna, A.}, \bibinfo{author}{Mart{\'\i}n, C.},
  \bibinfo{author}{Chen, J.}, \bibinfo{author}{Soler, E.},
  \bibinfo{author}{D{\'\i}az, M.}, \bibinfo{year}{2018}.
\newblock \bibinfo{title}{On blockchain and its integration with {I}o{T}.
  challenges and opportunities}.
\newblock \bibinfo{journal}{Future Generation Computer Systems}
  \bibinfo{volume}{88}, \bibinfo{pages}{173--190}.
\bibitem[{Roy et~al.(2022)Roy, Rahimi, Mitra and Kaluvakuri}]{roy2022efficient}
\bibinfo{author}{Roy, I.}, \bibinfo{author}{Rahimi, N.},
  \bibinfo{author}{Mitra, R.}, \bibinfo{author}{Kaluvakuri, S.},
  \bibinfo{year}{2022}.
\newblock \bibinfo{title}{Efficient secured public-key based data lookup and
  multicast protocols with anonymity in rc-based two-level hierarchical
  structured p2p network}, in: \bibinfo{booktitle}{Proceedings of 35th
  International Conference on Computer Applications in Industry and
  Engineering}, pp. \bibinfo{pages}{62--71}.
\bibitem[{Sandhu(1998)}]{sandhu1998role}
\bibinfo{author}{Sandhu, R.S.}, \bibinfo{year}{1998}.
\newblock \bibinfo{title}{Role-based access control}, in:
  \bibinfo{booktitle}{Advances in Computers}. \bibinfo{publisher}{Elsevier}.
  volume~\bibinfo{volume}{46}, pp. \bibinfo{pages}{237--286}.
\bibitem[{Servos and Osborn(2017)}]{servos2017current}
\bibinfo{author}{Servos, D.}, \bibinfo{author}{Osborn, S.L.},
  \bibinfo{year}{2017}.
\newblock \bibinfo{title}{Current research and open problems in attribute-based
  access control}.
\newblock \bibinfo{journal}{{A}{C}{M} {C}omputing {S}urveys (CSUR)}
  \bibinfo{volume}{49}, \bibinfo{pages}{1--45}.
\bibitem[{Shammar et~al.(2022)Shammar, Zahary and
  Al-Shargabi}]{shammar2022attribute}
\bibinfo{author}{Shammar, E.A.}, \bibinfo{author}{Zahary, A.T.},
  \bibinfo{author}{Al-Shargabi, A.A.}, \bibinfo{year}{2022}.
\newblock \bibinfo{title}{An attribute-based access control model for internet
  of things using hyperledger fabric blockchain}.
\newblock \bibinfo{journal}{Wireless Communications and Mobile Computing}
  \bibinfo{volume}{2022}, \bibinfo{pages}{6926408}.
\bibitem[{Singh et~al.(2024)Singh, Rani and Kumar}]{rbac_data_processing_2024}
\bibinfo{author}{Singh, J.}, \bibinfo{author}{Rani, S.},
  \bibinfo{author}{Kumar, V.}, \bibinfo{year}{2024}.
\newblock \bibinfo{title}{Role-based access control (rbac) enabled secure and
  efficient data processing framework for {I}o{T} networks}.
\newblock \bibinfo{journal}{International Journal of Communication Networks and
  Information Security} \bibinfo{volume}{16}, \bibinfo{pages}{91--103}.
\bibitem[{Thakare et~al.(2020)Thakare, Lee, Kumar, Nikam and
  Kim}]{thakare2020parbac}
\bibinfo{author}{Thakare, A.}, \bibinfo{author}{Lee, E.},
  \bibinfo{author}{Kumar, A.}, \bibinfo{author}{Nikam, V.B.},
  \bibinfo{author}{Kim, Y.G.}, \bibinfo{year}{2020}.
\newblock \bibinfo{title}{{P}{A}{R}{B}{A}{C}: Priority-attribute-based
  {R}{B}{A}{C} model for azure {I}o{T} cloud}.
\newblock \bibinfo{journal}{IEEE Internet of Things Journal}
  \bibinfo{volume}{7}, \bibinfo{pages}{2890--2900}.
\bibitem[{Wang et~al.(2017)Wang, Wang, Zhang and Cao}]{Wang2017}
\bibinfo{author}{Wang, J.}, \bibinfo{author}{Wang, H.}, \bibinfo{author}{Zhang,
  H.}, \bibinfo{author}{Cao, N.}, \bibinfo{year}{2017}.
\newblock \bibinfo{title}{Trust and attribute-based dynamic access control
  model for internet of things}, in: \bibinfo{booktitle}{2017 International
  conference on cyber-enabled distributed computing and knowledge discovery
  (CyberC)}, \bibinfo{organization}{IEEE}. pp. \bibinfo{pages}{342--345}.
\bibitem[{Wang et~al.(2021)Wang, Xu, Zhang, Sun and Benslimane}]{wang2021}
\bibinfo{author}{Wang, P.}, \bibinfo{author}{Xu, N.}, \bibinfo{author}{Zhang,
  H.}, \bibinfo{author}{Sun, W.}, \bibinfo{author}{Benslimane, A.},
  \bibinfo{year}{2021}.
\newblock \bibinfo{title}{Dynamic access control and trust management for
  blockchain-empowered {I}o{T}}.
\newblock \bibinfo{journal}{IEEE Internet of Things Journal}
  \bibinfo{volume}{9}, \bibinfo{pages}{12997--13009}.
\bibitem[{Xu et~al.(2020)Xu, He, Li, Jiang and Qin}]{xu2020bdssfa}
\bibinfo{author}{Xu, H.}, \bibinfo{author}{He, Q.}, \bibinfo{author}{Li, X.},
  \bibinfo{author}{Jiang, B.}, \bibinfo{author}{Qin, K.}, \bibinfo{year}{2020}.
\newblock \bibinfo{title}{Bdss-fa: A blockchain-based data security sharing
  platform with fine-grained access control}.
\newblock \bibinfo{journal}{IEEE Access} \bibinfo{volume}{8},
  \bibinfo{pages}{87552--87561}.
\bibitem[{Yaqoob et~al.(2017)Yaqoob, Ahmed, Hashem, Ahmed, Gani, Imran and
  Guizani}]{yaqoob2017iotarchitecture}
\bibinfo{author}{Yaqoob, I.}, \bibinfo{author}{Ahmed, E.},
  \bibinfo{author}{Hashem, I.A.T.}, \bibinfo{author}{Ahmed, A.I.A.},
  \bibinfo{author}{Gani, A.}, \bibinfo{author}{Imran, M.},
  \bibinfo{author}{Guizani, M.}, \bibinfo{year}{2017}.
\newblock \bibinfo{title}{{I}nternet of {T}hings {A}rchitecture: Recent
  advances, taxonomy, requirements, and open challenges}.
\newblock \bibinfo{journal}{IEEE Wireless Communications} \bibinfo{volume}{24},
  \bibinfo{pages}{10--16}.
\bibitem[{Yavari et~al.(2017)Yavari, Panah, Georgakopoulos, Jayaraman and van
  Schyndel}]{yavari2017scalable}
\bibinfo{author}{Yavari, A.}, \bibinfo{author}{Panah, A.S.},
  \bibinfo{author}{Georgakopoulos, D.}, \bibinfo{author}{Jayaraman, P.P.},
  \bibinfo{author}{van Schyndel, R.}, \bibinfo{year}{2017}.
\newblock \bibinfo{title}{Scalable role-based data disclosure control for the
  internet of things}, in: \bibinfo{booktitle}{2017 IEEE 37th International
  Conference on Distributed Computing Systems (ICDCS)},
  \bibinfo{organization}{IEEE}. pp. \bibinfo{pages}{2226--2233}.
\bibitem[{Yu et~al.(2022)Yu, Chen and Wang}]{yu2022blockchain}
\bibinfo{author}{Yu, H.}, \bibinfo{author}{Chen, T.}, \bibinfo{author}{Wang,
  J.}, \bibinfo{year}{2022}.
\newblock \bibinfo{title}{A blockchain-based access control mechanism for
  {I}o{T}}, in: \bibinfo{booktitle}{Proceedings of the 2022 6th International
  Conference on Electronic Information Technology and Computer Engineering},
  pp. \bibinfo{pages}{25--30}.
\bibitem[{Zaidi et~al.(2021)Zaidi, Shah, Khattak, Maple, Rauf, El-Sherbeeny and
  El-Meligy}]{zaidi2021attribute}
\bibinfo{author}{Zaidi, S.Y.A.}, \bibinfo{author}{Shah, M.A.},
  \bibinfo{author}{Khattak, H.A.}, \bibinfo{author}{Maple, C.},
  \bibinfo{author}{Rauf, H.T.}, \bibinfo{author}{El-Sherbeeny, A.M.},
  \bibinfo{author}{El-Meligy, M.A.}, \bibinfo{year}{2021}.
\newblock \bibinfo{title}{An attribute-based access control for {I}o{T} using
  blockchain and smart contracts}.
\newblock \bibinfo{journal}{Sustainability} \bibinfo{volume}{13},
  \bibinfo{pages}{10556}.
\bibitem[{Zhang et~al.(2018)Zhang, Kasahara, Shen, Jiang and Wan}]{Zhang2018}
\bibinfo{author}{Zhang, Y.}, \bibinfo{author}{Kasahara, S.},
  \bibinfo{author}{Shen, Y.}, \bibinfo{author}{Jiang, X.},
  \bibinfo{author}{Wan, J.}, \bibinfo{year}{2018}.
\newblock \bibinfo{title}{Smart contract-based access control for the
  {I}nternet of {T}hings}.
\newblock \bibinfo{journal}{IEEE Internet of Things Journal}
  \bibinfo{volume}{6}, \bibinfo{pages}{1594--1605}.
\bibitem[{Zhang et~al.(2025)Zhang, Wang, Jiang, Zhou, Xu and
  Liu}]{zhang2025efficient}
\bibinfo{author}{Zhang, Y.}, \bibinfo{author}{Wang, X.A.},
  \bibinfo{author}{Jiang, W.}, \bibinfo{author}{Zhou, M.}, \bibinfo{author}{Xu,
  X.}, \bibinfo{author}{Liu, H.}, \bibinfo{year}{2025}.
\newblock \bibinfo{title}{An efficient and secure data audit scheme for
  cloud-based ehrs with recoverable and batch auditing.}
\newblock \bibinfo{journal}{Computers, Materials \& Continua}
  \bibinfo{volume}{83}.

\end{thebibliography}
\section*{Acknowledgments}
We would thank all the collaborators who supported this research. A special thanks to the reviewers of Caine 2024.

\end{document}